\documentclass[useAMS,usenatbib]{mn2e}

\usepackage{natbib}

\usepackage{latexsym,graphicx}
\usepackage{color}
\usepackage{fixltx2e}
\usepackage{verbatim} \usepackage{float}
\usepackage{amsmath,amssymb}
\usepackage{times}

\usepackage{setspace}
\usepackage{rotating}
\usepackage{pdflscape}
\usepackage{subfig}

\def\apgt{\ {\raise-.5ex\hbox{$\buildrel>\over\sim$}}\ }
\def\aplt{\ {\raise-.5ex\hbox{$\buildrel<\over\sim$}}\ }

\let\oldhat\hat
\renewcommand{\hat}[1]{\oldhat{\mathbf{#1}}}

\title[Burst occurrence in MYSOs]{Burst occurrence in young massive stellar objects}

\author[D. M.-A.~Meyer et al.]
       {D. M.-A.~Meyer$^{1}$\thanks{E-mail: dmameyer.astro@gmail.com}, \textcolor{black}{E.~I.~Vorobyov}$^{2,3}$, \textcolor{black}{V.~G.~Elbakyan}$^{3}$, B.~Stecklum$^{4}$, J.~Eisl\" offel$^{4}$
       \newauthor and A.~M.~Sobolev$^{5}$ \\
       $^{1}$Astrophysics Group, School of Physics and Astronomy, University of Exeter, Exeter EX4 4QL, United Kingdom \\
       $^{2}$Department of Astrophysics, The University of Vienna, Vienna, A-1180, Austria \\
       $^{3}$Research Institute of Physics, Southern Federal University, Stachki 194, Rostov-on-Don, 344090, Russia \\        
       $^{4}$Th\" uringer Landessternwarte Tautenburg, Sternwarte 5, D-07778 Tautenburg, Germany \\ 
       $^{5}$Astronomical Observatory, Institute for Natural Sciences and Mathematics, Ural Federal University, Ekaterinburg, 620000, Russian Federation  \\        
       }

\begin{document}

\date{Received; accepted}

\maketitle
   
\label{firstpage}

\begin{abstract} 
Episodic accretion-driven outbursts are an extreme manifestation of accretion variability. 
It has been proposed that the development of gravitational instabilities in the 
proto-circumstellar medium of massive young stellar objects (MYSOs) can lead to such luminous 
bursts, when clumps of fragmented accretion discs migrate onto the \textcolor{black}{star}. 
We simulate the early evolution of MYSOs formed by the gravitational collapse of rotating $100\, \rm M_{\odot}$ 
pre-stellar cores and analyze the characteristics of the bursts that episodically accompany their  
strongly time-variable protostellar lightcurve. 
We predict that MYSOs spend $\approx 10^{3}\, \rm yr$ ($\approx 1.7\%$) of their modelled early 
$60\, \rm kyr$ experiencing eruptive phases, during which the peak luminosity exceeds the quiescent 
pre-burst values by factors from 2.5 to more than 40. Throughout \textcolor{black}{these} short time period\textcolor{black}{s}, they can 
acquire a substential fraction (up to $\approx 50\, \%$) of their zero-age-main sequence mass. 
\textcolor{black}{Our findings show that fainter bursts are more common than brighter ones.} 
%
%
We discuss our results in the context of the known bursting MYSOs, e.g. NGC6334I-MM1 and S255IR-NIRS3,  
and propose that these monitored bursts are part of a long-time ongoing series of eruptions, which might, in the future, 
be followed by other luminous flares.  
\end{abstract}



\section{Introduction}
\label{sect:intro}

Massive stars, i.e. stellar objects heavier than $8\, \rm M_{\odot}$, are well-established 
in their preponderant role in the cycle of matter of the 
interstellar medium of galaxies~\citep{langer_araa_50_2012}. Throughout successive stellar 
evolutionary phases, from the (post-)main sequence~\citep{meyer} to the supernova phase~\citep{meyer_mnras_450_2015}, massive stars are mainly 
responsible for chemically enriching their surroundings and driving turbulence, which further 
regulates local star formation. 
Gravitational interactions and/or mass-exchanges with close companions modify 
their internal structures and play an important role in the recent, realistic 
picture of massive stars evolution~\citep{sana_sci_337_2012,schneider_apj_805_2015}. 
\textcolor{black}{
Furthermore, observations of massive young stellar objects (MYSOs, i.e. protostars heavier than $8\, \rm M_{\odot}$ and having a bolometric 
luminosity $L_{\rm bol}\ge 5\times 10^{3}\, \rm L_{\odot}$) demonstrated that binarity can 
already exist during the early formation phase of high-mass stellar objects, see also~\citet{chini_424_mnras_2012,2013A&A...550A..27M,kraus_apj_835_2017}.  
Hence, the subsequent main sequence evolution of MYSOs in the Hertzsprung-Russel diagram is a 
function of mechanisms that are at work already since the initial stage of their formation phase. 
}
The evolution of massive stars starts well before the onset of their 
main sequence phase, at a formation epoch which has been shown to be highly sensitive to mass accretion 
from the protocircumstellar medium~\citep{hosokawa_apj_691_2009,haemmerle_585_aa_2016,haemmerle_458_mnras_2016,
haemmerle_602_aap_2017}.

Our knowledge of the formation of proto-O stars is mainly driven by observations. 
~\citet{fuente_aa_366_2001,testi_2003} and~\citet{cesaroni_natur_444_2006} noticed that massive star formation 
resembles any other low-mass star formation processes. Indeed their surroundings revealed 
the presence of converging accretion flows~\citep{keto_apj_637_2006}, jets~\citep{caratti_aa_573_2015,burns_mnras_467_2017,arXiv180102211B,purser_mnras_475_2018,2018arXiv180311413S}, 
bipolar radiation-driven bubbles filled by ionizing radiation~\citep{cesaroni_aa_509_2010,purser_mnras_460_2016} and collimated pulsed 
precessing jets~\citep[e.g. in Cepheus A, see][]{Cunningham_apj_692_2009,reiter_mnras_470_2017}. 
Numerically, global simulations of monolithic gravitational collapse of isolated pre-stellar cores 
predicted the formation of dense accretion discs~\citep{krumholz_apj_656_2007}, however, such models still 
clearly suffer from a lack of spatial resolution. Despite advances in supercomputing, it is difficult to fully 
resolve the central disc, from the outer infalling pre-stellar core material to the stellar surface, 
including the substructures, such as dense portions of spiral arms and eventually dense gaseous clumps 
developing with the arms. 
This is mainly due to the fact that their complex thermodynamic structure makes the so-called 
thin-disc approach less realistic than in the context of, e.g. low-mass star formation, and obliges to tackle the problem 
with three-dimensional models~\citep{1998MNRAS.298...93B,2002ApJ...569..846Y,peters_apj_711_2010,seifried_mnras_417_2011,
harries_mnras_448_2015,klassen_apj_823_2016,harries_2017}. 
Therefore, parameter studies are numerically not conceivable and justify analytic treatments of the problem~\citep{tanaka_apj_835_2017}. 
High-resolution, self-consistent simulations focusing on the early evolution of MYSOs revealed the 
existence of accretion-driven outbursts in the context of young massive stars caused by gravitational 
fragmentation of dense spiral arms followed by migration of gaseous clumps on to the forming 
star~\citep{meyer_mnras_464_2017}. Furthermore, some gaseous clumps can collapse before reaching the star, 
initiating secondary (low-mass) star formation and forming spectroscopic proto-binaries by direct disc 
fragmentation~\citep{meyer_mnras_473_2018}. 
These findings orient further investigations on the evolution of massive protostars to their inner 
proto-circumstellar medium, where mass accretion on to MYSOs occurs.

\begin{figure}
        \hspace{-3mm}%
        \includegraphics[width=0.475\textwidth]{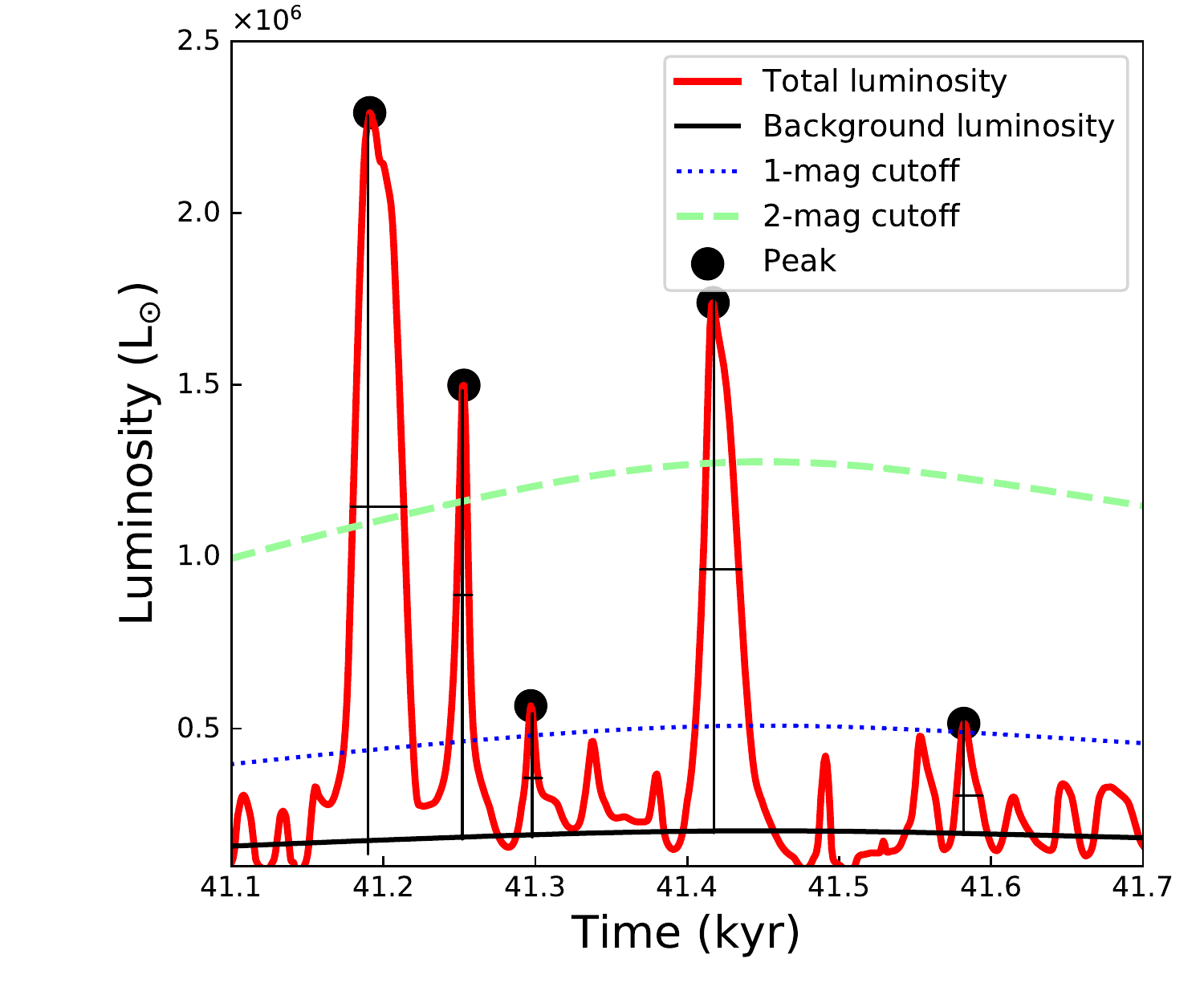}
        \caption{ 
                 Illustration of the burst filtering method on a series of peaks. 
                 On each bursts, their prominence and duration at half maximum are indicated. 
                 }      
        \label{fig:illustration}  
\end{figure}

\textcolor{black}{This} is our current picture of the formation of massive stars: theoretical models and available 
observational data indicate that it is nothing but a scaled-up version of 
low-mass star formation, retouched by corrections accounting for extreme photospheric conditions via 
radiation feedback~\citep{richling_aa_327_1997,rosen_jcp_330_2017} and its coupling to magnetic fields~\citep{seifried_mnras_417_2011}. 
However, the opaque environment in which they form keeps direct disc observation of deeply-embedded proto-O-typed stars still 
out of reach of modern instrumentation, although good candidates have been 
identified~\citep{johnston_apj_813_2015,ilee_mnras_462_2016,forgan_mnras_463_2016,cesaroni_aa_602_2017}. 
A more systematic campaign with ALMA failed in finding evidence of fragmented disc structures within a radius 
of $1000\, \rm au$ from young protostars~\citep{beuther_aa_603_2017}, that were numerically predicted~\citep{meyer_mnras_473_2018}.  
%

While a growing number of Keplerian discs around MYSOs is directly detected~\citep{johnston_apj_813_2015,ilee_mnras_462_2016,forgan_mnras_463_2016}, some of them 
including signs of substructures potentially linked to gravitational instability~\citep{2018arXiv180410622G}, 
the fragmentation scenario proposes an explanation that bridges the gap between strong variability in the lightcurve of 
MYSOs (an observable) and a precursor mechanism - disc fragmentation at $\sim 100$ au from the growing protostar. 
It is the vivid manifestation of the effectiveness of gravitational non-linear instabilities in the radiatively 
cooled protostellar surroundings~\citep{gammie_apj_462_1996} of new-born hot stars, that is regulated by the 
mechanical heating of dense spiral arms~\citep{roger_mnras_423_2012}. 
Accretion bursts caused by disk fragmentation followed by inward migration of dense clumps onto the star have 
the advantage to be well-understood in the context of young low-mass 
stars~\citep{vorobyov_apj_633_2005,vorobyov_apj_805_2015,machida_mnras_413_2011,zhao_mnras_473_2018,nayakshin_mnras_426_2012}.  
%
%
In the high-mass context, bursts constitute a formidable insight into the recent past 
of the circumprotostellar medium, allowing us to probe the not-yet directly 
observable inner disc regions. Coincidently, such disc-mediated bursts have 
recently been observed, e.g. from the MYSOs S255IR 
NIRS\,3~\citep{fujisawa_atel_2015,stecklum_ATel_2016,caratti_nature_2016}  or 
from NGC6334I-MM1~\citep{2017arXiv170108637H}.

\begin{figure*}
        \centering
        \begin{minipage}[b]{ 0.9\textwidth}
                \includegraphics[width=1.0\textwidth]{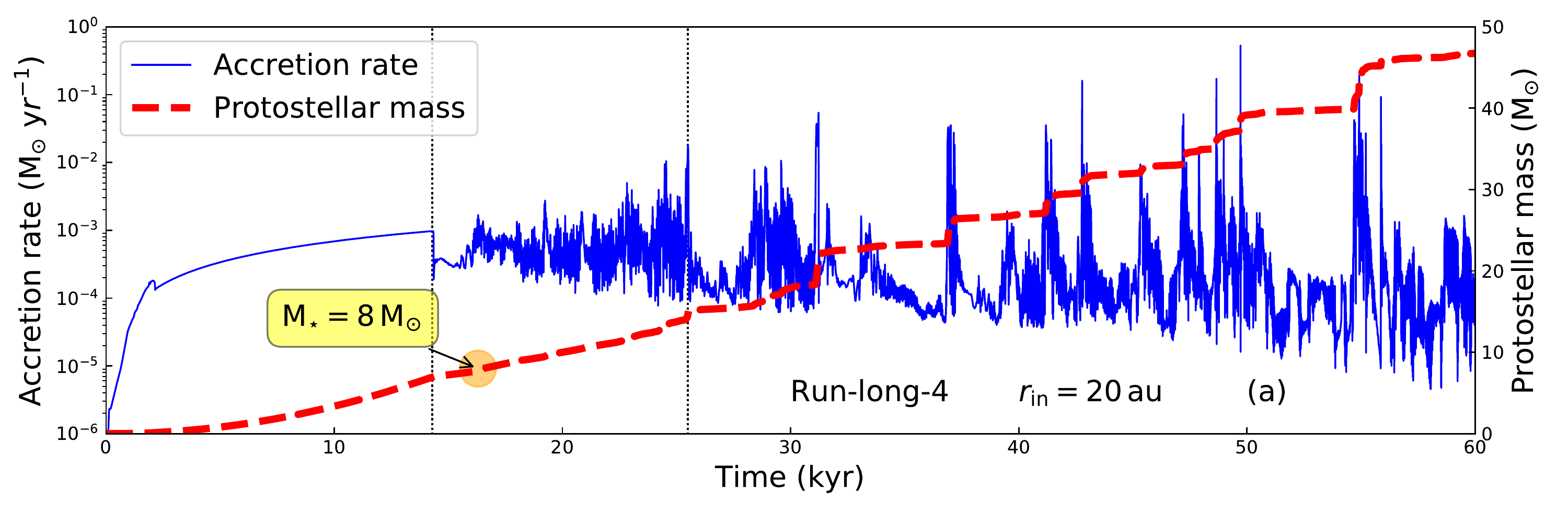}
        \end{minipage} \\   
        \begin{minipage}[b]{ 0.9\textwidth}
                \includegraphics[width=1.0\textwidth]{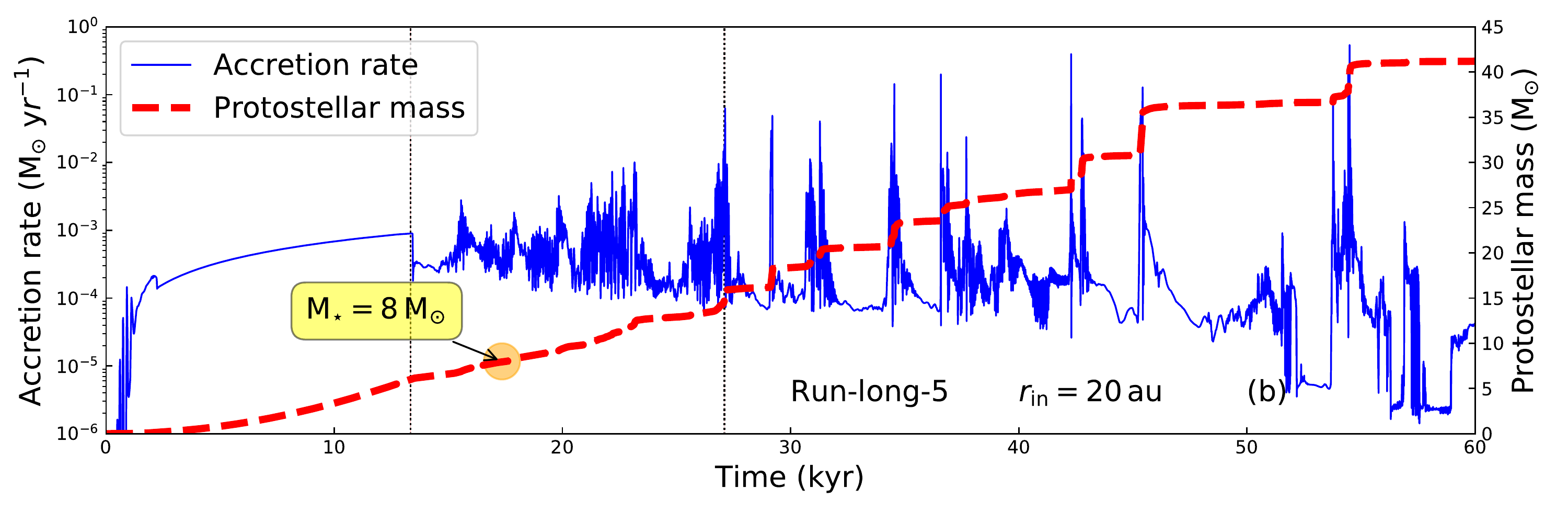}
        \end{minipage} \\              
        \centering
        \begin{minipage}[b]{ 0.9\textwidth}
                \includegraphics[width=1.0\textwidth]{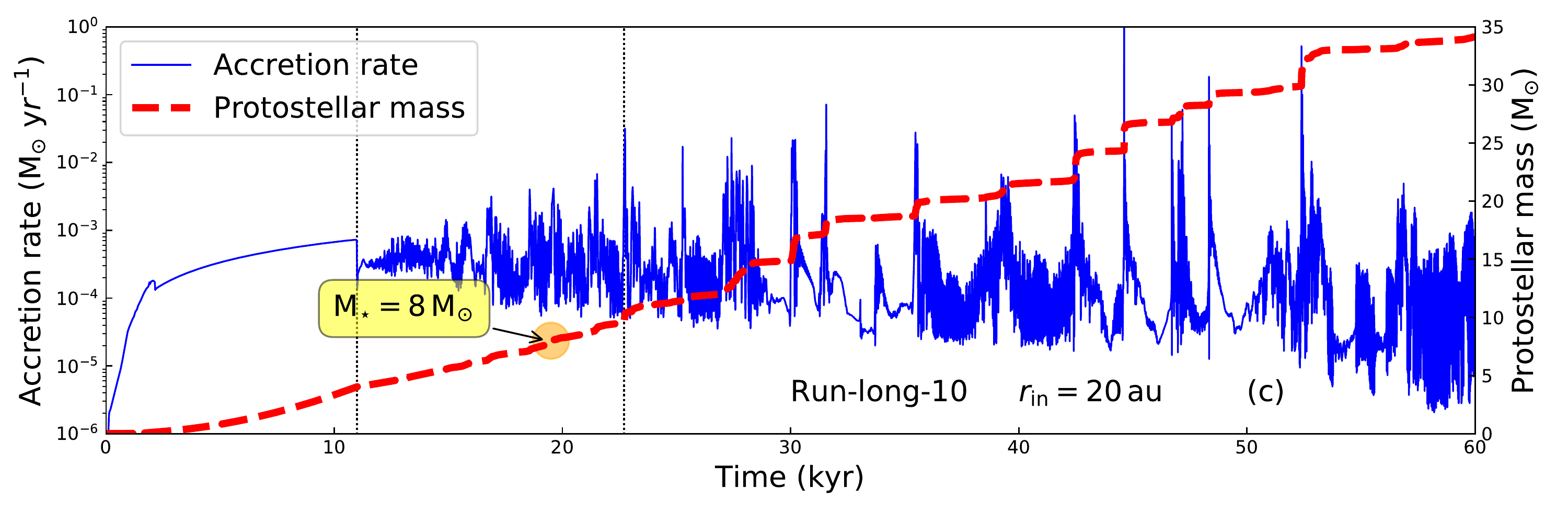}
        \end{minipage} \\   
        \begin{minipage}[b]{ 0.9\textwidth}
                \includegraphics[width=1.0\textwidth]{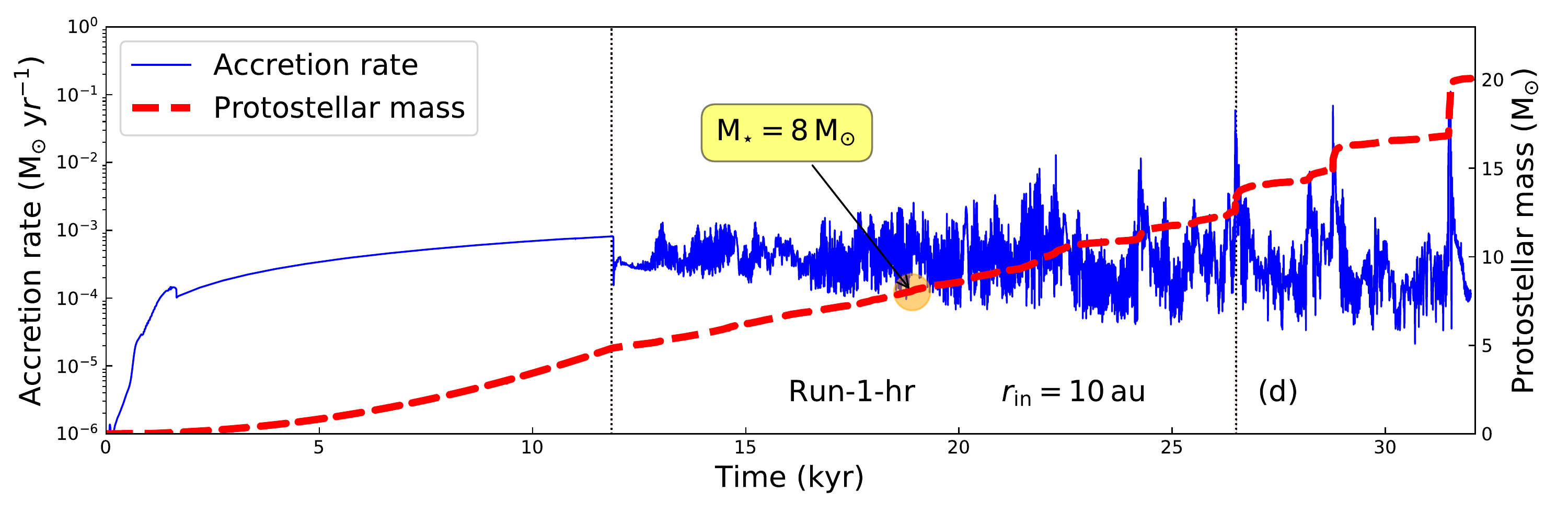}
        \end{minipage} \\        
        \caption{ 
                 Mass accretion rate (thin solid line, in $\rm M_{\odot}\, \rm yr^{-1}$) 
                 \textcolor{black}{and protostellar mass evolution (thick dashed line, in $\rm M_{\odot}$)}
                 in our simulations of $100\, \rm M_{\odot}$ solid-body-rotating pre-stellar cores 
                 with $\beta$=$4\%$ (a), $5\%$ (b), $10\%$ (c) and Run1-hr (d).
                 }      
        \label{fig:gen_plots_1}  
\end{figure*}

This work aims at studying the properties and occurrence of accretion-driven outbursts from massive stars. 
We perform three-dimensional \textcolor{black}{gravito-radiation-hydrodynamical} simulations of collapsing $100\, \rm M_{\odot}$ rotating pre-stellar cores 
to model the long-term evolution of gravitationally-unstable accretion discs around MYSOs. 
We extract from those simulations both the highly-variable accretion rate histories and luminosity 
curves interspersed with episodic bursts. The bursts are generated by the 
accretion of dense material from spiral arms in the disk and/or by the \textcolor{black}{inward} migration of fragments as described 
in~\citet{meyer_mnras_473_2018}. This picture is well-supported by observations of the spiral filament that is feeding the 
candidate disc around the young massive stellar object MM1-Main~\citep{maud_467_mnras_2017}, by the 
observation of a young stellar object fed by a gaseous clump in the double-cored system G350.69-0.49~\citep{chen_apj_835_2017}. 
We separate the bursts of our modelled MYSOs using the method developed 
in the context of FU-Orionis bursts~\citep{vorobyov_apj_805_2015,2018arXiv180106707V}. 
The intensity and duration of individual bursts are analysed and compared with each other. We evaluate their occurrence 
during the early pre-main sequence phase of stellar evolution and estimate 
the amount of time our MYSOs spend experiencing accretion bursts. 
%

This study is organized as follows. In Section 2, we review the methods and initial conditions that we used to 
perform numerical hydrodynamical simulations of the circumstellar medium of our massive protostars. 
The obtained total internal luminosities of young massive protostars, 
exhibiting violent luminous spikes, are analysed in order to study the  
burst occurence throughout their early pre-main sequence star's live. The accretion rate evolution onto the MYSOs, 
the analysis of their luminosity histories and their occurence are presented in Section 3, \textcolor{black}{and 
we investigate the effects of stellar inertia onto the burst activity in Section 4.} In Section 5, we 
detail the different types of accretion bursts happening throughout the early formation phase of MYSOs, investigate 
the effects of the burst phenomenon in the growth of young massive protostars, further discuss our results in 
the context of observations. Finally, we conclude in Section 6.


\begin{table}
	\centering
	\caption{
	Initial characteristics of our $100\, \rm M_{\odot}$ solid-body-rotating pre-stellar cores. 
	The table gives the inner sink cell radius $r_{\rm in}$, the ratio of rotational-to-gravitational 
	energy $\beta$, final simulation time $t_{\rm end}$ \textcolor{black}{and indicates if the stellar inertia 
	is included in the models}.   
	}
	\begin{tabular}{lcccr}
	\hline
	${\rm {Models}}$             & $r_{\rm in}$ ($\rm au$)   &  $\beta$ ($\%$)  &  $t_{\rm end}$ ($\rm kyr$) & $\textcolor{black}{{\rm Wobbling}}$ \\ 
	\hline    
	${\rm Run-long-4}^{(a)}$       & $20$                        &  $4$             &  $60$             &   $no$         \\  
	${\rm Run-long-5}^{(a)}$       & $20$                        &  $5$             &  $60$             &   $no$         \\  
	${\rm Run-long-10}^{(a)}$      & $20$                        &  $10$            &  $60$             &   $no$         \\  	
	${\rm Run 1-hr}^{(b)}$         & $10$                        &  $4$             &  $32$             &   $no$         \\
	${\rm Run-without}^{(a)}$      & $12$                        &  $4$             &  $40$             &   $no$         \\	
	${\rm Run-with}^{(a)}$         & $12$                        &  $4$             &  $40$             &   $yes$        \\
	\hline    
	\end{tabular}
\label{tab:models}\\
\footnotesize{ ${(a)}$This work, ${(b)}$~\citet{meyer_mnras_473_2018} }\\
\end{table}

\section{Methods}
\label{sect:methods}

In this section, we provide a concise description of the \textcolor{black}{gravito-radiation-hydrodynamical} simulations utilised to 
derive the mass transport rate onto our forming high-mass stars and we briefly summarise the subsequent 
analysis method of their total luminosity history.

\begin{table*}
	\centering
	\caption{
	Summary of burst characteristics. 
	$N_{\mathrm{bst}}$ is the number of bursts at a given magnitude cut-off.  
	$L_{\mathrm{max}}/L_{\mathrm{min}}/L_{\mathrm{mean}}$ are the maximum, minimum and mean burst luminosities, respectively. 
	Similarly, $\dot{M}_{\mathrm{max}}/\dot{M}_{\mathrm{min}}/\dot{M}_{\mathrm{mean}}$ are the maximum, minimum and mean 
	accretion rates through the central sink cell and 
	$t_{\mathrm{bst}}^{\mathrm{max}}$/$t_{\mathrm{bst}}^{\mathrm{min}}$/$t_{\mathrm{bst}}^{\mathrm{mean}}$
	are the maximum, minimum and mean bursts duration, while 
	$t_{\mathrm{bst}}^{\mathrm{tot}}$ is the integrated bursts duration time throughout the star's live. 
	}
        \begin{tabular}{lcccccr}
        \hline  
        Model &  & $N_{\mathrm{bst}}$  & $L_{\mathrm{max}}/L_{\mathrm{min}}/L_{\mathrm{mean}}$   ($10^{5}\, \rm L_{\odot}\, \rm yr^{-1}$) & $\dot{M}_{\mathrm{max}}/\dot{M}_{\mathrm{min}}/\dot{M}_{\mathrm{mean}}$ ($\rm M_{\odot}\, \rm yr^{-1}$) & $t_{\mathrm{bst}}^{\mathrm{max}}$/$t_{\mathrm{bst}}^{\mathrm{min}}$/$t_{\mathrm{bst}}^{\mathrm{mean}}$ $\rm (yr)$  & $t_{\mathrm{bst}}^{\mathrm{tot}}$ $\rm (yr)$\tabularnewline
        \hline
        \multicolumn{7}{c}{\textbf{1-mag cutoff}}\tabularnewline
Run-long-4\%  &    & 34 & 14.7 / 0.56  / 7.39 & 0.018 / 0.003 / 0.007 &  39 / 5 / 14 & 471\tabularnewline
Run-long-5\%  &   & 21 & 15.1 / 0.067 / 4.4  & 0.019 / 0.001 / 0.008 & 88 / 9 / 26 & 553\tabularnewline
Run-long-10\% &   & 49 & 13.1 / 0.054 / 3.78 & 0.023 / 5.13$\times10^{-4}$ / 0.006 & 94 / 5 / 16 & 790\tabularnewline
Run-hr        &  & 24 & 2.9  / 0.06  / 0.54 & 0.029 / 0.001 / 0.008 & 100 / 5 / 51 & 1231\tabularnewline
        \textbf{Total all models} &   & \textbf{128} & \textbf{15.1 / 0.054 / 4.03} & \textbf{0.029 / 5.13$\times10^{-4}$ / 0.007} & \textbf{100 / 5 / 27} & \textbf{761}\tabularnewline
	\multicolumn{7}{c}{\textbf{-}}\tabularnewline
{\rm Run-without}  &   & 9 & 4.31 / 0.11  / 1.49 & 0.0177 / 0.0022 / 0.0081   & 107 / 9 / 38  & 340\tabularnewline
{\rm Run-with} &   & 6 & 3.60 / 0.123 / 1.19 & 0.0262 / 0.0024 / 0.0131   & 66  / 9 / 29  & 173\tabularnewline        
	\multicolumn{7}{c}{\textbf{2-mag cutoff}}\tabularnewline
Run-long-4\%  &   & 22 & 45.6 / 10.5 / 25.7 & 0.036 / 0.012 / 0.022 & 55 / 3 / 13 & 285\tabularnewline
Run-long-5\%  &   & 6  & 35.7 / 6.4$ $ / 15.3 & 0.049 / 0.010 / 0.027 & 17 / 6 / 12 & 69\tabularnewline
Run-long-10\% &   & 13 & 28 / 0.15 / 7.65 & 0.049 / 0.003 / 0.021 & 74 / 6 / 20 & 264\tabularnewline
Run-hr            &   & 3  & 4.26 / 1.59 / 2.76 & 0.059 / 0.035 / 0.050 & 61 / 30 /44 & 131\tabularnewline
        \textbf{Total all models} &  & \textbf{44} & \textbf{45.6 / 0.15 / 12.9} & \textbf{0.059 / 0.003 / 0.03} & \textbf{  74 / 3 / 22  } & \textbf{187}\tabularnewline
	\multicolumn{7}{c}{\textbf{-}}\tabularnewline
{\rm Run-without}  &   & 5  & 8.05 / 1.90 / 4.64 & 0.0479 / 0.0241 / 0.0343 & 34 / 5 / 18 & 88\tabularnewline	
{\rm Run-with}    &   & 5  & 9.80 / 1.57 / 4.12 & 0.0567 / 0.0175 / 0.0397 & 33 / 7 / 21 & 105\tabularnewline        
        \multicolumn{7}{c}{\textbf{3-mag cutoff}}\tabularnewline
Run-long-4\%  &    & 5 & 65.7 / 13.5 / 48.8 & 0.054 / 0.035 / 0.044 & 29 / 4 / 12 & 60\tabularnewline   
Run-long-5\%  &    & 2 & 37.4 / 4.99 / 21.2 & 0.062 / 0.045 / 0.053 & 35 / 8 / 22 & 43\tabularnewline
Run-long-10\% &    & 3 & 38.5 / 9.24 / 23.2 & 0.071 / 0.038 / 0.056 & 8 / 3 / 6 & 17\tabularnewline
Run-hr            &    & 2 & 17.9 / 5.28 / 11.6 & 0.111 / 0.069 / 0.090 & 51 / 50 / 51 & 101\tabularnewline
        \textbf{Total all models} &  & \textbf{12} & \textbf{65.7 / 4.99 / 26.2} & \textbf{0.111 / 0.035 / 0.061} & \textbf{51 / 3 / 23} & \textbf{55}\tabularnewline
	\multicolumn{7}{c}{\textbf{-}}\tabularnewline
{\rm Run-without}  &    & 2 & 12.2 / 8.55 / 10.4 & 0.105 / 0.083 / 0.094  & 6  / 4 / 5 & 10\tabularnewline        
{\rm Run-with}    &    & 2 & 17.8 / 16.2 / 17.0 & 0.121 / 0.059 / 0.0895 & 11 / 7 / 9 & 18\tabularnewline         
        \multicolumn{7}{c}{\textbf{4-mag cutoff}}\tabularnewline
Run-long-4\%  &    & 5 & 745 / 140.5 / 308 & 0.524 / 0.093 / 0.226 & 9 / 2 / 5 & 27\tabularnewline
Run-long-5\%  &    & 8 & 644.3 / 37.5 / 221.3 & 0.438 / 0.080 / 0.215 & 10 / 2 / 4 & 32\tabularnewline
Run-long-10\% &    & 4 & 432.4 / 100 / 261.9 & 0.925 / 0.101 / 0.424 & 7 / 2 / 4 & 14\tabularnewline
Run-hr            &    & - & - & - & - & -\tabularnewline
        \textbf{Total all models} &  & \textbf{17} & \textbf{745 / 37.5 / 263.7} & \textbf{0.925 / 0.080 / 0.288} & \textbf{10 / 2 / 4} & \textbf{24}\tabularnewline
	\multicolumn{7}{c}{\textbf{-}}\tabularnewline
{\rm Run-without}  &    & - & -                  & -                  & -         & -\tabularnewline        
{\rm Run-with}    &    & 1 & 51.2 / 51.2 / 51.2 & 0.16 / 0.16 / 0.16 & 4 / 4 / 4 & 4\tabularnewline       
        \hline    
        \end{tabular}
\label{tab:1}
\end{table*}

\subsection{\textcolor{black}{Numerical model}}
\label{sect:hydro}

The \textcolor{black}{numerical} simulations are \textcolor{black}{performed} using a midplane-symmetric computational domain, that we 
initialise with a $100\, \rm M_{\odot}$ solid-body-rotating pre-stellar core of uniform temperature 
$T_{\rm c}=10\, \rm K$ and of \textcolor{black}{an initially spherically symmetric} density distribution profiled as $\rho(r)\propto r^{\beta_{\rho}}$, with $\beta_{\rho}=-3/2$ and 
$r$ the radial coordinate. 
Its inner radius $r_{\rm in}$ constitutes a semi-permeable sink cell fixed onto the origin of the domain 
and its outer radius, assigned to outflow boundary conditions, is located at $R_{\rm c}=0.1\, \rm pc$. 
The grid mesh maps the domain $[r_{\rm in},R_{\rm c}]\times[0,\pi/2]\times[0,2\pi]$ 
made of $N_{\rm r}=138\times\,N_{\rm \theta}=21\times\,N_{\rm \phi}=138$ grid zones minimum. It expands 
logarithmically along the radial direction $r$, goes as a cosine in the polar direction $\theta$ and is uniformly 
spaced along the azimuthal direction $\phi$. 
Such grid intrinsically saves computing resources while high-spatially resolving the inner region of 
the midplane with a reduced total number of grid zones.  
We follow the gravitational collapse of the pre-stellar core and the early formation and evolution of the 
subsequent circumstellar accretion disc that surrounds the protostar. 
We calculate the material loss $\dot{M}$ through the sink cell as directly being the accretion rate onto the 
protostar, whose properties \textcolor{black}{(e.g., stellar radius and photospheric luminosity)} are time-dependently 
updated using the pre-calculated protostellar evolutionary tracks of~\citet{hosokawa_apj_691_2009}. 
\textcolor{black}{Hence, our estimate of the protostellar radius are in accordance with the calculations of~\citet{hosokawa_apj_691_2009}, 
as in~\citet{meyer_mnras_464_2017,meyer_mnras_473_2018}}. 
The different models are therefore characterised by the \textcolor{black}{radius} of the sink-cell $r_{\rm in}$ and the spin of the 
pre-stellar core that is parametrised by its ratio of rotational-by-gravitational energy $\beta$. 
\textcolor{black}{
Two additional simulations are performed in order to test the validity of our assumption consisting in neglecting the stellar 
inertia~\citep{regaly_aa_601_2017,Hirano_2017Science}. The comparison is effectuated in our Section~4. 
}
We use a slightly larger sink-cell radius ($r_{\rm in}=20\, \rm au$) than in our 
pilot study~\citep[see][]{meyer_mnras_464_2017}, which allows us to reach longer integration times 
$t_{\rm end}$ without dealing with dramatic time-step restrictions in the innermost grid zones. Such models, 
whose characteristics are summarised in Tab.~\ref{tab:models}, permit to obtain a large burst population, \textcolor{black}{which}  
is more suitable for our study.

\begin{figure*}
        \centering
        \begin{minipage}[b]{ 0.8\textwidth}  \centering
                \includegraphics[width=1.0\textwidth]{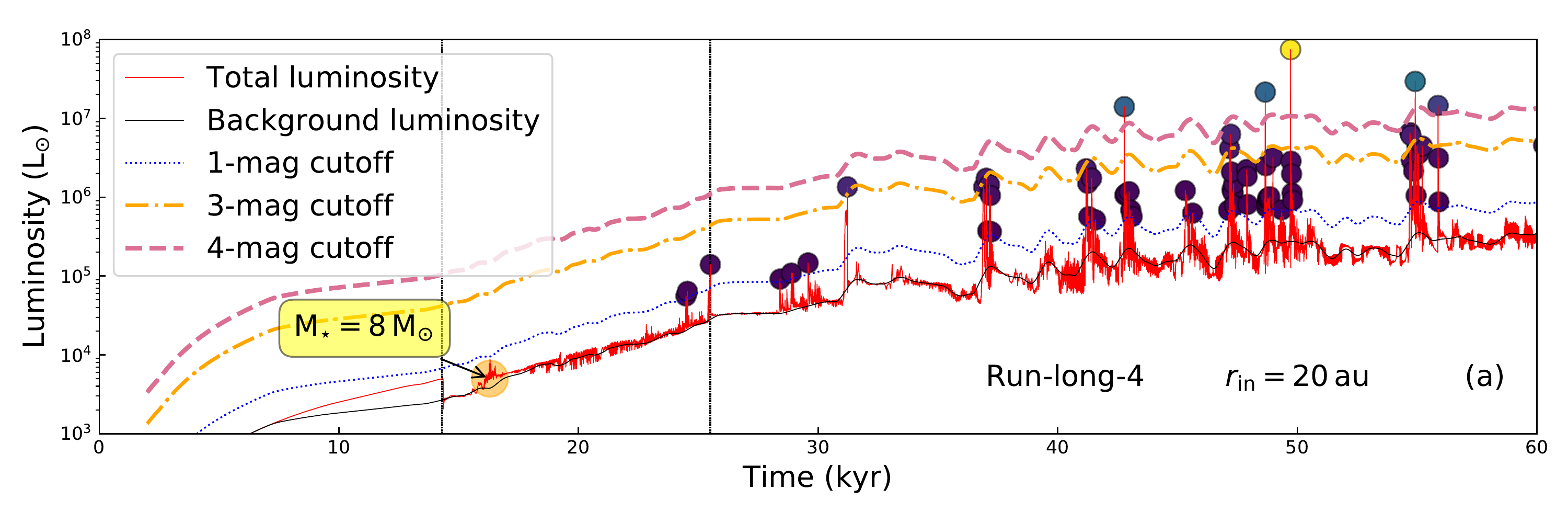}
        \end{minipage} \\   
                \centering
        \begin{minipage}[b]{ 0.8\textwidth}  \centering
                \includegraphics[width=1.0\textwidth]{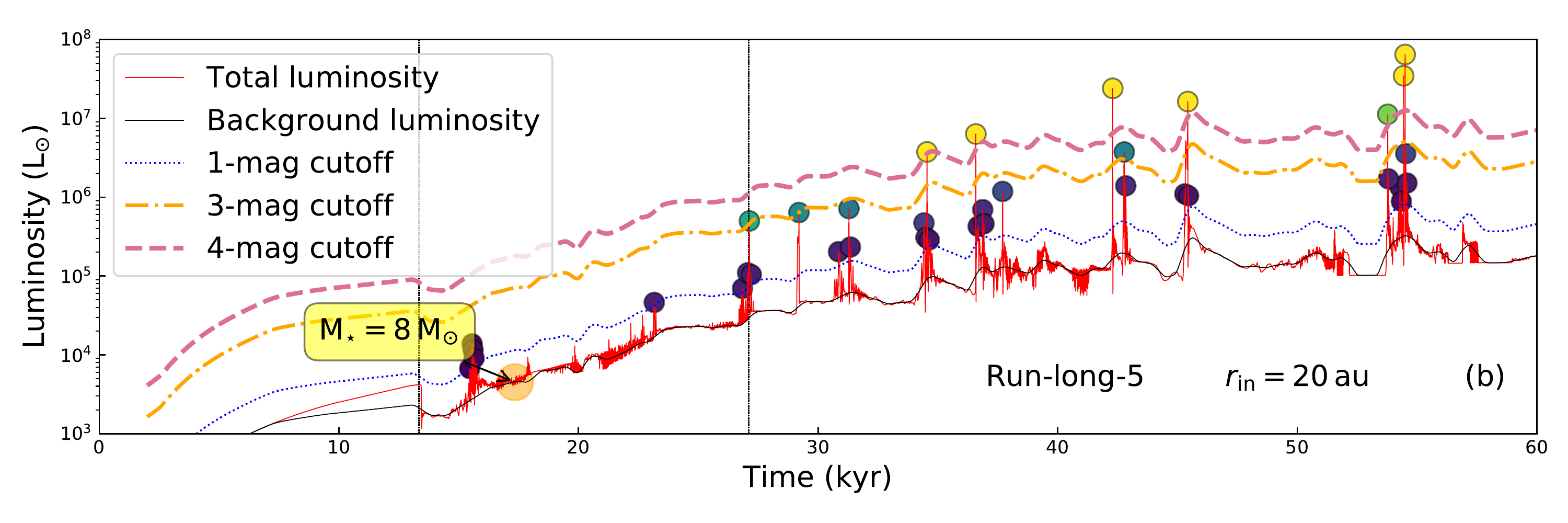}
        \end{minipage} \\              
        \centering
        \begin{minipage}[b]{ 0.8\textwidth}  \centering
                \includegraphics[width=1.0\textwidth]{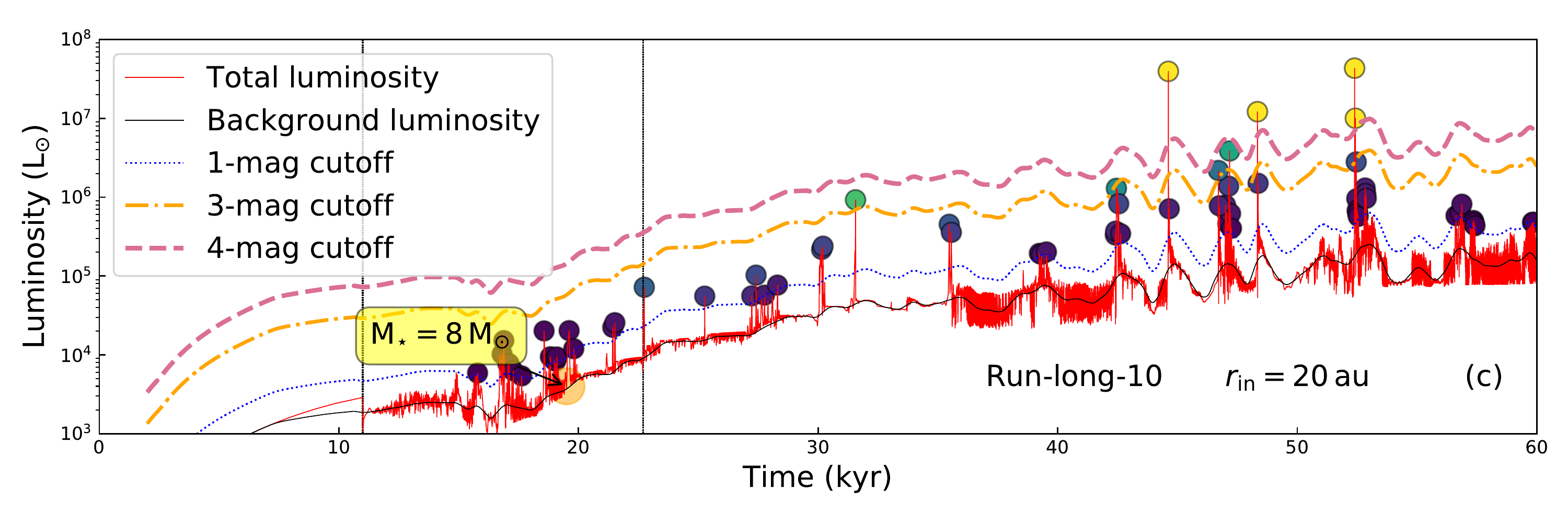}
        \end{minipage} \\ 
                \centering
        \begin{minipage}[b]{ 0.8\textwidth}  \centering
                \includegraphics[width=1.0\textwidth]{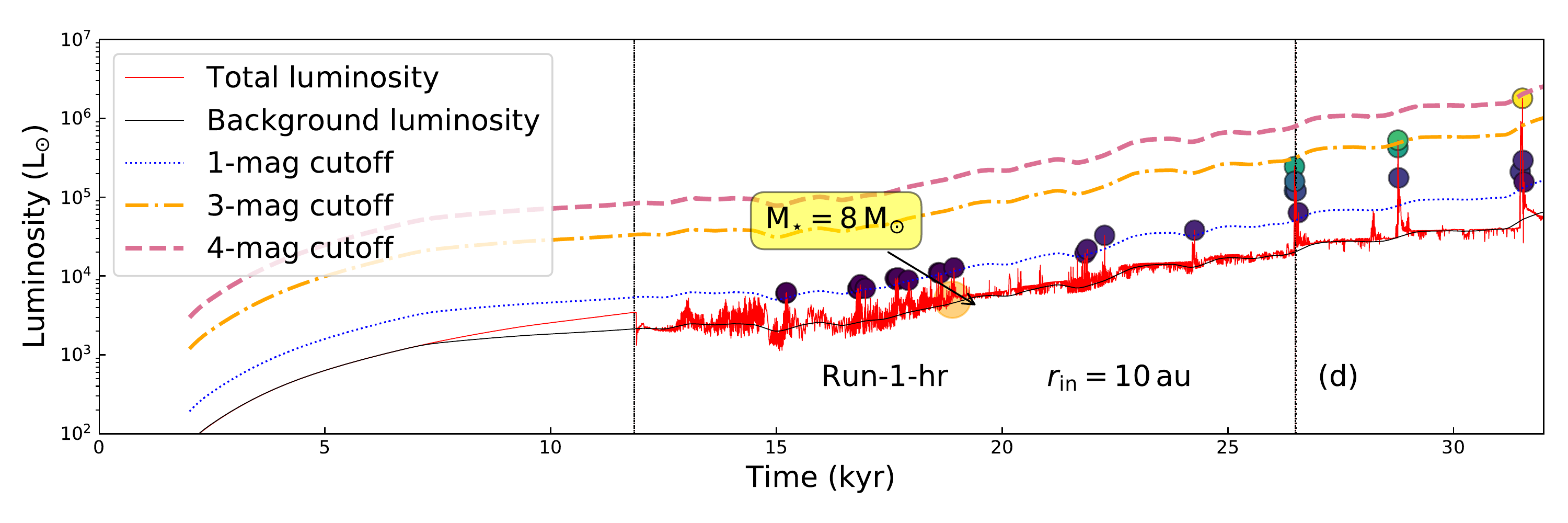} 
        \end{minipage} \\    
                \centering
        \begin{minipage}[b]{ 0.8\textwidth}  \centering
                \includegraphics[width=1.0\textwidth]{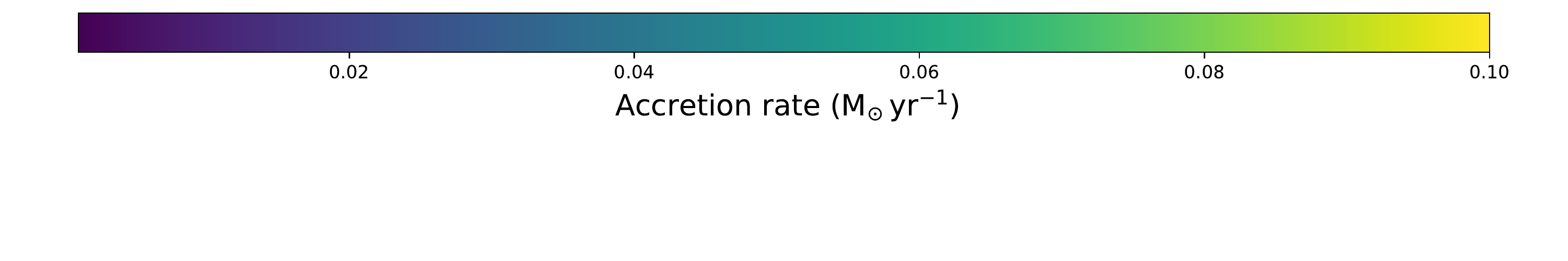}  
        \end{minipage}   
        \caption{ 
                 Total luminosity (thin solid red line, in $\rm L_{\odot}$) as a function of time (in $\rm kyr$) 
                 of our models with $\beta$=$4\%$ (a), $5\%$ (b), $10\%$ (c) and Run1-hr (d). 
                 \textcolor{black}{The panels show the background luminosity (thin solid black line) and the dotted lines indicate the 
                 luminosity greater than 2.5, $\approx 6.3$ and $\approx 39.0$ times the background luminosity (1,2- and 4-mag cutoff), respectively.} 
                 The color bar indicates the accretion rate for the filled circles. 
                 }      
        \label{fig:gen_plots_2}  
\end{figure*}

We solve the equations of gravito-radiation-hydrodynamics with the 
{\sc pluto} code\footnote{http://plutocode.ph.unito.it/}~\citep{mignone_apj_170_2007,migmone_apjs_198_2012}. 
Direct proto-stellar irradiation feedback and radiation transport in the accretion disc are taken into account 
within the gray approximation with the scheme 
of~\citet{kolb_aa_559_2013}\footnote{http://www.tat.physik.uni-tuebingen.de/~\,pluto/pluto\_radiation/} adapted in the fashion 
of~\citet{meyer_mnras_473_2018}. 
This two-fold algorithm ray-traces photon packages from the protostellar surface and flux-limited 
diffuses their propagation into the disc. Although other sophisticated methods have 
been developed for non-uniform Cartesian coordinate systems~\citep{klassen_apj_823_2016,rosen_jcp_330_2017}, 
it allows us to accurately treat both the inner heating and the outer cooling of our irradiated discs~\citep{vaidya_apj_742_2011}.  
Equivalent radiation-hydrodynamics methods have also been presented in 
e.g.~\citet{commercon_aa_529_2011},~\citet{flock_aa_560_2013} and~\citet{bitsch_aa_564_2014}. 
The opacity description as well as the estimate of the local dust properties are similar as in~\citet{meyer_mnras_473_2018}, 
where gas and dust temperature are calculated assuming the equilibrium between the silicate grains temperature 
and the total radiation field. 
We model the stellar gravity by calculating the total gravitational potential of the central protostar and include the  
self-gravity of the gas\footnote{https://shirano.as.utexas.edu/SV.html} with the method of~\citet{black_apj_199_1975} 
by solving the Poisson equation using the PETSC library\footnote{https://www.mcs.anl.gov/petsc/}. 
As in the other papers of this series, we neglect \textcolor{black}{turbulent} viscosity by assuming that the most efficient angular momentum 
transport mechanism are the gravitational torques generated once a self-gravitating disc has formed after the 
collapse~\citep[see also][]{hosokawa_2015}. 
We refer the reader interested in further reading about the numerical method to~\citet{meyer_mnras_473_2018}.

\subsection{Internal luminosities of accreting protostars} 
\label{sect:stat}

The internal luminosity history of our accreting young stars is calculated as the sum of their photospheric 
luminosity $L_{\star}$, interpolated from the protostellar tracks of~\citet{hosokawa_apj_691_2009}, plus 
the accretion luminosity $L_{\rm acc} = G M_{\star} \dot{M} / 2R_{\star}$, where $G$ is the 
universal gravitational constant, $M_{\star}$ is the stellar mass, $\dot{M}$ is the accretion rate onto the protostar 
and $R_{\star}$ is the protostellar radius, respectively. 
Using pre-calculated tracks is a good compromise between the direct (but computationally-costly) coupling 
of the gravito-radiation-hydrodynamics simulations to a stellar evolution code, which updates in real time the stellar 
properties by solving the stellar structure and the simple treatment of $L_{\star}$ as a function of the 
stellar effective temperature. 
We analyse the total luminosity histories using four models, namely runs 
Run-long-4\%, Run-long-5\% and Run-long-10\%, plus the model Run1-hr of~\citet{meyer_mnras_473_2018}. 
While Run-long-4\%, Run-long-5\% and Run-long-10\% form a homogeneous ensemble of simulations that assume similar 
grid resolution and sink-cell radius, model Run1-hr differs by its smaller sink-cell radius and higher grid 
resolution (see Tab.~\ref{tab:models}). 
The main limitation of such simulations, in addition to spatial resolution, is the radius of 
the inner sink cell, which should be as small as possible. The maximum radius of our sink cell is 20 au, which is 
smaller than in some other studies~\citep[up to $50\, \rm au$, see in][]{hosokawa_2015}. Smaller sinks are very 
computationally expensive due to strong limitations imposed by the Courant timestep condition. We note that the 
numerical convergence of runs with a grid resolution as ours was demonstrated in Section~5 of~\citet{meyer_mnras_473_2018}.  
Our overall goal consists in identifying the accretion bursts, their numbers, characteristics and occurrence, 
in order to determine the time that young massive stars spend bursting during their early formation phase, and the 
mass they accrete while experiencing those luminous events.

We use the statistical method developed in the context of accreting low-mass protostars 
in~\citet{vorobyov_apj_805_2015,2018arXiv180106707V}. It aims at discriminating between the background secular variability 
accounting for anisotropies in the accretion flow generated by the presence of enrolled spiral arms, and 
the various luminosity bursts generated by the accretion of dense gaseous clumps. It can be 
summarised as follows. 
A background luminosity $L_{\rm bg}$ is calculated by filtering out all accretion-driven events such 
that a smooth function of time is obtained. The quantity $L_{\rm bg}$ is taken as being 
$L_{\rm acc}+L_{\star}$ if the instantaneous accretion rate $\dot{M}(t)$ is smaller than 
$\dot{M}_{\rm crit} = 5 \times 10^{-4}\, \rm M_{\odot}\, \rm yr^{-1}$ and 
as $L_{\rm acc} \times \dot{M}_{\rm crit}/\dot{M}(t) + L_{\star}$ otherwise, which has the  
effect to filter out all strong accretion bursts from the stellar internal luminosity. 
We then derive the intensities and durations of so-called 1-mag, 2-mag, 3-mag, and 4-mag accretion-driven 
outbursts. In particular, by the 1-mag outburst we mean all outbursts with luminosity greater 
than 2.5 times the background luminosity (1-mag cutoff), but is lower than $2.5^2$ times the background 
luminosity (2-mag cutoff). The 2-, 3-, and 4-mag bursts are defined accordingly, with an exception of 
the 4-mag bursts having no upper limit (Table~\ref{tab:1}). 
The filtering method additionally insures that (i) the accretion-driven bursts duration is short enough so that slow 
increases and decreases of the total internal stellar luminosity are not mistaken as outbursts and that (ii) low-intensity 
kinks of less than 1-mag present in the lightcurves are also not interpreted as accretion outbursts, because they may 
be produced by boundary effects. Fig.~\ref{fig:illustration} illustrates the burst filtering method on a series 
of successive accretion peaks on which their prominence (luminous intensity with respect to $L_{\rm bg}$) and duration 
(width at half-maximal prominence) are plotted. 
Statistical quantities such as variance and standard deviations of the collection of selected bursts 
are finally derived (Table~\ref{tab:2}). We refer the reader interested in further details about the 
bursts analysis method to~\citet{vorobyov_apj_805_2015} and~\citet{2018arXiv180106707V}.


\section{Episodic accretion-driven bursts}
\label{sect:bursts}

In this section, we present the accretion rate histories of our set of simulations
and we describe their corresponding protostellar luminosities.
Then, we analyse the burst characteristics in order to extract their properties such 
as their duration and occurence.

\begin{figure}
	\centering
        \begin{minipage}[b]{ 0.45\textwidth}
                \includegraphics[width=1.0\textwidth]{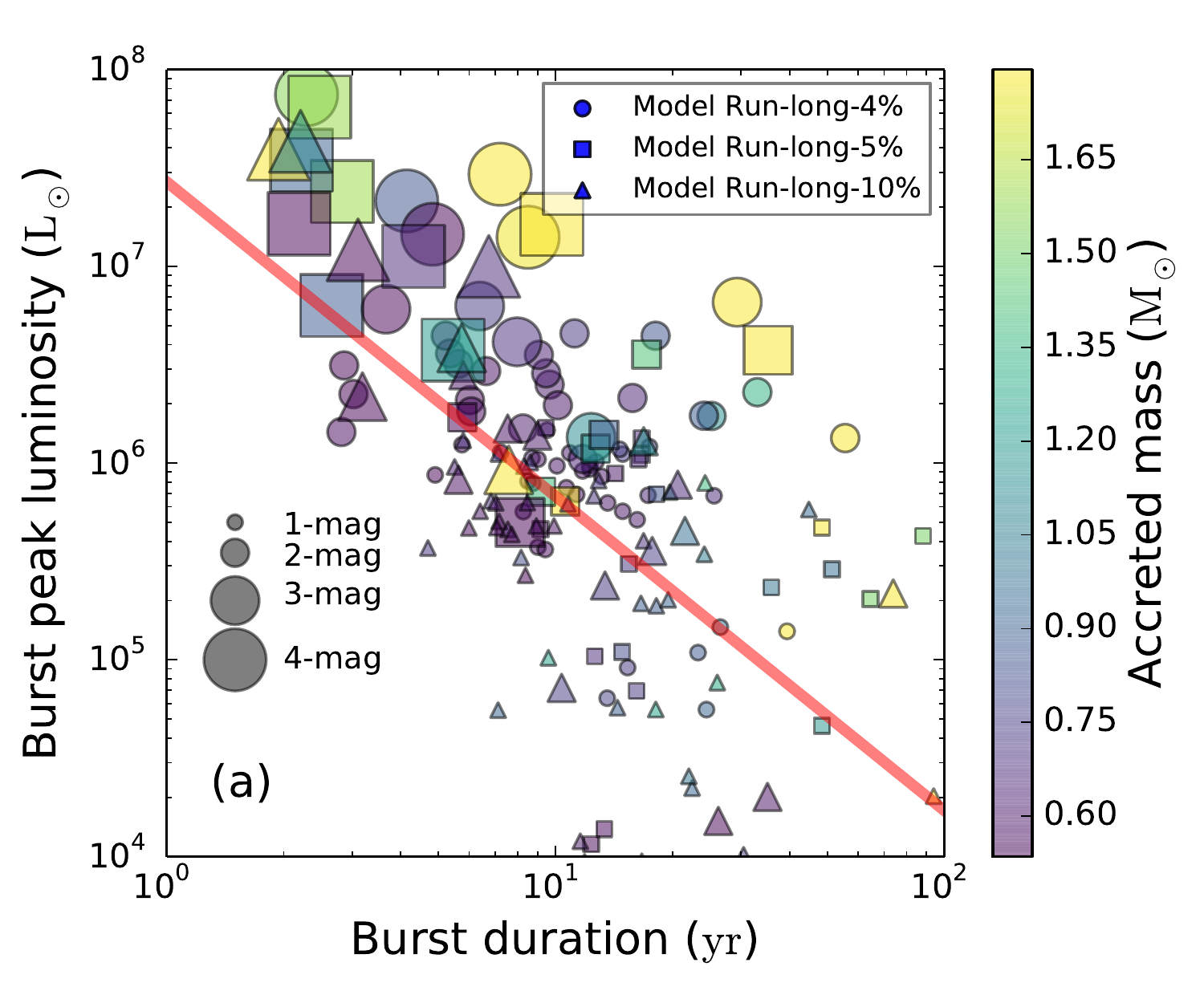}
        \end{minipage}  \\
        \begin{minipage}[b]{ 0.45\textwidth}
                \includegraphics[width=1.0\textwidth]{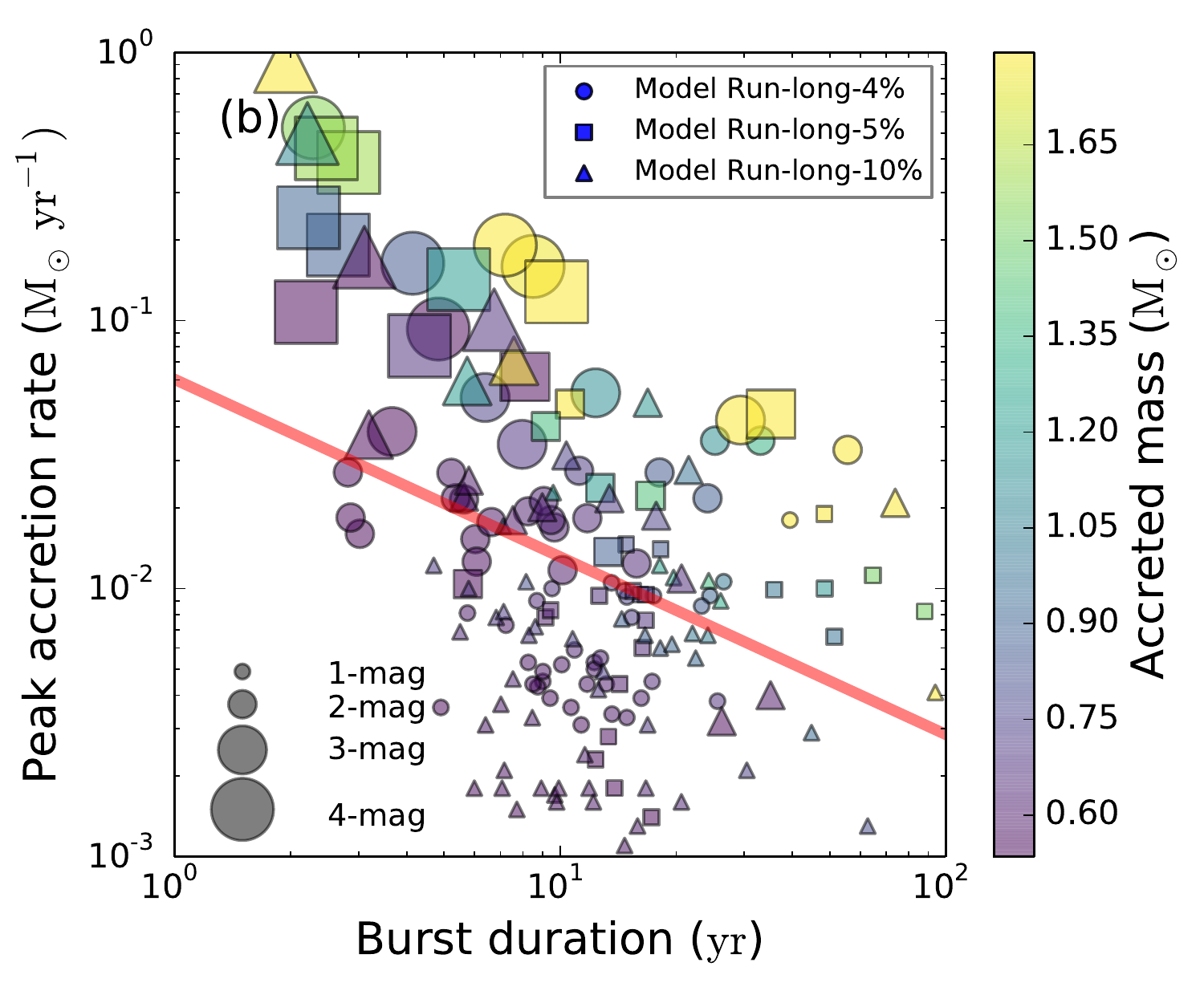}
        \end{minipage}  
        \caption{ 
                Correlation between burst duration versus peak luminosity (a) and burst duration versus 
                peak accretion rate (b) for each individual bursts, with colours representing the mass 
                accreted by the protostar during the bursts.   
                 }      
        \label{fig:correlation_1}  
\end{figure}

\subsection{Accretion rate histories}
\label{sect:rates}

Fig~\ref{fig:gen_plots_1} shows the accretion rate evolution (in $\rm M_{\odot}\, \rm yr^{-1}$) 
onto our protostars formed out of pre-stellar cores with initial rotational-to-gravitational 
energy ratios $\beta$=$4\%$ (Run-long-4, panel a), $5\%$ (Run-long-5, panel b), 
$10\%$ (Run-long-10, panel c) and Run1-hr (panel d). 
The effects of accretion on the protostar's growth is explicitly illustrated in the figure by plotting 
the evolution of the protostellar mass (thick dotted red line, in $\rm M_{\odot}$) over the 
accretion rate histories (thin solid blue lines).  Additionally, an orange dot marks the time instance when  
the young star becomes a massive object ($\rm M_{\star}\ge8\,\rm M_{\odot}$). 
The dotted black vertical lines separate different sequences of the young star's life characterised 
by distinct accretion regimes. The first vertical line corresponds to the onset of disc formation, 
when the free-fall of the envelope material onto the star ceases and the star starts gaining its mass 
via accretion from the circumstellar disc. The second vertical line marks the beginning 
of the first sudden increase of the accretion rate that is associated to a $\ge2$-mag \textcolor{black}{burst} 
(see Section~\ref{sect:luminosities}).
These two vertical lines highlight distinct protostellar mass accretion regimes corresponding to 
several different accretion phases.

First, the collapse of the parent pre-stellar core generates an infall of material through the sink cell, 
producing the initial increase of the mass flux 
through the inner boundary. During the \textcolor{black}{initial} $\approx 2\, \rm Myr$, the mass accretion rate increases from 
$\approx 10^{-6}$ to $\approx 10^{-4}\, \rm M_{\odot}\, \rm yr^{-1}$ and a further increase  
of the collapsing gas radial velocity happens over the next $\approx\, 10\, \rm kyr$, increasing $\dot{M}$ from 
$\approx 10^{-4}$ to slightly less than $\approx 10^{-3}\, \rm M_{\odot}\, \rm yr^{-1}$. 
The latter value is the typical rate at which MYSOs are predicted to accrete~\citep{hosokawa_apj_691_2009}. 
The time instance of disc formation is a function of the initial spin of the pre-stellar 
core, see also~\citet{meyer_mnras_473_2018}. When the disc forms (first vertical line of 
Figs.~\ref{fig:gen_plots_1}a-d), moderate variability appears immediately in the accretion flow 
as a result of anisotropies in the density and velocity fields close to the MYSOs. The  
values of $\dot{M}(t)$ evolve such that the variation amplitude gradually augments up to spanning 
over almost 2 orders of magnitudes, from $\approx 10^{-4}$ to $\approx 10^{-2}\, \rm M_{\odot}\, \rm yr^{-1}$ 
(Fig.~\ref{fig:gen_plots_1}b). 
A slight change in the slope of the mass evolution is associated with the beginning of this second, 
variable-accretion phase. All our protostars become heavier that 
$8\, \rm M_{\odot}$ and therefore enter the massive regime of stellar masses in this phase (see 
orange dot). Although the increase of the stellar mass is still smooth (thick dotted red line), step-like 
augmentations due to the fast accretion of dense material by the sink cell happen, e.g. at times 
$\approx 23\, \rm kyr$ in model Run-long-5 and $\approx 24\, \rm kyr$ in model Run-1-hr 
(thick red dotted line of Fig.~\ref{fig:gen_plots_1}b,d), when the protostellar mass has 
reached $\approx 10\, \rm M_{\odot}$. 
Those accretion spikes grow in number and intensity as a function of time, because the discs 
surrounding the MYSOs, while growing in size, develop complex morphologies by adopting the 
filamentary structure of self-gravitational rotating discs. 
A more detailed description of the early disc structure evolution of massive irradiated 
self-gravitating discs can be found in~\citet{meyer_mnras_473_2018}.

The variations amplitude in the accretion rate continues to increase during about $10$$-$$15\, \rm kyr$ 
after disc formation, until the MYSOs enter the next evolution phase marked by the second dotted vertical 
line. For each model, such events are highlighted in Fig.~\ref{fig:gen_plots_1} with the second vertical line. 
Without stopping the baseline accretion (at the average rate of $\approx 10^{-3}\, \rm M_{\odot}\, \rm yr^{-1}$), 
this evolution phase is regularly interspersed with strong accretion spikes which generate luminosity outbursts via 
the mechanism highlighted in~\cite{meyer_mnras_464_2017}. The essence of this mechanism is gravitational fragmentation 
of disc’s spiral arms, which extend up to radii $\approx 100$$-$$1000$ au, followed by inward migration and infall of gaseous 
clumps on the protostar.
%
%
At this stage, the mass of the protostar grows essentially due to those strong accretion events 
and the mass gained by accretion during the quiescent phase is negligible (see also discussion in Section~\ref{sect:mass_gained}). 
\textcolor{black}{The difference} in the spike occurrence is only a function of the initial gravitational-to-kinetic energy $\beta$ 
(our Table~\ref{tab:models}), as it is the sole parameter by which models Run-long-4, Run-long-5 
and Run-long-10 differ (Fig.~\ref{fig:gen_plots_1}a-c). 
At $t_{\rm end}=60\, \rm kyr$, the MYSOs have masses of about $\approx 47$, $\approx 42$ and 
$\approx 34\, \rm M_{\odot}$, respectively. Their growth should continue up to the exhaustion of disc's envelope mass reservoir. 
When \textcolor{black}{the envelope} no more fuel\textcolor{black}{s} the star-disc system, both infall rate onto the disc and accretion rate onto the protostar 
should gradually decrease and stop (not modeled in our simulations) as it happens in the context of low-mass star 
formation~\citep{vorobyov_apj_713_2010,vorobyov_aa_605_2017}.

\subsection{Protostellar luminosities}
\label{sect:luminosities}

Fig~\ref{fig:gen_plots_2} plots the total, bolometric luminosity of the MYSOs evolution (photospheric plus accretion 
luminosity, in $\rm L_{\odot}$) as a function of time (in $\rm kyr$) of our protostars 
generated by the collapse of pre-stellar cores of initial kinetic-by-gravitational energy 
ratio $\beta$=$4\%$ (Run-long-4, panel a), $5\%$ (Run-long-5, panel b), $10\%$ 
(Run-long-10, panel c) and Run1-hr (panel d). The thin dotted red line represents the total 
luminosity, the black solid line corresponds to the background luminosity $L_{\rm bg}$,  
while the dotted thin blue line, dashed dotted orange line and dashed violet lines show 
the limit of the 1-, 3- and 4-magnitudes cutoff with respect to $L_{\rm bg}$. The overplotted 
dots are the  peak accretion rates for the burst which peak luminosity is at least $2.5$ times brighter 
than their quiescent pre-flare protostellar luminosity. Shown in the 
$0.01$$-$$0.1$ $\rm M_{\odot}$ range, their colour corresponds 
to the peak's accretion rate (in $\rm M_{\odot}\, \rm yr^{-1}$). 
As in Fig~\ref{fig:gen_plots_1}, 
the orange dot \textcolor{black}{indicates the time instance} when the star becomes a massive object ($M_{\star}\ge8\, \, \rm M_{\odot}$) 
and the dotted black vertical lines separate the above described mass regimes: 
the onset of disc formation, and the beginning of the above described 
spiked accretion phase.

The initial free-fall gravitational collapse of the pre-stellar cores corresponds to 
a luminosity phase $<10^{3}\, \rm L_{\odot}$ which last $\approx 10\, \rm kyr$. 
As demonstrated in Fig.~7 of~\citet{meyer_mnras_473_2018}, the protostars still have a negligible accretion luminosity:  
because of still relatively small stellar mass and/or large stellar radius, their luminosity is governed by their 
photospheric component $L_{\rm bol} \approx L_{\star} \le 10^{4}\, \rm L_{\odot}$ (Figs.~\ref{fig:gen_plots_2}a-d). 
Stellar objects in such phase are likely not observable, as not bright enough 
to produce photons able to escape their opaque embedded host core\textcolor{black}{s}. 
Once the circumstellar disc forms (first black vertical line), the luminosity adopts a time-dependent 
behaviour which reflects the protostellar accretion rate history (Figs.~\ref{fig:gen_plots_1}a). 
%
%
%
%
All our models exhibit a spectrum of burst magnitudes from 1-mag bursts to 4-mag bursts, except for model 
Run-1-hr, which lacks the 4-mag bursts (perhaps due to shorter simulation times, $\approx 35\, \rm kyr$, as compared to 
other models). However, the total number of the bursts and the number of the burst of a certain magnitude 
vary from model to model.

%
Indeed, a finer grid resolution 
resolves better the Jeans length. That, in its turn, allows us to treat more realistically   
the disc gravitational instabilities and reveal smaller disc fragments, \textcolor{black}{a subset of which} of 
will end \textcolor{black}{up} migrating onto the protostar and \textcolor{black}{generating} (milder intense) bursts. 
As an example, our model Run-long-4 
has a single 1-mag burst (Fig.~\ref{fig:gen_plots_2}a), whereas our model Run-1-hr exhibits 9 of 
them over roughly the same time interval (between the two vertical black lines, see  
Fig.~\ref{fig:gen_plots_2}d). A more thorough discussion of the resolution-dependence of 
numerical investigations on disc fragmentation can be found in~\citet{meyer_mnras_473_2018}. 
Similarly, bursts are more frequent at higher $\beta$-ratio, e.g. Run-long-4 has a single 1-mag 
burst but Run-long-10 has more than a dozen (Fig.~\ref{fig:gen_plots_2}a,c). This illustrates 
the effects of the initial conditions on the disc dynamics. 
We also notice that, in the case of our fast-rotating models, bursts develop earlier in 
time, e.g. models Run-long-10 and Run-long-4 have their first 1-mag burst at a time 
$\approx 15\, \rm kyr$ and $\approx 26\, \rm kyr$, respectively. 
%

\begin{figure}  
	\centering
        \begin{minipage}[b]{ 0.45\textwidth}  
                \includegraphics[width=1.0\textwidth]{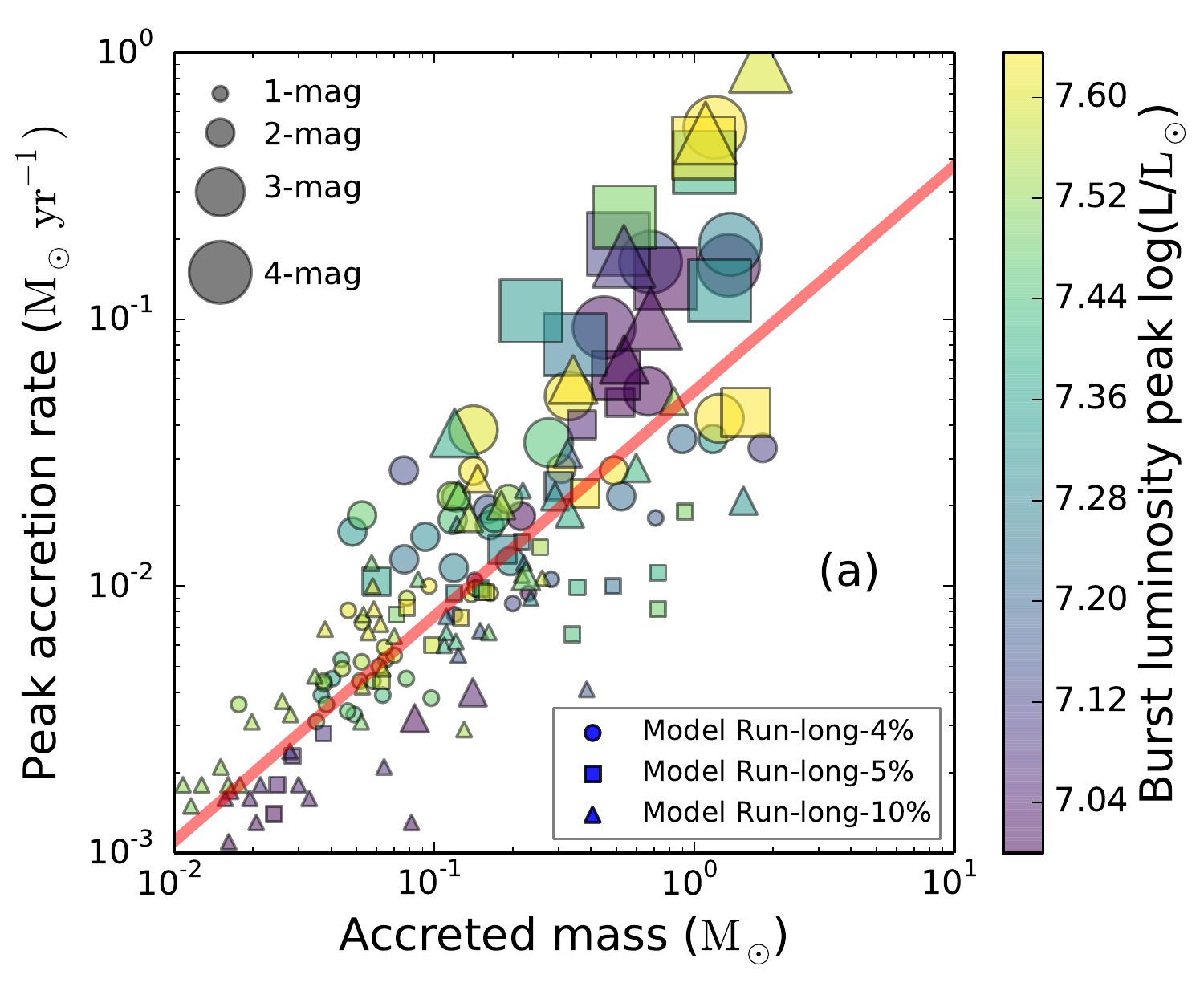}
        \end{minipage}\\   
        \begin{minipage}[b]{ 0.45\textwidth}
                \includegraphics[width=1.0\textwidth]{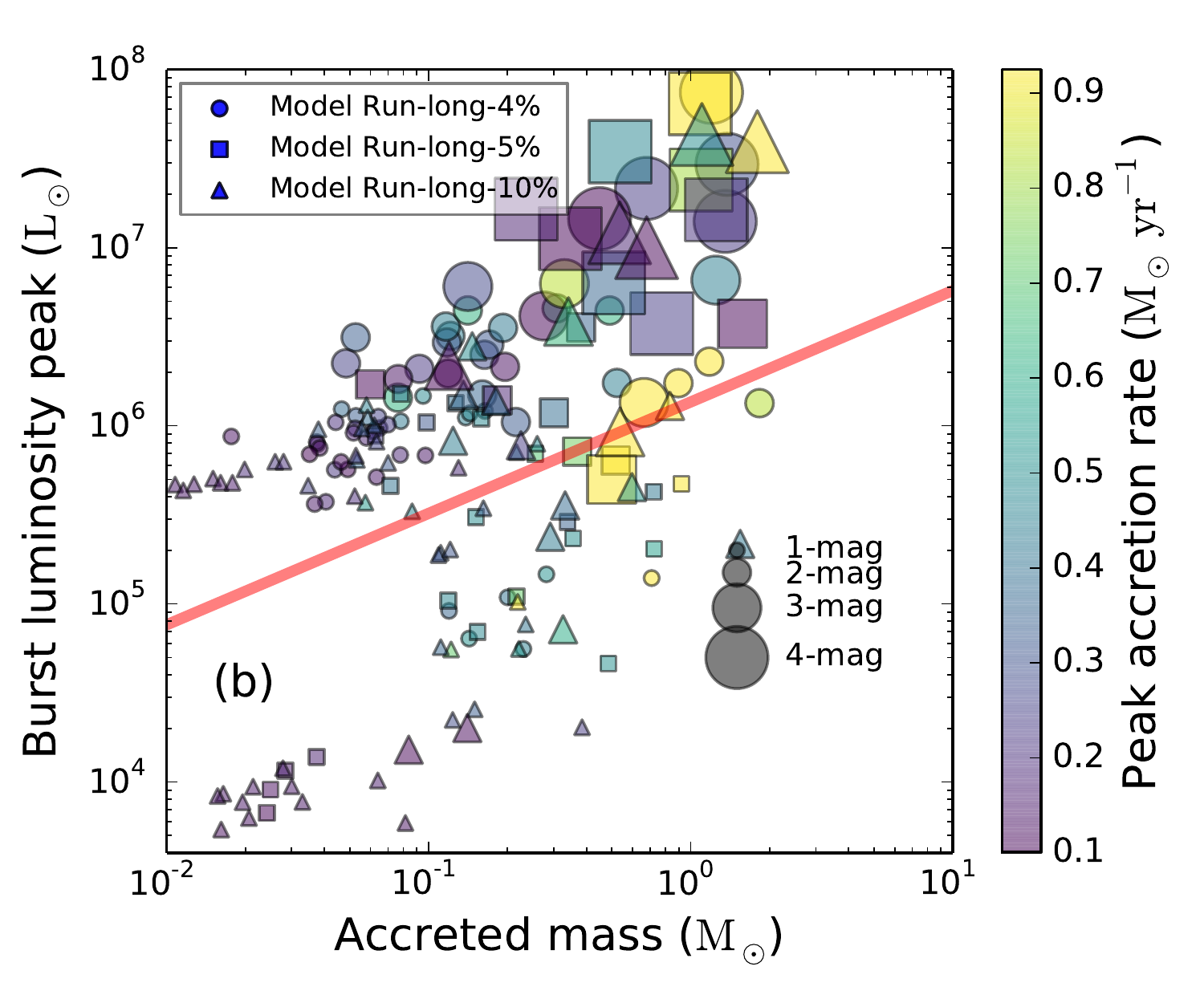}
        \end{minipage}        
        \caption{ 
                Correlation between accreted mass versus peak accretion rate (a) and 
                accreted mass versus peak luminosity (b) for each individual burst, 
                with colours representing the peak luminosity (a) and the peak accretion rate (b), respectively. 
                 }      
        \label{fig:correlation_2}  
\end{figure}

After the first strong ($\ge\, 2$-mag) burst, the variations of the total protostellar luminosity gradually 
increases as the accretion luminosity overwhelms the photospheric contribution to the total luminosity. 
This is tied to the rising complexity of the disc pattern: the multiple spiral arms and gaseous 
clumps forming in the disc experience a chaotic motion and violently interact by formation, 
destruction, merging and migration in the star's surroundings~\citep{meyer_mnras_473_2018}. 
The number of bursts per unit time consequently increases after $\approx 30\, \rm kyr$ 
and the protostellar lightcurve is shaped as a forest of accretion-driven peaks. The burst intensity is not 
a monotonic function of time as high-intensity bursts (e.g., 3- and 4-mag) can be interspersed by lower-intensity 
ones (e.g., 1- and 2-mag) or by quiescent burstless phases. Bursts often appear grouped as a 
collection of different magnitude bursts on a rather small time interval, see e.g. our model Run-long-4 at 
times $45$$-$$50\, \rm kyr$. 
%
%
%


\textcolor{black}{
Note that accretion-driven bursts can appear either isolated or clustered. Single bursts are provoked by the 
infall of a gaseous clump which is detached from its parent spiral arm and experiences a migration towards 
the central MYSO, see mechanism described in~\citet{meyer_mnras_464_2017}. They correspond to the accretion 
of relatively large amount of compact circumstellar material over rather short timescales ($\sim 10\, \rm yr$), 
which produces a unique high-luminosity burst. 
The rapid successive accretion of several migrating clumps from different parent arms, the merging of 
clumps into inhomogeneous gaseous structures or even the migration of clumps that separate an inner 
portion of spiral arms into two segment make the accretion pattern more complex and can produce 
different a type of lightcurves exhibiting lower-intensity, clustered bursts, in a manner consistent 
with the description of~\citet{vorobyov_apj_805_2015} in low-mass star formation. 
}

\subsection{Bursts characteristics}
\label{sect:bursts_prop_1}

We summarise the analysis of our population of accretion-driven bursts in our 
Table~\ref{tab:1}. For each of the four models, it reports the number of the 
bursts with a specific magnitude (1-, 2-, 3-, and 4-mag) and also the total number 
of the bursts. It also indicates the maximum, minimum and mean 
values of the burst luminosity ($L_{\mathrm{max}}$, $L_{\mathrm{min}}$, 
$L_{\mathrm{mean}}$, in $10^{5}\, \rm L_{\odot}\, \rm yr^{-1}$), the maximum, minimum and mean values of the burst  
peak accretion rate through the sink cell ($\dot{M}_{\mathrm{max}}$, $\dot{M}_{\mathrm{min}}$, 
$\dot{M}_{\mathrm{mean}}$,  in $\rm M_{\odot}\, \rm yr^{-1}$) and the maximum, minimum and mean values of the burst duration 
($t_{\mathrm{bst}}^{\mathrm{max}}$, $t_{\mathrm{bst}}^{\mathrm{min}}$, $t_{\mathrm{bst}}^{\mathrm{mean}}$, in $\rm yr$), 
respectively. Finally, it provides the integrated duration of the bursts $t_{\rm bst}$ (in $\rm yr$), 
i.e., the time that the protostar in each model spends in the burst phase during the 
initial $60\, \rm kyr$ of evolution. For each cut-off magnitude, the bold line reports 
the averaged values of those quantities.

The number of low-intensity accretion-driven bursts is much higher than that of high-intensity bursts, e.g. 
the integrated number of 1-mag bursts in our set of models is $N_{\rm bst}=128$ while it is $N_{\rm bst}=44$ and 
$N_{\rm bst}=12$ for the 2- to 4-mag bursts, respectively. This is because the low-rate accretion variability, provoked 
by accretion of dense portions of spiral arms, is more frequent than the high-rate variability caused by the 
infalling gaseous clumps. 
%
Our simulations with $\beta=4\, \%$ and $10\, \%$ have more 1-, 2- and 3-mag bursts than in our simulations with 
$\beta=5\, \%$, however, the opposite trend is seen for the 4-mag bursts. Thus, our results do not support the interpretation of the 
simplistic picture of discs exhibiting an efficiency of the gravitational instability directly proportional to the initial 
spin of the pre-stellar core and gradually increasing with $\beta$. 
%
The peak luminosities of 1-mag bursts range from $L_{\rm min}=0.054\times 10^{5}\, \rm L_{\odot}$ to 
$L_{\rm max}=15.1\times 10^{5}\, \rm L_{\odot}$, whereas they vary by 2 and 1 orders of magnitudes for our 
2- and 3-mag bursts, exhibiting $L_{\rm min}=0.15\times 10^{5}$ and $4.99\times 10^{5}\, \rm L_{\odot}$ to reach 
$L_{\rm max}=45.6\times 10^{5}$ and $65.7\times 10^{5}\, \rm L_{\odot}$, respectively. Therefore, some 1-mag bursts can 
be more luminous (in bolometric luminosity) than some 2- or 3-mag bursts, depending on the corresponding pre-burst 
background stellar luminosity. The model Run-1-hr does not have 4-mag bursts because, due to timestep restrictions, we 
could not run it for \textcolor{black}{longer integration times}.

\begin{figure}
	\centering 
        \begin{minipage}[b]{ 0.45\textwidth}
                \includegraphics[width=1.0\textwidth]{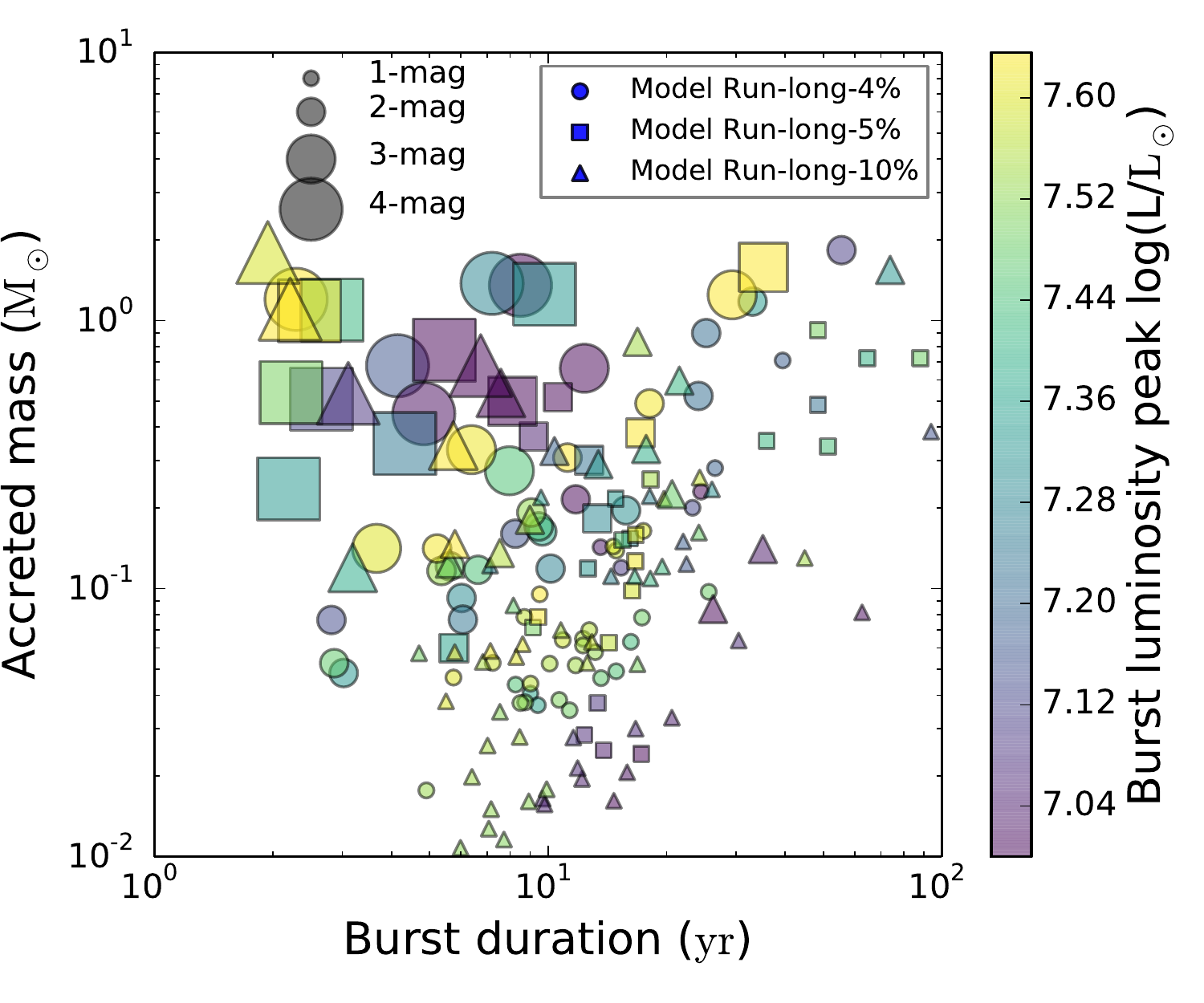}
        \end{minipage}  
        \caption{ 
                Correlation between burst duration versus accreted mass  
                for each individual bursts, with colours representing the burst peak luminosity.
                 }      
        \label{fig:correlation_3}  
\end{figure}

Figs.~\ref{fig:correlation_1} to~\ref{fig:correlation_3} show several correlation plots of the various burst 
characteristics of a homogeneous subset of three of our models ($r_{\rm in}=20\, \rm au$). Fig.~\ref{fig:correlation_1}   
displays \textcolor{black}{the burst peak luminosity $L_{\rm burst}$ (in $L_{\odot}$) as a function of the burst duration $t_{\mathrm{bst}}$} 
with a colour coding giving information about the accreted mass per burst (see panel (a), in $M_{\odot}$) 
and \textcolor{black}{the peak accretion rate $\dot{M}$ (in $\rm M_{\odot}\, \rm yr^{-1}$) as a function of the 
burst duration $t_{\mathrm{bst}}$} with a colour coding giving information about the accreted mass 
per burst (see panel (b), in $M_{\odot}$). 
On each plots the symbols and their sizes are functions of the burst magnitude (smallest symbols 
for 1-mag bursts and largest symbols for 4-mag bursts) and specify the models, to which each burst 
belongs (circles for our Run-long-4$\%$, squares for Run-long-5$\%$ and triangles for Run-long-10$\%$). 
The overplotted line is a fit of all bursts together. 

\begin{table*}
	\centering
	\caption{
	Mass (in $\mathrm{M_{\odot}}$) and proportion (in $\%$) of final protostellar mass accreted as 
	a function of the protostellar brightness at the moment of the accretion.  
	}
        \begin{tabular}{lcccccr}
        \hline  
	${\rm {Models}}$       & $\rm M_{\star}$   &  $\rm L_{\rm tot} \sim\, \rm L_{\rm bg}$  &  $1$$-$$\rm mag$ & $2$$-$$\rm mag$ & $3$$-$$\rm mag$ & $4$$-$$\rm mag$ \\ 
        \hline
	\multicolumn{7}{c}{\textbf{Accreted mass (in $\rm M_{\odot}$)}}\tabularnewline
Run-long-4\%  &  47.33  &  25.06  &  5.28  &  10.04        &  2.59  & 4.36\tabularnewline
Run-long-5\%  &  41.57  &  21.48  &  5.48  &  $\,\,$4.97   &  5.20  & 4.44\tabularnewline
Run-long-10\% &  34.15  &  19.35  &  6.39  &  $\,\,$3.66   &  0.93  & 3.82\tabularnewline
Run-hr        &  20.07  &  15.25  &  1.13  &  $\,\,$1.88   &  1.74  & 0.07\tabularnewline 
        \textbf{Mean ($r_{\rm in}=20\, \rm au$)} & \textbf{41.02}  & \textbf{21.96} & \textbf{5.72} & \textbf{6.22} & \textbf{2.91} & \textbf{4.21}\tabularnewline
	\multicolumn{7}{c}{\textbf{-}}\tabularnewline
{\rm Run-without}     &  24.76  &  18.64  &  3.01  &  $\,\,$2.46   &  0.63  & 0.00\tabularnewline	
{\rm Run-with}        &  24.35  &  15.46  &  3.76  &  $\,\,$3.63   &  1.19  & 0.29\tabularnewline        
        %
        \multicolumn{7}{c}{\textbf{Proportion of final protostellar mass (in $\%$)}}\tabularnewline
Run-long-4\%  &  100  &  52.94  &  11.17         &  21.21        &  $\,\,$5.47   &  $\,\,$9.21\tabularnewline
Run-long-5\%  &  100  &  51.66  &  13.18         &  11.95        &  12.51        &  10.70\tabularnewline
Run-long-10\% &  100  &  56.65  &  18.70         &  10.73        &  $\,\,$2.74   &  11.18\tabularnewline
Run-hr        &  100  &  75.97  &  $\,\,$5.60    &  $\,\,$9.36   &  $\,\,$8.70   &  $\,\,$0.37\tabularnewline
        \textbf{Mean ($r_{\rm in}=20\, \rm au$)} & \textbf{100}  & \textbf{53.75} & \textbf{14.35}  & \textbf{14.63}  & \textbf{6.91}  & \textbf{10.36}\tabularnewline
	\multicolumn{7}{c}{\textbf{-}}\tabularnewline
{\rm Run-without}     &  100  &  75.30  &  12.19         &  $\,\,$9.93   &  $\,\,$2.55   &  $\,\,$0.00\tabularnewline
{\rm Run-with}        &  100  &  63.49  &  15.45         &  14.94        &  $\,\,$4.89   &  $\,\,$1.21\tabularnewline        
        \hline    
        \end{tabular}
\label{tab:2}
\end{table*}

Fig.~\ref{fig:correlation_1}a,b show that 4-mag bursts have $L_{\rm tot}>3\times 
10^{6}\, \rm L_{\odot}$ and a duration time $<10\, \rm yr$ while 1-mag bursts 
are located in the region with $L_{\rm tot}<3\times 10^{6}\, \rm L_{\odot}$ and 
$>4\, \rm yr$. The same distribution appears regardless of the model conditions. 
The overplotted line fitting the burst population highlights this 
general trend $-$ luminous bursts are shorter in duration, even if our  
bursts diversity scatters in bolometric luminosity by spanning over several 
orders of magnitude at a given burst duration. 
Therefore, the burst peak luminosity (Fig.~\ref{fig:correlation_1}a) and maximal accretion rate 
(Fig.~\ref{fig:correlation_1}b) globally decrease with the burst duration (thick red lines in Fig~4). 
The more efficient mass-accreting bursts are mostly \textcolor{black}{3- and 4-mag}, short-lived and 
some high-luminosity events or long-lived 1-mag bursts 
(see colour scale of Fig.~\ref{fig:correlation_1}a). Such bursts are typically associated to 
the largest peak accretion rates, located in the short-duration and 
high-luminosity region of the figure (Fig.~\ref{fig:correlation_1}b). 
This statistical discrepancy between the global correlation between peak 
accretion mass and peak accretion rate is already visible in our 
Table~\ref{tab:1} and illustrates the variety of bursts duration produced for a 
given increase of the MYSOs's bolometric luminosity with respect to their 
pre-bursts luminosity. 
%


Fig.~\ref{fig:correlation_2}a shows the accreted mass (in $\rm M_{\odot}$ as a function of 
the bursts peak accretion rate (in $\rm M_{\odot}\, \rm yr^{-1}$) 
and Fig.~\ref{fig:correlation_2}b plots the accreted mass as a function of the
the bursts peak luminosity (in $\rm L_{\odot}$), respectively. Additionally, the 
colour coding indicates the burst peak luminosity (a) and the peak accretion rate during the bursts (b). 
Our models reveal a good correlation between the peak accretion rate and the 
accreted mass of their bursts (see thick red line).  
%
Fig.~\ref{fig:correlation_2}a is a consequence of the typical Gaussian-like shape of the accretion 
profiles, whose integral as a function of time (i.e. the accreted mass) is proportional to the peak 
accretion rate, whereas Fig.~\ref{fig:correlation_2}b is a natural outcomes of the used 
prescription to estimate $L_{\rm acc} \propto \dot{M}$, low-luminosity deviations appearing because 
of occasional, rapid fluctuations of the stellar radius~\citep{hosokawa_apj_691_2009}. 
%
%
%
Finally, Fig.~\ref{fig:correlation_3} plots for the 
burst duration (in $\rm yr$) as a function of the accreted mass during the corresponding  
burst (in $\rm M_{\odot}$)  with a colour coding giving information about the burst luminosity 
(in $\rm L_{\odot}$). 
This figure clearly illustrates our model predictions regarding luminosity bursts in MYSOs: 
more luminous bursts (3- and 4-mag) accrete more mass than their less luminous counterparts 
(1- and 2-mag), although the former are shorter in duration than the latter.
%
%
Accordingly, the different burst population do not overlap in the 
figure and organise in different, adjacent regions (Fig.~\ref{fig:correlation_3}b). 
%

\section{The effect of stellar motion}
\label{sect:inertia}

\textcolor{black}{
The role of stellar motion on the disc dynamics and bursts properties is 
examined in this section. We first introduce the mechanism responsible for stellar motion and its 
implementation in our code, and then compare two simulations of equivalent initial 
conditions and included physics, but considered without and with stellar wobbling. 
}

\subsection{Stellar inertia and its implementation in the {\sc PLUTO} code}
\label{sect:inertia1}

\textcolor{black}{
When hydrodynamical simulations of circumstellar discs are run using a curvilinear coordinate system, 
the sink cell representing the star and its direct surrounding is attached to the origin of the computational 
domain. The natural response of the star to the changes in the mass distribution of its circumstellar disc, 
when the disc barycenter does not coincide with the origin of the domain, is therefore neglected. 
Taking stellar motion into account may become important when the initial axisymmetry of the rigidly-rotating pre-stellar core is broken 
by the development of dense spiral arms and heavy gaseous clumps in the disc and their 
gravitational pull on the star is not mutually canceled. 
\textcolor{black}{
This mechanism has been first revealed in the analytical study on eccentric gravitational instabilities in discs around low-mass protostars of~\citet{adam_apj_347_1989}. 
It is} intrinsically taken into account in Lagragian simulations such as the 
smooth hydrodynamical models (SPH) of, e.g.~\citet{bonnell_mnras_271_1994,rice_mnras_364_2005,meru_mnras_406_2010,lodato_mnras_413_2011} 
as well as in simulations performed with three-dimensional Cartesian 
codes, see e.g.~\citet{krumholz_apj_656_2007,kratter_apj_708_2010,commercon_apj_742_2011,klassen_apj_823_2016,rosen_mnras_463_2016}. 
~\citet{regaly_aa_601_2017} and~\citet{2018arXiv180607675V} studied the effect of stellar wobbling 
on the disc evolution and migration of gaseous clumps and protoplanets using grid-based codes 
in the polar coordinates. They found that stellar motion may affect notably the migration speeds 
of gaseous clumps and protoplanets in the disc, but the final disc masses and sizes and also disc 
propensity to fragment were affected to a lesser extent.
%
%
Since high-mass protostars are surrounded by massive discs~\citep{johnston_apj_813_2015,ilee_mnras_462_2016,forgan_mnras_463_2016,cesaroni_aa_602_2017} 
in which heavy substructures (gaseous clumps, low-mass companions and/or asymmetric spiral arms) can form by fragmentation in the disc~\citep{meyer_mnras_473_2018}, 
the question is therefore how much those asymmetries may affect the total gravitational potential of the protostar-disc system during their rather short pre-main-sequence 
phase ($\sim \rm kyr$), see Section~\ref{sect:inertia3}. 
}

\textcolor{black}{
Our implementation of the stellar motion into the {\sc PLUTO} code for the spherical coordinate system consists in 
calculating the resultant gravitational force applied by the disc to the star using Cartesian projections and 
it is inspired from~\citet{Hirano_2017Science}. The gravitational force applied by the massive disc onto the growing protostar reads, 
\begin{equation}
	\delta \vec{F}_\mathrm{disc/\star} =  -G \frac{ \delta M_\mathrm{disc}(r) M_{\star} }{ r^{2} } \vec{e}_\mathrm{r}, 
	\label{eq:force_cell}
\end{equation}
where $\delta M_\mathrm{disc}(r)$ is the mass of a grid zone located in the disc, $r$ is the distance 
of the considered cell to the central protostar and $\vec{e}_\mathrm{r}$ is the unit vector in the radial direction. 
The projection of the force $\vec{F}_\mathrm{disc/\star}$ onto the Cartesian directions 
($\vec{e}_\mathrm{x}$,$\vec{e}_\mathrm{y}$,$\vec{e}_\mathrm{z}$) are, 
\begin{equation}
	\delta F_\mathrm{disc/\star}^{x} = \delta \vec{F}_\mathrm{disc/\star}^{x} \cdot \vec{e}_\mathrm{x} = \delta | \vec{F}_\mathrm{disc/\star} | \cos(\phi) \sin(\theta), 
	\label{eq:delta_force_disc1}
\end{equation}
\begin{equation}
	\delta F_\mathrm{disc/\star}^{y} = \delta \vec{F}_\mathrm{disc/\star}^{y} \cdot \vec{e}_\mathrm{y} = \delta | \vec{F}_\mathrm{disc/\star} | \sin(\phi) \sin(\theta), 
	\label{eq:delta_force_disc2}
\end{equation}
and, 
\begin{equation}
	\delta F_\mathrm{disc/\star}^{z} = \delta \vec{F}_\mathrm{disc/\star}^{z} \cdot \vec{e}_\mathrm{z} = \delta | \vec{F}_\mathrm{disc/\star} | \cos(\theta), 
	\label{eq:delta_force_disc3}
\end{equation}
where $\phi$ and $\theta$ represent the azimuthal and polar angles, respectively. We integrate the Cartesian 
components of the gravitational backaction of the circumstellar disc onto the protostar. Therefore, the 
Cartesian components of the total force $\vec{F}_\mathrm{disc/\star}$ read,  
\begin{equation}
	F_\mathrm{disc/\star}^{x} = \int_\mathrm{disc} \delta F_\mathrm{disc/\star}^{x}, 
	\label{eq:force_disc1}
\end{equation}
\begin{equation}
	F_\mathrm{disc/\star}^{y} = \int_\mathrm{disc} \delta F_\mathrm{disc/\star}^{y}, 
	\label{eq:force_disc2}
\end{equation}
and
\begin{equation}
	F_\mathrm{disc/\star}^{z} = \int_\mathrm{disc} \delta F_\mathrm{disc/\star}^{z}, 
	\label{eq:force_disc3}
\end{equation}
where the disc is evaluated as the gas that lies within the innermost envelope located at radii 
$\le 3000\, \rm  au$ of the computational domain~\citep{meyer_mnras_473_2018}. Finally, we recover, 
\begin{equation}
	\vec{F}_\mathrm{disc/\star} = \sqrt{ (F_\mathrm{disc/\star}^{x})^{2} +  (F_\mathrm{disc/\star}^{y})^{2} + (F_\mathrm{disc/\star}^{z})^{2} }  \vec{e}_\mathrm{r}, 
	\label{eq:force_cartesian_1}
\end{equation}
which is the total force exerted by the disc onto the star. In our midplane-symmetric picture, 
\begin{equation}
	F_\mathrm{disc/\star}^{z}=0, 
	\label{eq:Fz}
\end{equation}
and $\vec{F}_\mathrm{disc/\star}$ reduces to, 
\begin{equation}
	\vec{F}_\mathrm{disc/\star} = \sqrt{ (F_\mathrm{disc/\star}^{x})^{2} +  (F_\mathrm{disc/\star}^{y})^{2}  }  \vec{e}_\mathrm{r}. 
	\label{eq:force_cartesian_2}
\end{equation}
As we keep the protostar fixed to the origin of the computational domain, the effective force that we apply 
to each grid cell is, 
\begin{equation}
	\vec{F}_\mathrm{wobbling} =  \vec{F}_\mathrm{\star/\mathrm{disc}}  = - \vec{F}_\mathrm{disc/\star}, 
	\label{eq:wobbling}
\end{equation}
that has the effect to translate the disc solution around the star by the opposite vector of the force 
exerced by the disc onto it. The effect of the wobbling force is finally introduced as an additional 
acceleration term,
\begin{equation}
	\vec{g}^{'}= \frac{ \vec{F}_\mathrm{wobbling} }{ M_{\star}},  
	\label{eq:acceleration}
\end{equation}
in the gravity solver of the {\sc pluto} code. In spherical coordinates, the different components of $\vec{g}^{'}$ read,
\begin{equation} \begin{split}
	g^{'}_{r}            = \frac{ 1 }{ M_{\star}} \Big[   F_\mathrm{disc/\star}^{x} \sin(\theta) \cos(\phi)  +  F_\mathrm{disc/\star}^{y} \sin(\theta) \sin(\phi) \\ +  F_\mathrm{disc/\star}^{z} \cos(\theta)  \Big], 
	\label{eq:acc_disc1}
\end{split} \end{equation} 
\begin{equation} \begin{split}
	\vec{g}^{'}_{\phi}   = \frac{ 1 }{ M_{\star}} \Big[   F_\mathrm{disc/\star}^{x} \cos(\theta) \cos(\phi)  +  F_\mathrm{disc/\star}^{y} \cos(\theta) \sin(\phi) \\ -  F_\mathrm{disc/\star}^{z} \sin(\theta)  \Big], 
	\label{eq:acc_disc2}
\end{split} \end{equation}
and,
\begin{equation}
	\vec{g}^{'}_{\theta} = \frac{ 1 }{ M_{\star}} \Big[   -F_\mathrm{disc/\star}^{x} \sin(\phi)  +  F_\mathrm{disc/\star}^{y} \cos(\phi)   \Big],  
	\label{eq:acc_disc3}
\end{equation}
respectively. As described by~\citet{regaly_aa_601_2017}, this method induces boundary effects 
at the outer extremities of the computational domain, when the stellar motion is so pronounced that the corresponding 
translation of the whole solution by $\vec{F}_\mathrm{wobbling}$ gradually provokes cavities that open at the outer 
edges of the pre-stellar core. This is an issue meriting future numerical investigations and code development, however,  
it does not affect our comparison tests because we run simulations with high-mass cores that collapse faster 
($\sim 40\, \rm kyr$) than in the context of solar-mass pre-stellar cores. 
In the section below, we compare two simulations of gravitational collapse and disc formation of similar initial 
conditions but considered with and without stellar motion. 
}


\begin{figure*}
        \centering
        \begin{minipage}[b]{ 0.72\textwidth}  \centering
                \includegraphics[width=1.0\textwidth]{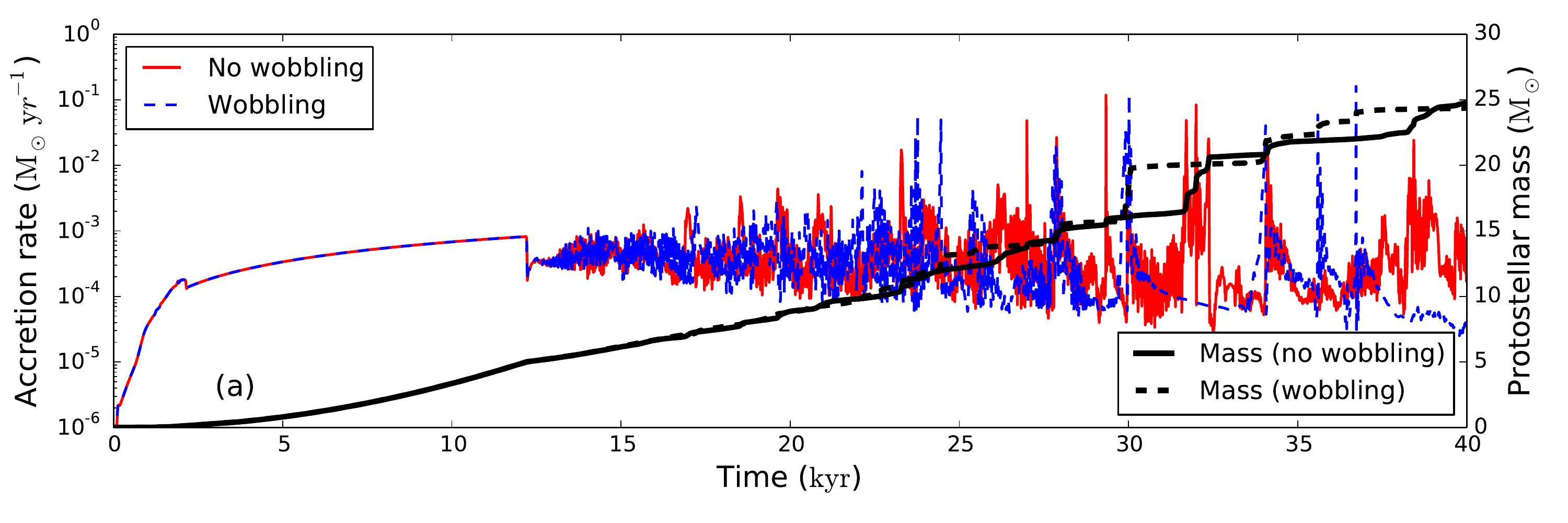}
        \end{minipage} \\   
                \centering
        \begin{minipage}[b]{ 0.7\textwidth}  \centering
                \includegraphics[width=1.0\textwidth]{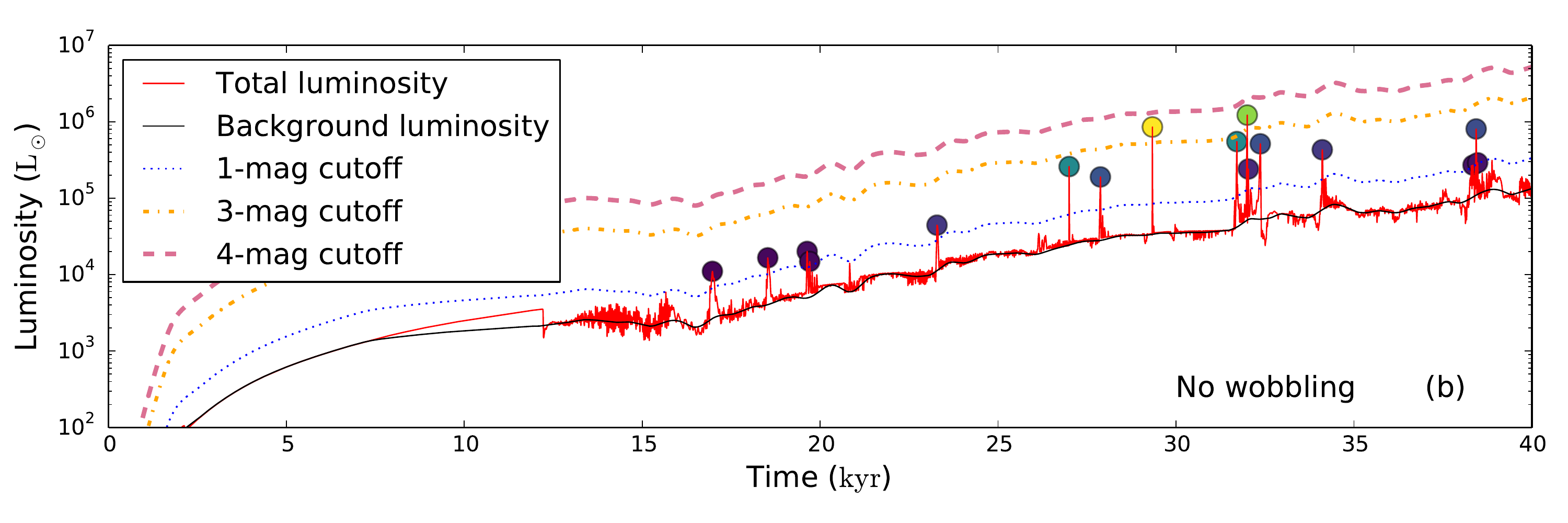}
        \end{minipage} \\   
                \centering
        \begin{minipage}[b]{ 0.7\textwidth}  \centering
                \includegraphics[width=1.0\textwidth]{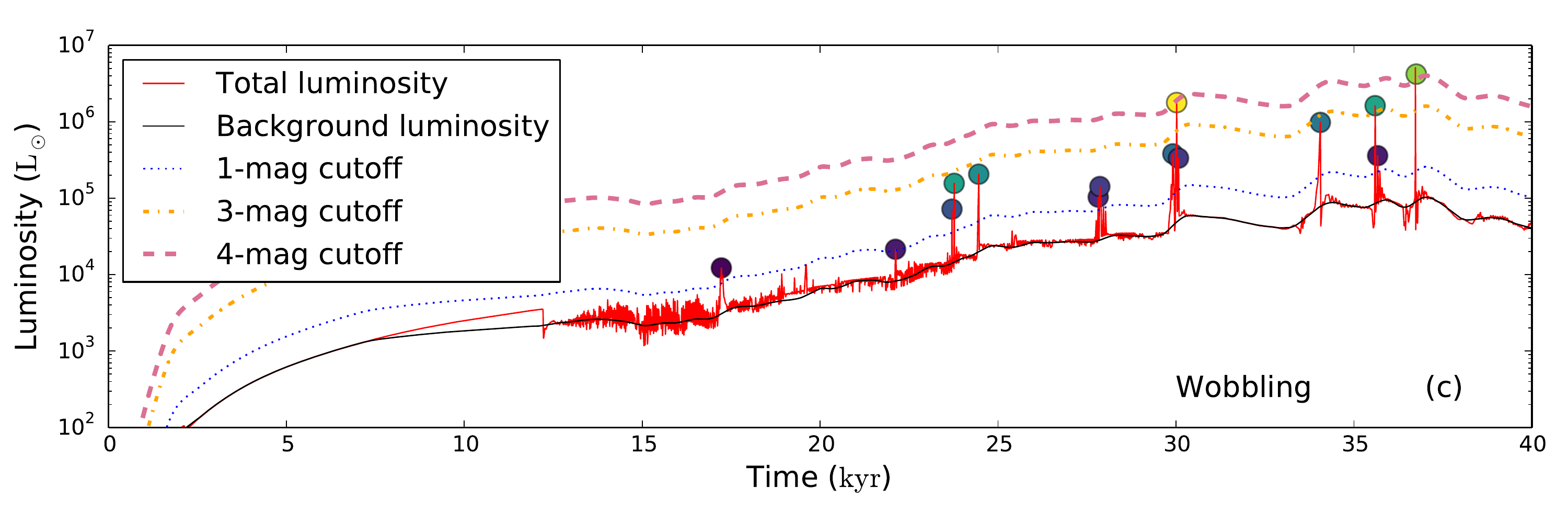}
        \end{minipage} \\   
                \centering
        \begin{minipage}[b]{ 0.7\textwidth}  \centering
                \includegraphics[width=1.0\textwidth]{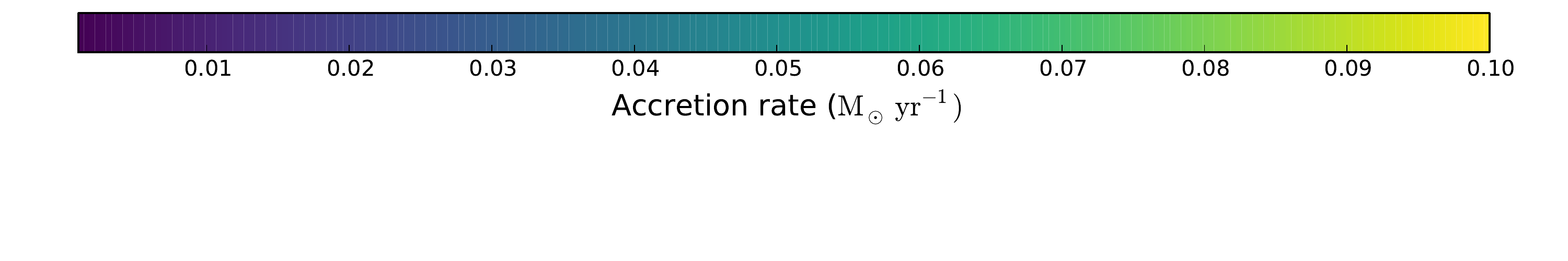}
        \end{minipage}    
        \caption{ 
                \textcolor{black}{
                Comparison of mass accretion rates and luminosities in our comparison 
                model ($\beta=4\%$ and $r_{\rm in}=12\, \rm au$) with and without stellar wobbling. 
                Panel (a) shows the mass accretion rate (thin lines, in $\rm M_{\odot}\, \rm yr^{-1}$) 
                and protostellar masses (thick lines, in $\rm M_{\odot}$) in our simulations 
                without (solid lines) and with (dashed lines) stellar wobbling. 
                Panels (b) and (c) present the total luminosities (thin red solid lines, in $\rm kyr$), 
                background luminosities (thin solid black lines) and luminosity cutoffs 
                (the dotted and dashed lines) for the model without stellar wobbling 
                (middle panel) and with stellar wobbling (bottom panel). The color bar 
                indicates the accretion rate for the filled circles in panels (b) and (c).
                 }
                 }      
        \label{fig:comparison}  
\end{figure*}

\subsection{Effects of stellar wobbling on mass accretion}
\label{sect:inertia3}

\textcolor{black}{
To compare the effects of stellar motion on the accretion disc and protostellar properties, 
we carry out two additional simulations initialised with a $100\, \rm M_{\odot}$ pre-stellar core, rigidly-rotating 
with $\beta=4\, \%$ and having a sink cell with a radius of $r_{\rm in}=12\, \rm au$. We use a small sink cell 
radius in order to follow the clumps migration at radii $\le\, 20\, \rm au$. This, however, reduces the timestep of the 
simulations and the timescale over which we can perform the model runs (see Run-1-hr). Hydrodynamics and radiation 
transport are treated as in the other runs of this study (see Table~\ref{tab:models}). We additionally 
include the effects of the stellar motion in one run (Run-with), whereas the other simulation neglects it (Run-without). 
Fig.~\ref{fig:comparison}a shows the evolution of the accretion rate onto the protostar (in $\rm 
M_{\odot}\, \rm yr^{-1}$) in our comparison simulations as a function of time (in $\rm yr$). The 
Figure distinguishes between the model without (red solid line) and with (blue dashed line) stellar wobbling. The 
mass of the protostar is also plotted for the run without (thick dotted black line) 
and with (thick solid black line) wobbling. 
As the initial conditions are similar, the time of the free-fall collapse is the same in both runs 
($\approx\, 12\, \rm kyr$) and moderate accretion variability onto the MYSOs develops in the accretion flow 
similarly after the onset of the disc formation phase. 
In both cases (without and with stellar motion), the accretion rate is interspersed with episodic accretion spikes 
arising in addition to the baseline accretion rate of a few $\times 10^{-4}\, \rm M_{\odot}\, \rm yr^{-1}$. 
This demonstrates that, the stellar motion does not prevent the production of accretion-driven outbursts through 
the disc-fragmentation-based mechanism described in~\citet{meyer_mnras_464_2017} and~\citet{meyer_mnras_473_2018} 
and also has little effect on the final mass of the MYSOs ($\approx 25\, \rm M_{\odot}$).   
}

\textcolor{black}{
The thin red lines in Fig.~\ref{fig:comparison}b,c show the total luminosity of the MYSOs, i.e. the sum of the photospheric 
and accretion luminosity (in $L_{\odot}$) as a function of time (in $\rm kyr$) of both comparison 
models. In addition, the black solid lines show the background luminosity $L_{\rm bg}$, while the dotted thin blue line, dashed dotted orange line and dashed violet 
lines show the limits of the 1-, 3- and 4-magnitudes cutoffs with respect to $L_{\rm bg}$, respectively. 
The overplotted dots represent the peak accretion rates for each individual burst, the peak luminosity of which is at least 2.5 
times brighter than their quiescent pre-flare luminosity and they are coloured as a function of the  
peak accretion rate, shown in the $0.001$-$0.1\, \rm M_{\odot}\, \rm yr^{-1}$ range. 
As in our other models, the time evolution of the luminosity is a direct function of the mass accretion 
rate (Fig.~\ref{fig:comparison}a). In both comparison models (without and with wobbling), variations in the 
luminosity begins $t\approx 12\, \rm kyr$ when the accretion luminosity ceases to be negligible as compared 
to the photospheric luminosity. 
The protostellar accretion rate history then further affects the total luminosity of the protostars  
by producing variations of increasing intensity in their lightcurves, according to the mechanism 
described above in this work and in~\citet{meyer_mnras_464_2017}. Moreover, we 
recover the initial development of mild accretion-driven bursts with a peak accretion rate 
$\le 10^{-3}\, M_{\odot}\, \rm yr^{-1}$ at $\approx 17\, \rm kyr$ (1,2-mag bursts) 
that precede stronger flares with a peak accretion rate $\ge 10^{-1}\, M_{\odot}\, \rm yr^{-1}$ 
at times $\ge 29\, \rm yr$ (3,4-mag bursts). We report in Table~\ref{tab:1} 
the bursts statistics and properties (peak luminosity, peak accretion rate, duration). 
}

\textcolor{black}{
The similitude between the two comparison models goes beyond a simple general, 
qualitative aspect of their accretion rate histories and luminosity curves, but it also concerns 
the properties of the bursts themselves. 
The number of bursts is similar in both models. For example, we obtain the same number of bursts for the 
2- and 3-mag flares and find only a slight difference for the 1-mag bursts (9 in comparison to 6 flares 
when wobbling is included. We also note the presence 
of a 4-mag burst in the model with stellar wobbling. In both cases, the bursts have durations of 
the same order of magnitude, with 
$t_{\rm bst}^{\rm mean}=38\, \rm yr$ (without) and $29\, \rm yr$ (with), 
$t_{\rm bst}^{\rm mean}=18\, \rm yr$ (without) and $21\, \rm yr$ (with) and  
$t_{\rm bst}^{\rm mean}=5\, \rm yr$ (without) and $9\, \rm yr$ (with), for the 
1-, 2- and 3-mag bursts, respectively. Additionally, one can note that the total time 
the model with stellar wobbling spends in the burst phase ($300\, \rm yr$, representing 
$\approx 0.75\, \%$ of the modelled $40\, \rm kyr$) is slightly shorter than that without wobbling ($438\, \rm yr$, 
representing $\approx 1.1\, \%$ of the modelled $40\, \rm kyr$) while the accrete mass is larger (see below). 
The mean luminosity and mean peak accretion rates of the two models are equivalent. 
For example, for the 2-mag bursts $L_{\rm mean} \approx 4.64 \times 10^{5}$  
and $\dot{M}_{\rm mean} \approx 0.0343\, \rm M_{\odot}\, \rm yr^{-1}$ in the model without stellar wobbling 
and $L_{\rm mean} \approx 4.12 \times 10^{5}\, \rm L_{\odot}\, \rm yr^{-1}$  
and $\dot{M}_{\rm mean} \approx 0.0397\, \rm M_{\odot}\, \rm yr^{-1}$ in the model with stellar wobbling. 
Moreover, both models exhibit bursts happening either isolated or in clusters, which strengthens the 
similitude between the two comparison simulations. 
}

\textcolor{black}{
More importantly, the analysis of the fractions of final mass accreted as a function of the protostellar 
luminosity (Table~\ref{tab:2}) shows that the disc response to stellar motion  
does not dramatically affect our conclusions in terms of the accreted mass throughout the eruptive episodes. 
In both comparison models, the star accretes $\approx 24\, \rm M_{\odot}$ over $40\, \rm kyr$ and gains 
$18.64\, \rm M_{\odot}$ ($\approx 75\, \%$) and $15.46\, \rm M_{\odot}$ ($\approx 63\, \%$) of its final 
mass when $L_{\rm tot} \simeq L_{\rm bg}$, which is in accordance with what was found in our model Run-1-hr that 
was modelled over a similar timescale ($35\, \rm kyr$). 
Our results show that the model with stellar wobbling accretes more mass through bursts as compared to the simulation 
without wobbling, especially during the 2- and 3-mag bursts for which the accreted mass is doubled compared 
to the model without wobbling. Interestingly, our model with wobbling 
exhibits a 4-mag burst while the model without stellar wobbling does not (at least during the first 
$40\, \rm kyr$), as our simulation Run-long-4 with $\beta=4\, \%$ but $r_{\rm in}=20\, \rm au$ has its 
first 4-mag burst at $\approx 43\, \rm kyr$). 
This mean that \textcolor{black}{our} models with stellar motion form massive clumps earlier in the evolution that the 
models without wobbling. We conclude that the burst characteristics are only weakly sensitive to stellar motion. 
%
} 


\section{Discussion}
\label{sect:discussion}

In this section, we present the limitation and caveats of our model, estimate the burst occurence during 
the early protostellar formation phase and discuss the connection between bursts and protostellar 
mass growth. Furthermore, we review our knowledge of observations of several monitored bursts of 
massive protostars.

\subsection{Model limitation}
\label{sect:model}

The assumptions of our models are globally similar to those of~\citet{meyer_mnras_464_2017}, where we describe 
several improvements which could be added to our model. These improvements include making the initial conditions 
more realistic and by improving the grid geometry/resolution. 
Indeed, our parent pre-stellar cores are assumed to be \textcolor{black}{initially} spherically-symmetric, while high-mass stars also form in 
filamentary structures~\citep[see, e.g.][]{banerjee_mnras_373_2006}. 
The midplane-symmetry imposed in our simulations could be improved in favour of full three-dimensional protostellar 
disc models. 
Moreover, although the logarithmically-expanding grid in the radial direction permits fulfilling the 
Truelove criterion~\citep{truelove_apj_495_1998}, which warrants the correct treatment of self-gravity 
within the disc, a full convergence of the entire disc structure is not reached (if possible at all) 
and higher resolution models would allow us to investigate the clumps trajectories in the very inner disc. 
Nevertheless, we show in~\citet{meyer_mnras_473_2018} that our accretion rate histories qualitatively 
converge as the numerical resolution is increased.

As an improvement compared to the previous works in this series, the longer integration times of our 
simulations up to $60\, \rm kyr$ follow better the disc evolution, the aftermath of gravitational 
instabilities and/or clump migration in the protostellar lightcurves. 
Even longer runs should be conceived, e.g. in order to determinate the zero-age-main sequence of the 
MYSOs and see whether the discs modify their morphology and/or orientation as around evolved low-mass 
protostars~\citep{matsumoto_apj_839_2017}. Additionally, the huge parameter space should be explored 
within a large grid of simulations. 
One should nevertheless keep in mind that such improvements will make the simulations more computationally-intensive, 
which is at the moment not affordable and remains out of reach of modern supercomputing capabilities.  
%
Apart from the above mentioned caveats, the sink cell size \textcolor{black}{may affect} our results. 
%
%
%
A smaller size of the sink cell allows to further follow the clump migration, internal evolution and probable 
\textcolor{black}{stretching} and fragmentation. 
%
Using a smaller sink cell dramatically decreases the time-step of the simulations, which in turn 
makes them even more costly. A more thorough discussion of our model limitations can be 
found in~\citet{meyer_mnras_473_2018}.

\begin{figure}     
        \centering
        \begin{minipage}[b]{ 0.49\textwidth} 
        \centering
                \includegraphics[width=1.0\textwidth]{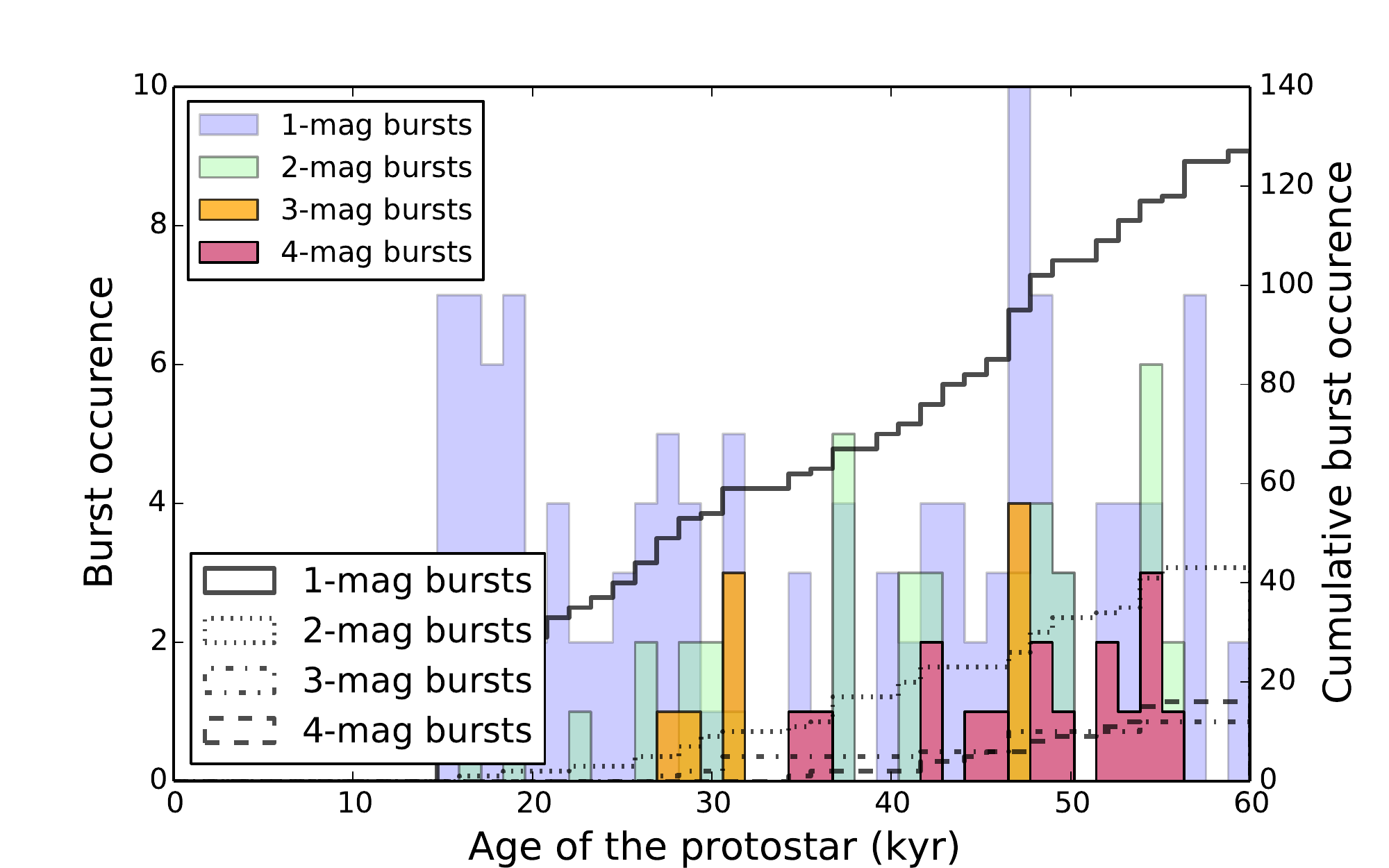}  
        \end{minipage}   
        \caption{ 
                 Bursts occurence as a function of the age of the protostar. 
                 }      
        \label{fig:hist1}  
\end{figure}

\subsection{Burst occurence}
\label{sect:bursts_prop_2}

Fig.~\ref{fig:hist1} plots the burst occurrence as a function of the 
\textcolor{black}{age of the protostar. More specifically, the histograms report the number of the bursts per time interval 
for different burst magnitudes (1-, 2-, 3-, and 4-mag). The lines show the cumulative burst occurrence 
for different burst magnitudes as a function of protostellar age. The colour} coding is similar as in 
Figs.~\ref{fig:correlation_1}-\ref{fig:correlation_3}. As indicated in Table~\ref{tab:1}, our 
MYSOs burst incidence is 128, 44, 12 and 17 for 1-mag (solid line), 2-mag (dotted line), 3-mag (dashed dotted line) and 
4-mag bursts (dashed line), respectively. Initially, the bursts are mostly 1-mag bursts up 
to $\approx 30\, \rm yr$, although run-long-5 has already some 3-mag bursts at a time $\approx 27\, \rm kyr$ 
(Fig.~\ref{fig:gen_plots_1}b) and run-long-10 shows 2-mag bursts at a time $\approx 20\, \rm kyr$ 
(Fig.~\ref{fig:gen_plots_1}c). As the protostars evolve towards the main sequence phase, the burst 
activity strongly increases, with a accumulation of 3-mag and 4-mag bursts starting from $\approx 40\, \rm kyr$, 
when our protostars experience a period of rapid and violent eruption (Fig.~\ref{fig:gen_plots_1}a) during 
which the probability to observe such phenomenon is the larger.

\begin{figure}
        \centering
        \begin{minipage}[b]{ 0.49\textwidth} 
        \centering
                \includegraphics[width=1.0\textwidth]{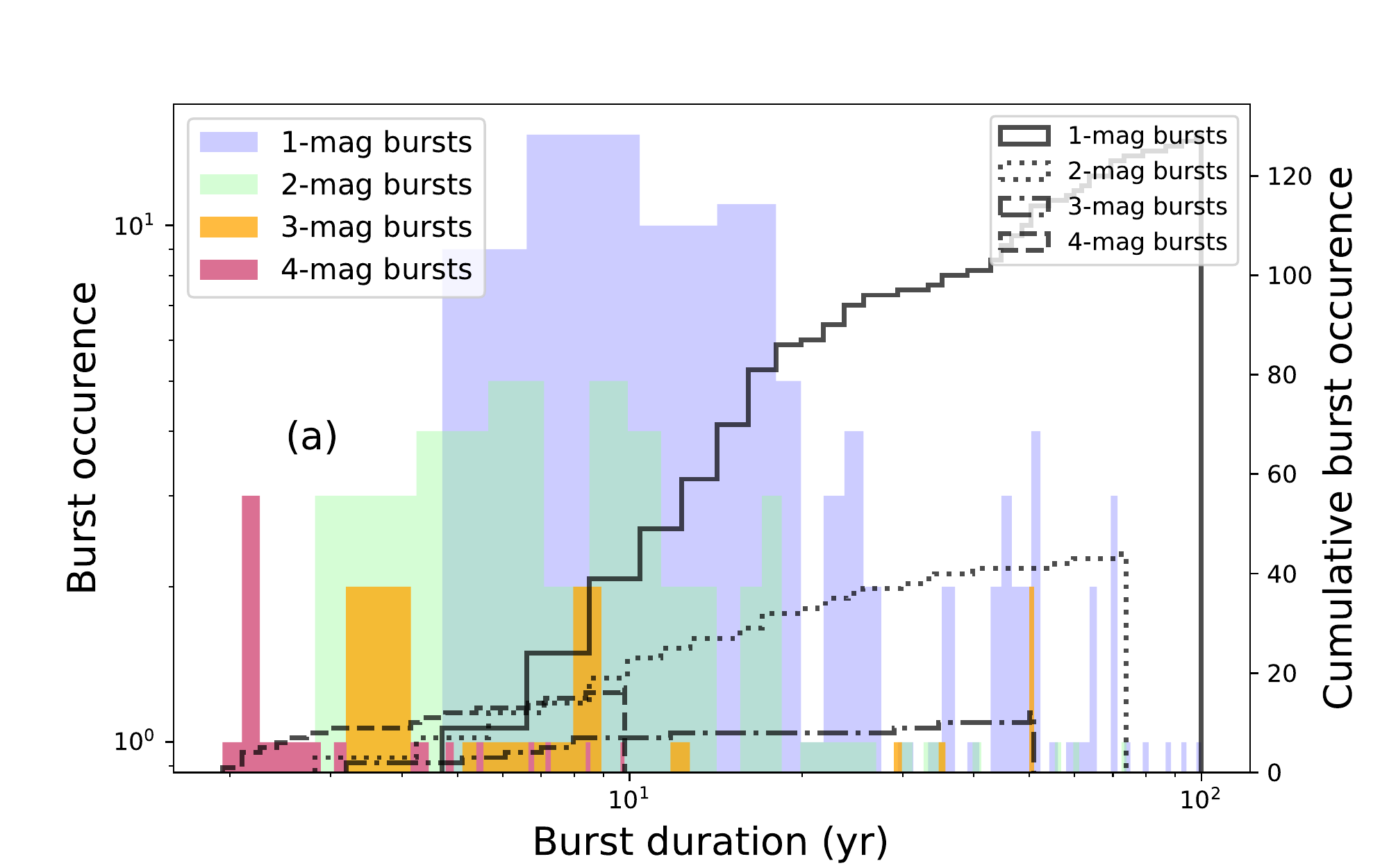}
        \end{minipage} \\
        \centering
        \begin{minipage}[b]{ 0.49\textwidth} 
        \centering
                \includegraphics[width=1.0\textwidth]{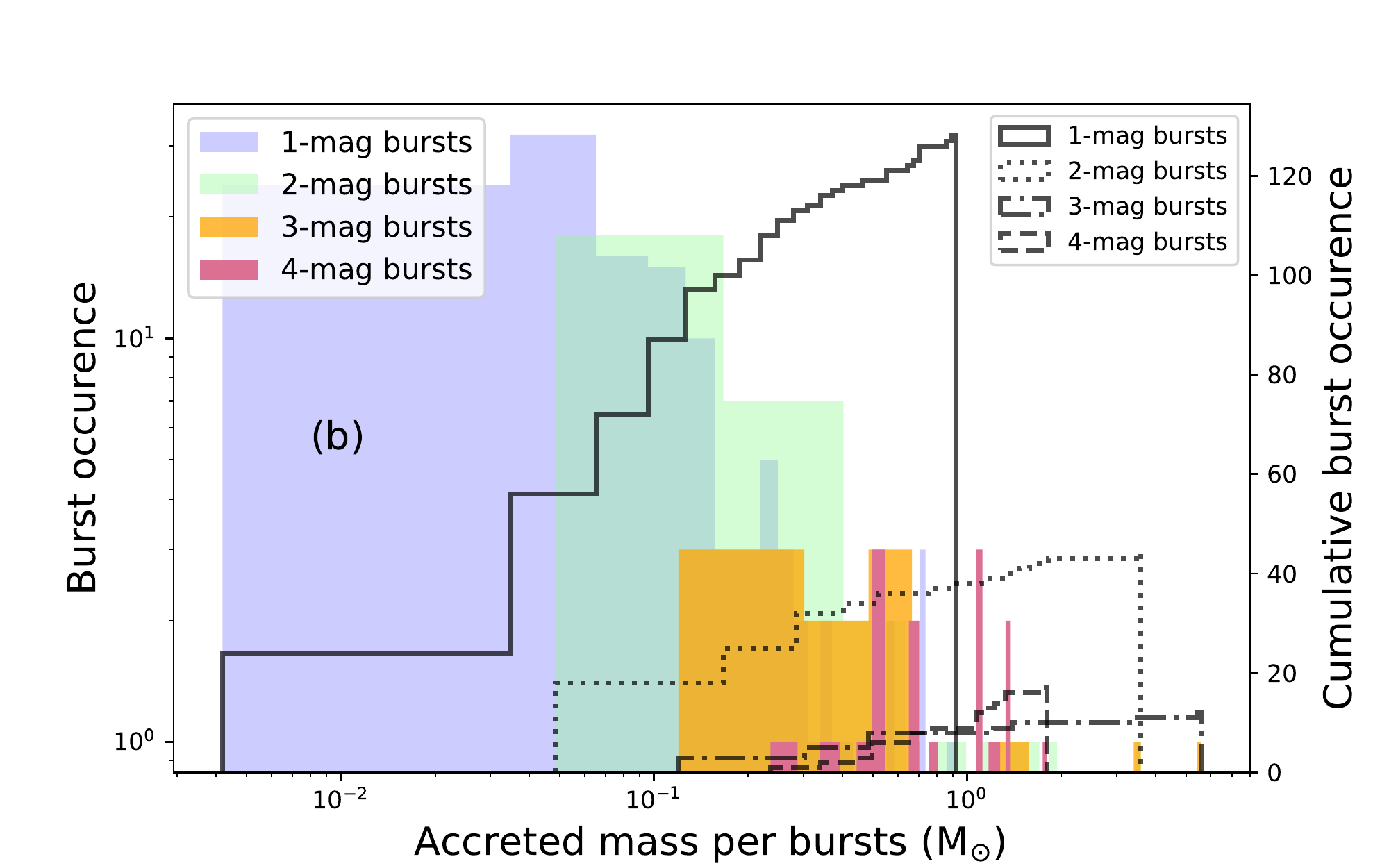}
        \end{minipage} \\              
        \centering
        \begin{minipage}[b]{ 0.49\textwidth} 
        \centering
                \includegraphics[width=1.0\textwidth]{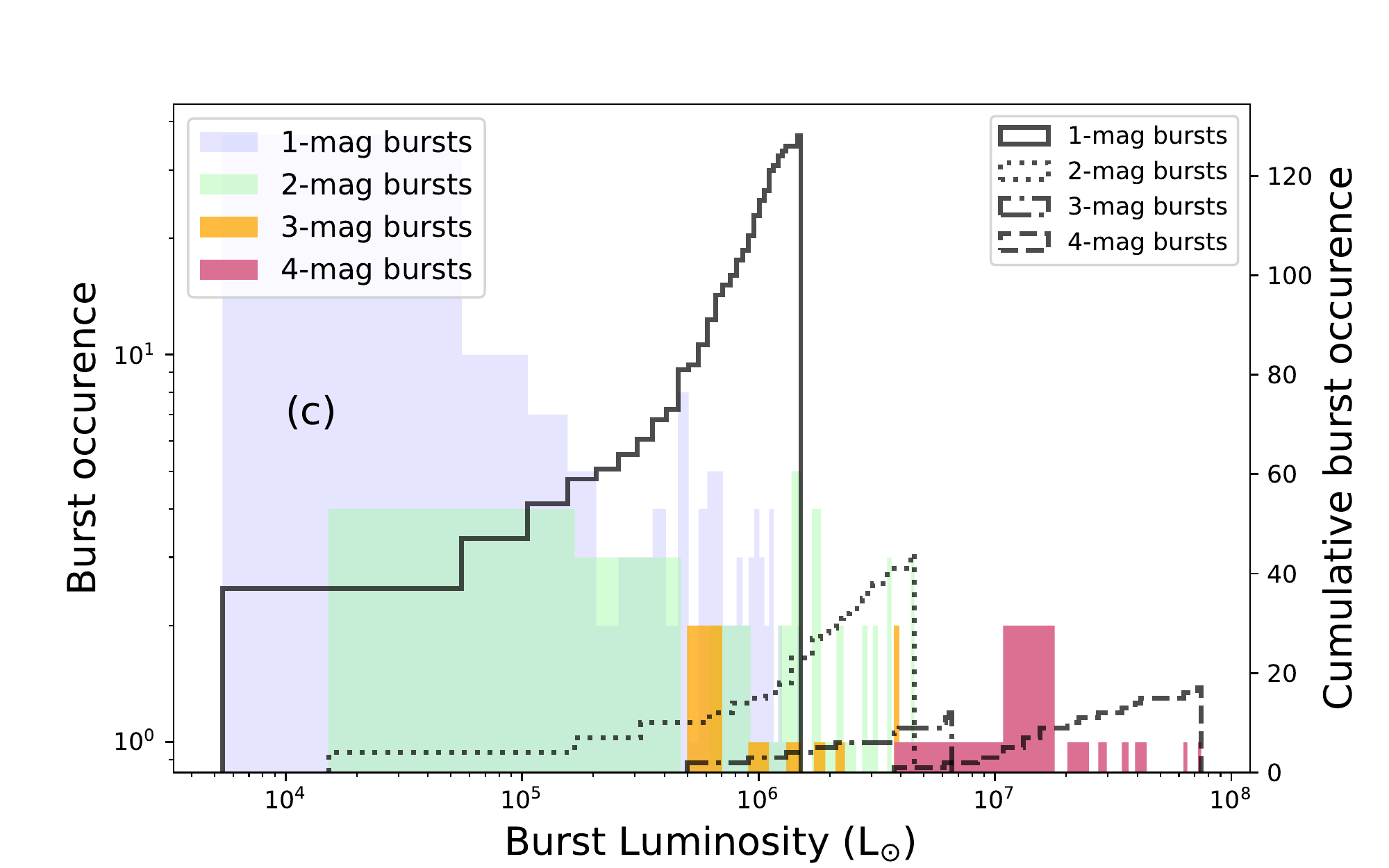}
        \end{minipage} \\        \centering
        \begin{minipage}[b]{ 0.49\textwidth} 
        \centering
                \includegraphics[width=1.0\textwidth]{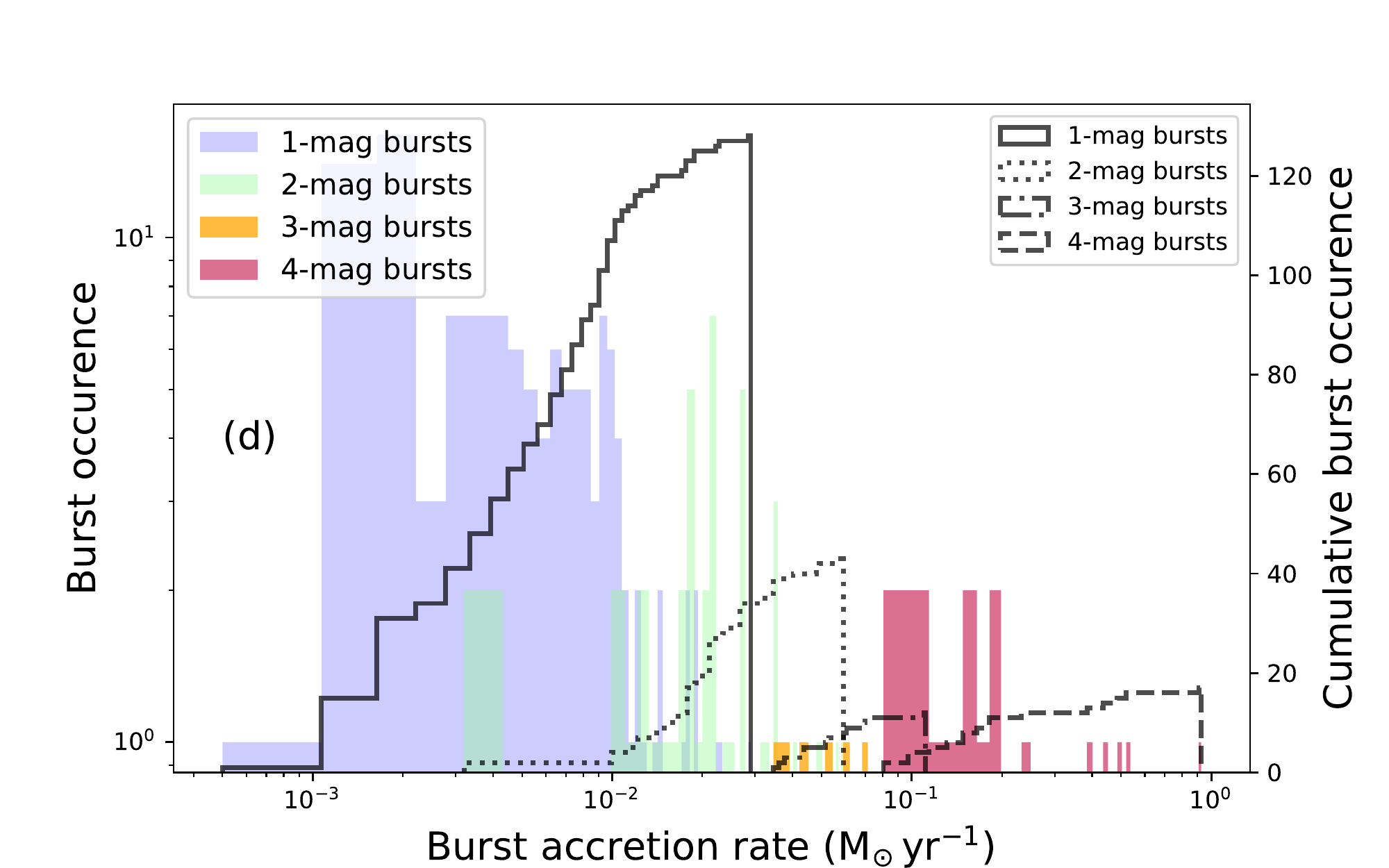} 
        \end{minipage}     
        \caption{ 
                 Burst occurence as a function of the burst duration (a), 
                 accreted mass per burst (b), burst peak luminosity (c) and 
                 burst peak accretion rate (d). 
                 }      
        \label{fig:hist2}  
\end{figure}

Fig.~\ref{fig:hist2} present\textcolor{black}{s} the number of bursts and the cumulative number of bursts as a function of 
the burst duration (a), accreted mass per bursts (b), burst peak luminosity (c) and burst accretion 
rate (d). The burst duration remains within the $2$-$100\, \rm yr$ limits, \textcolor{black}{with a} moderate dependence 
on the initial pre-stellar core spin, as a result of the convergence of our simulations. 
%
A higher-resolution model run-hr has about twice more bursts as run-long-4, despite similar initial 
conditions since the former better resolves and further follows the second collapse of the migrating 
clumps because of its smaller sink cell radius $r_{\rm in}$. As the higher-magnitude bursts are 
caused by more massive and compact clumps, these bursts are evidently shorter in time 
($t_{\mathrm{bst}}^{\mathrm{mean}}\approx\, 4\, \rm yr$) than lower-magnitude ones 
($t_{\mathrm{bst}}^{\mathrm{mean}}\approx\, 27\, \rm yr$, which are caused by more 
diffuse and extended clumps. 
The stronger the bursts the smaller their duration and therefore the less probable it is 
to observe such events of MYSOs, which raises the question of their possible observability. 
Our lower amplitude burst (1- and 2-mag) characteristics are nevertheless in rather good agreement 
with observations of bursts from massive protostars (Section~\ref{sect:obs}).

The accreted mass spans a range from a few Jupiter masses to a few solar masses and 
the accreted mass during the outburst of S255IR NIRS\,3 
lies in this range of values~\citep{caratti_nature_2016}. 
This distribution of accreted masses reflects the apparent variety of disk fragments, e.g. 
collapsing clumps or overdense portion of spiral arms formed via disk gravitational instability 
and fragmentation and migrated in the inner disc region. The strongest accretion 
events may be associated to the formation of low-mass companions to the central MYSOs. 
In~\citet{meyer_mnras_473_2018}, within the so-called disc fragmentation scenario for the 
formation of close/spectroscopic binaries around proto-OB stars, we showed the formation 
possibility of such objects, when migrating massive clumps may both lose their diffuse 
envelope and further contract into their center to form a secondary low-mass protostellar core. 
Such an event would then produce a companion and a strong (4-mag) burst from a single infalling dense 
clump. Since we do not consider this effect in the current study, the actual bursts may be of 
slightly lower amplitude than our model predicts. Interestingly, the number 
of 4-mag bursts predicted in our model (from 5 to 8) is consistent with the occurrence of close 
binaries around OB stars~\citep{chini_424_mnras_2012,2013A&A...550A..27M,2014ApJS..213...34K}.

\begin{figure*}
    \begin{minipage}[b]{ 0.85\textwidth}  
	\centering
        \includegraphics[width=1.0\textwidth]{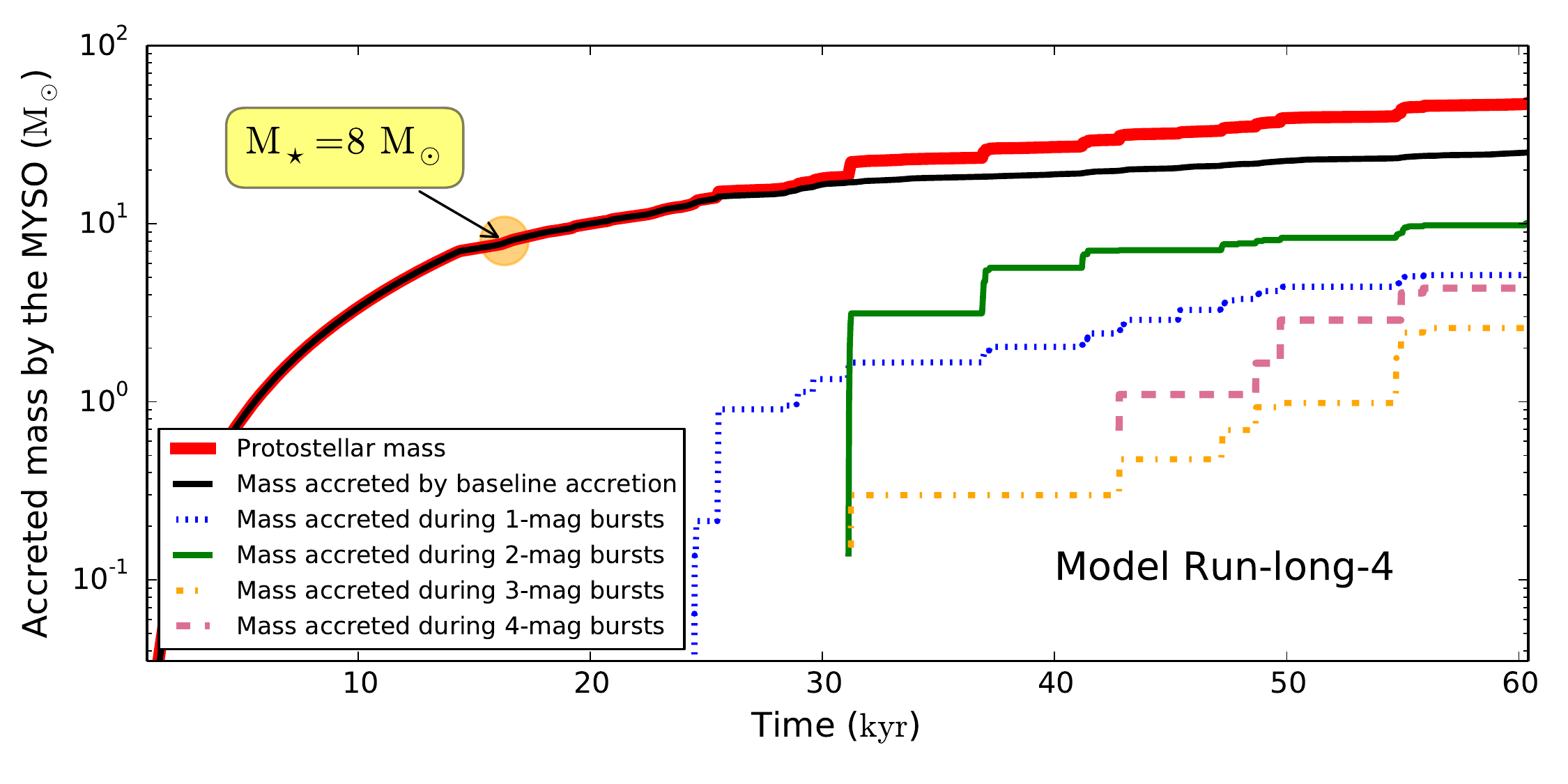}
        \end{minipage}           
        \caption{ 
                 Protostellar mass growth as a function of the star's total luminosity.
                 }      
        \label{fig:mass_accretion_details}  
\end{figure*}

The burst luminosity distribution clearly shows that 1-mag and 2-mag bursts are by far the more frequent 
(Fig.~\ref{fig:hist2}c), with a luminosity peak $\approx 10^{5}$-$10^{6}\, \rm L_{\odot}$. Other bursts 
of higher luminosity are not as common as their fainter counterparts, and, if some 3-mag bursts 
can reach luminosity peaks $\ge 10^{6}\, \rm L_{\odot}$, only 4-mag bursts can reach luminosities 
$\ge 10^{7}\, \rm L_{\odot}$. 
As was stated above, these strong bursts might be simultaneously associated with the formation of binaries 
and, thus, may be of slightly fainter intensity. Consequently and given their rareness, one should not 
expect to observe bursts from MYSOs more luminous than $\sim 10^{7}\, \rm L_{\odot}$, if at all. 
The last histogram (Fig.~\ref{fig:hist2}d) plots the burst peak accretion rate distribution. It underlines the correlation between 
$L_{\rm max}$ and $\dot{M}_{\rm max}$, already depicted in Fig.~\ref{fig:correlation_2}b: the stronger 
the burst the higher the peak accretion rate of circumstellar material. Most bursts peak at 
$\simeq 10^{2}\, \rm M_{\odot}\, \rm yr^{-1}$, while only a minor fraction of them exceed 
$\simeq 10^{-1}\, \rm M_{\odot}\, \rm yr^{-1}$. 
%

\subsection{Protostellar mass growth by episodic accretion}
\label{sect:mass_gained}

Fig.~\ref{fig:mass_accretion_details} plots the mass evolution of the protostar (in $\rm 
M_{\odot}$) in our Run-long-4 model as a function of time (in $\rm yr$). The Figure distinguishes 
between the total accreted mass of the protostar (thick solid red line) and the mass accreted 
when the protostar is in the quiescent phase (thin solid black line), expericing 1-mag (thin dotted 
blue line), 2-mag (thin dashed green line), 3-mag (solid orange line) and 4-mag (thin dotted dashed 
purple line) bursts. 
%
During the free-fall gravitational collapse and up to the end of the smooth accretion phase, the 
protostar grows exclusively by quiescent accretion driven by gravitational torques that are 
generated by anisotropies in the inner disc flow. In this case, variations in stellar brightness 
remains smaller than 1-mag, i.e. increase by a maximal factor of $\approx 2.5$ compared to 
$L_{\rm bg}$. When the young star enters the high-mass regime ($\ge 8\, \rm M_{\odot}$, orange circle), 
no clear bursts has yet happened, although accretion bursts start to occur in the models with a 
higher $\beta$ (Fig.~\ref{fig:gen_plots_2}).  
During the burst phase of accretion, more and more mass is accreted while the protostar experiences 
$\ge\, 2$-mag bursts. As was already discussed in our Table~\ref{tab:1}, the higher the burst 
intensity, the later the burst appear and the rarer they are, as testifies the decreasing number of 
step-like escalation of the mass histories at times $\ge 35\, \rm kyr$. The Figure also depicts the 
larger proportion of the 1- and 2-mag bursts compare to the less frequent 3- and 4-mag ones. All our 
models have similar behaviour.

In Table~\ref{tab:2}, we summarise the stellar mass accreted in the quiescent mass and during the bursts 
(for bursts of different magnitude) and express the results in terms of mass and percentage of the final 
protostellar mass, respectively. Our model Run-long-4 accretes $\approx\, 25.06\, \rm M_{\odot}$ in the 
quiescent phase and $\approx\, 5.28\, \rm M_{\odot}$, $\approx\, 10.04\, \rm M_{\odot}$, $\approx\, 2.59\, \rm M_{\odot}$ 
and $\approx\, 4.36\, \rm M_{\odot}$ while experiencing 1- to 4-mag bursts, respectively. 
%
%
By performing an average of the different values over our models with $r_{\rm in}=20\, \rm AU$, we find 
that protostars gain, during their early $60\, \rm kyr$, about $5.72$, $6.22$, $2.91$ and 
$4.21\, \rm M_{\odot}$ while experiencing 1-mag, 2-mag, 3-mag, and 4-mag bursts, respectively. 
In other words, a fraction of $\approx 53.75\, \%$ of the final mass is gained by quiescent accretion (when 
$L_{\rm tot}\simeq\, L_{\rm bg}$) while about $14.35\, \%$, $14.63\, \%$, $6.91\, \%$ and $10.36\, \%$ 
of the final mass is accreted during 1-mag to 4-mag accretion-driven events. 
Over the first $60\, \rm kyr$ of their life, as a direct consequence of the rareness of the strongest 
flares, the total time the MYSOs spend in 1-mag bursts ($\approx\, 1.27\, \%$ of $t_{\rm end}$) is longer 
than the cumulative time of all higher intensity bursts ($\approx\, 0.43\, \%$ of $t_{\rm end}$), which implies  
a very rare observation probability. The total burst time $\approx 1027\, \rm yr$ is $\approx\, 1.7\, \%$ of the 
$60\, \rm kyr$, but it permits the accretion of a significative fraction of about half ($\approx 46.25\, \%$) 
of their final mass, which demonstrates the importance and non-negligibility of the accretion bursts in 
the understanding of the early evolution and mass growth of MYSOs.  

\subsection{Comparison with observations}
\label{sect:obs}

The discovery and analysis of episodic bursts from massive protostars is 
hampered by various circumstances. According to the initial stellar mass 
function, these objects are much rarer than their low-mass counterparts and 
thus, on average, more distant. Their protostellar evolution occurs at much 
higher pace compared to low-mass young stellar objects, so they remain deeply 
embedded during most of this evolutionary period. This renders the discovery of 
burst-induced photometric variability impossible at optical and difficult at 
infrared wavelengths in contrast to smaller-mass YSOs~\citep{2013MNRAS.430.2910S}. Since the most massive young star dominates the emission 
of embedded clusters, attempts were made to trace its variability in scattered 
light~\citep{stecklum_2017a}. A remedy comes from the frequent association of 
high-mass young stellar objects with masers. In particular methanol Class~II 
masers are a reliable signpost of massive star 
formation~\citep{breen_mnras_435_2013}. These masers arise from molecules which 
are released from the ice mantles of grains to the gas phase by heating of the 
dust which surrounds the massive protostar. The maser excitation can be 
collisional, e.g. in shocks of bipolar outflows, or due to radiative pumping, as 
in the case of the 6.7\,GHz Class~II methanol 
masers~\citep{sobolev_aa_324_1997}. Thus, the temporary increase in luminosity 
due to an accretion burst may cause flares of radiatively pumped masers. 
However, other circumstances affect the maser strength as well. In fact, many of 
these masers show pronounced variability which can be even 
periodic~\citep{goedhart_mnras_339_2003}. The non-linear maser response in the 
non-saturated regime makes it difficult to judge the increase of the mid-IR 
radiation. Nevertheless, to some extent, the variability of Class II methanol 
masers may be considered to reflect luminosity variations of 
MYSOs~\citep{szymczak_mnras_474_2018}. Indeed, recent observations confirmed 
this view for two objects, described in the following.

First evidence for bursts from massive protostars was gained from the case of 
V723 Carinae~\citep{tapia_mnras_446_2015,tapia_mnras_448_2015}, a 
${\sim}$10\,M$_\odot$ MYSO with a post-burst luminosity of 
$4{\times}10^3$\,L$_{\odot}$. It was fainter than $K_s\!\simeq$\,16.6 in 1993 
and found to be three magnitudes brighter in 2003~\citep{tapia_mnras_367_2006}. 
The object reached its maximum $K_s$ brightness of 12.9 in 2004 February and 
shows erratic variability of up to $\Delta K_s=2$ since then. Its infrared 
colours resemble those of Class I objects and indicate the presence of very 
substantial extinction. Because of the sparse pre-burst IR photometry, no 
conclusion on the luminosity increase during the burst could be drawn. At the 
time of its discovery, V723 Carinae was the most luminous, most massive, most 
deeply embedded, and possibly youngest eruptive young variable.

This situation changed in 2015 when a flare of the 6.7\,GHz methanol maser in 
the S255-IR star forming region was announced~\citep{fujisawa_atel_2015}. 
Immediate NIR follow-up imaging disclosed that NIRS3, a $\sim$20\,M$_\odot$ MYSO 
representing the most luminous member of the embedded cluster, brightened by 
$\Delta K_s\!=2.5$ compared to the latest archival data from 
2009~\citep{stecklum_ATel_2016}. A thorough investigation of the NIRS3 burst was 
possible thanks to the observational advancement since the V723 Carinae event 
and the availability of spectroscopic as well as photometric pre-burst data. 
Major results include the discovery of spectral features which are typical for 
young low-mass eruptive variables - however three orders of magnitude stronger, 
the discovery of the light echo from the burst which allows to reconstruct its 
history, and an increase of the bolometric luminosity from 
2.9$\pm^{1}_{0.7}$)$\times$10$^4$\,L$_\odot$ to 
(1.6$\pm^{0.4}_{0.3}$)$\times$10$^5$\,L$_\odot$ which points to an accretion 
rate of (5$\pm$2)$\times$10$^{-3}$\,M$_\odot$\,yr$^{-1}$ 
\citep{caratti_nature_2016}. The burst was accompanied by a spatial 
rearrangement of 6.7\,GHz methanol masers \citep{moscadelli_aa_600_2017} at 
subluminal speed~\citep{stecklum_2017b} as well as enhanced outflow activity 
traced by an increase of the radio continuum emission, delayed by a about one 
year~\citep{cesaroni_aa_612_2018}.

\textcolor{black}{
Notably, there is evidence from the distribution of molecules around NIRS3 that 
its nearest vicinity may be considerably inhomogeneous~\citep{zinchenko_apj_810_2015,zinchenko_aa_606_2017}  
and may be subject to gravitational instability. Various authors reported about four 
bursts from NIRS3, including the recent one, within the last $\sim 7000$ 
years~\citep{wang_aa_527_2011,zinchenko_apj_810_2015,burns_mnras_460_2016}. The arrangement 
of bow shocks in the outflows in this region seems to support multiple 
outbursts~\citep{zinchenko_apj_810_2015}, although most recent ALMA data shows that the 
situation is complicated and the youngest of the bow shocks is produced by 
another source in the region, SMA2~\citep{2018IAUS..332..270Z}. 
}

This seems to indicate 
that 2015 burst is part of a series of eruptive events and, probably, more are 
to come in the future. Such remark can be applied to the other observed bursts 
from MYSOs, and, more generally, by analogy with low-mass star formation, we 
propose that episodic bursts from massive protostars are produced in groups of 
successive flares of different, possibly decreasing, magnitude. 
Some properties of the burst of S255-IR \textcolor{black}{derived in~\citet{caratti_nature_2016}} are consistent with our results. 
Its luminosity increased by a factor $3.6$-$5.1$ making it a 2-mag burst, 
which are amongst the most common bursts we predict. Its accretion rate of 
(5$\pm$2)$\times$10$^{-3}$\,M$_\odot$\,yr$^{-1}$ fits the lower limit of
accretion rates that we predict for 2-mag bursts (3$\times$10$^{-3}$\,M$_\odot$\,yr$^{-1}$). 
\textcolor{black}{
Note that this observed accretion rate has been estimated using a canonical
value for the protostellar radius which corresponds the standard radius of a
massive star of the mass of NIRS3 at the zero-age- main-sequence time. As
indicated in~\citet{caratti_nature_2016}, the accretion rate would be larger
if one assumes that the protostar is bloating. Indeed, if the radius is larger
more mass has to be accreted to produce a similar burst energy. One should also
keep in mind that our model accretion rates were derived neglecting the
formation of close (spectroscopic) companions to the MYSOs, which have been
shown to be consistent with the disc fragmentation and accretion-driven burst
scenario in the context of massive protostars~\citep{meyer_mnras_473_2018}. Hence, the
observed accretion rate, which corresponds well to our minimal predictions, is
probably a lower bound to its real value, and our predictions may be slightly
overestimated.
}

\textcolor{black}{
While the observed and model burst durations remain within the same order
of magnitude, a difference seems to be present nevertheless. The value of 
$1.5\, \, \rm yr$~\citep{caratti_nature_2016} is shorter than the lower limit of our
predictions of $3.0\, \, \rm yr$ for that kind of 2-mag bursts (our Table.~\ref{tab:1}). However,
very recent results obtained by monitoring the maser emission associated with
NIRS3 solves this discrepancy. The radio observations of~\citet{2018arXiv180707334S} 
show that the methanol maser flare of S255IR started in February 2014 already,
i.e. before the estimated date of~\citet{caratti_nature_2016}. This
difference is primarily due to their simplified initial analysis of light echo
which neglected multiple scattering. Therefore, the estimated burst duration of
$1.5\, \rm yr$ represents a lower bound. The radio observations point to an effective
burst length in the order of $\approx 2.0\, \, \rm yr$ which brings the burst of S255-IR closer
to our minimal prediction of $3.0\, \, \rm yr$.
}

\textcolor{black}{
By coincidence another MYSO accretion event was detected
at about the same time when NIRS3 was bursting.} However, 
the burst from NGC6334I-MM1 could be observed at 
mm/submm wavelengths only~\citep{2017arXiv170108637H}. From the quadruple rise 
of the dust continuum emission a sustained luminosity surge by a factor of 
70$\pm$20 was derived. \textcolor{black}{It suggests that the burst from NGC6334IMM1
was a major one which possibly lasts longer than that from NIRS3 which ceased 
after $\sim 2.0\, \rm yr$. This conjecture is also supported by the duration of 
the flare in different maser transitions, which lasts longer than 3 years~\citep{macleoad_MNRAS_478_2018}. 
Similar to NIRS3, this event was also accompanied by 
strong flares of $6.7\, \rm GHz$ methanol and other maser transitions of various molecules, including new 
components which emerged due to the burst~\citep{2018arXiv180102141H,macleoad_MNRAS_478_2018}. High resolution 
ALMA and VLA data revealed the presence of an outflow with a dynamic age of about $170\, \rm yr$, which suggests that 
there was another recent accretion burst. A decrease in the water maser emission around MM1 also suggests the 
presence of high infrared radiation density in this region~\citep{2018arXiv180904178B}. The analysis of 
the maser data provides evidence that there were two earlier maser flares, which hints at 
the recurrent nature of the flares in this object~\citep{macleoad_MNRAS_478_2018}.  
}


\section{Conclusion}
\label{sect:cc}

In this study, we have numerically explored the mass accretion history of a series of massive 
protostars forming \textcolor{black}{during} the gravitational collapse of $100\, \rm M_{\odot}$ 
pre-stellar clouds. Our three-dimensional gravito-radiation-hydrodynamics simulations follow the formation 
and early ($60\, \rm kyr$) evolution of gravitationally-unstable circumstellar discs that surround 
young protostars in the spirit of~\citet{meyer_mnras_464_2017,meyer_mnras_473_2018}. 
%
%
Massive fragments that form in the outer disk through gravitational fragmentation regularly generate 
sudden increases of the accretion rate when migrating towards the protostar and produce violent 
accretion-driven luminosity bursts accompanying the background variability of the protostellar lightcurves. 
We analyzed the characteristics, the properties and the occurrence of those bursts. 
All our MYSOs have accretion rate histories exhibiting a 
gradual increase in the accretion rate variability, from a smooth, burstless accretion phase during the 
initial gravitational collapse, through a variable accretion phase with $\dot{M}$ varying around the canonical value of 
$10^{-3}\, \rm M_{\odot}\, \rm yr^{-1}$~\citep{hosokawa_apj_691_2009} in the early $10\, \rm kyr$ of the disk 
evolution, and finally to a violent accretion phase with multiple luminous bursts. 
We classified the bursts in terms of their peak luminosity. More specifically, we defined 1-mag, 2-mag, 
3-mag, and 4-mag luminosity bursts as those having the peak luminosity factors $2.512$, 
$2.512^2\approx\, 6.3$, $2.512^3\approx\, 15.9$, and $2.512^4\approx\,  39$ higher 
than the luminosity in the quiescent phase.
\textcolor{black}{
Interestingly, the inclusion of stellar motion in the simulation slightly accelerates the 
fragmentation of the disc and the development of strong bursts during the early phase 
($\le 40\, \rm kyr$) of its evolution. 
However, the absence/presence in the numerical setup of the effects of the stellar motion  
neither strongly affected nor dramatically modified the burst occurrence and the main burst 
characteristics.
}
%

The main prediction of this study is the scaling between burst magnitude and flare duration: 
the more intense (3- and 4-mag) the burst, the rarer its occurrence and the shorter ($\simeq\, \rm yr$) 
its duration, and so is the associated observation probability. 
The burst occurrence increases as a function of the protostellar age, at least within the computed 
time limit of $60\, \rm kyr$. The same is true for the peak stellar luminosity during the burst.  
Very luminous (3- and 4-mag) outbursts are generated at later times ($\ge\, 40\, \rm kyr$), i.e. they 
are less frequent than more fainter flares associated with moderate accretion events. 
%
Our analysis demonstrates that the most luminous bursts accrete at 
higher rates and over smaller timescales than fainter and longer ones. 
While our MYSOs exhibit about a few dozen of 1-mag bursts over their early pre-main sequence phase, only about 
$4$-$8$ violent 4-mag bursts develop over the pre-main sequence phase. Their occurrence is consistent with the 
number of close/spectrosopic companions to the MYSOs~\citep{2013A&A...550A..27M,2014ApJS..213...34K}, with which 
such violent events may be simultaneously associated~\citep{meyer_mnras_473_2018}.  
%
We note that our limited number of simulations does no exhibit yet clear trends in terms of burst properties 
and/or occurrence as a function of the initial properties (rigidly-rotating cores with kinetic-to-gravitational 
energy ratios $\beta\, \approx\, 4\, $ to $10\%$), although bursts tend to appear slightly sooner as $\beta$ increases. 
%
More numerical investigations such as a parameter study exploring the pre-stellar core internal 
structure and providing us with more statistics, are required to address this question.

Although the total time our protostars spend in the burst phase represents only a small fraction 
($\approx\, 1.7\, \%$) of their early formation phase, the mass accreted during the bursts constitutes a 
significant fraction (up to $50\, \%$) of their final mass. The strongest, 4-mag bursts 
account for ($\approx\, 0.1\, \%$) of the early $60\, \rm kyr$ pre-main sequence life of MYSOs. 
While not being frequent, we should nevertheless expect modern observation facilities to record 
signatures of more and more similar bursts in the years to come.
%
%
Finally, we discuss our results in the light of our knowledge of observations of flares from MYSOs, 
i.e. the flare of the $10\, \rm M_{\odot}$ massive Class I eruptive variable V723 Carinae~\citep{tapia_mnras_367_2006}, the disk-mediated 
burst observed from the young massive stellar object S255IR NIRS\,3~\citep{caratti_nature_2016} and 
the eruption in the massive protostellar system NGC6334I-MM1~\citep{2017arXiv170108637H}.  
According to our study, we suggest that \textcolor{black}{the burst activity} of S255IR NIRS~3 and NGC6334I-MM1 
may have just started and it might further evolve as a series of flares.  
\textcolor{black}{
This conjecture is supported by the observational data hinting at a possible recurrent nature of the flares 
in these sources. 
}
Our study constitutes a step towards the understanding of the burst phenomenon 
in the high-mass star context, and further work as well as getting more observational data 
are necessary to accurately compare our theoretical predictions to measures and properly 
understand the mechanisms of bursts from young high-mass stars.

%
%


\section*{Acknowledgements}

\textcolor{black}{The authors thank the anonymous referee for their useful advice and suggestions which 
greatly improved the manuscript. } 
D.~M.-A.~Meyer thanks T.~Hosokawa for kindly sharing his pre-main sequence stellar evolutionary tracks
\textcolor{black}{and for discussions on stellar motion}, and W.~Kley for his expertise in 
radiation-hydrodynamics with the {\sc pluto} code. 
%
%
%
%
\textcolor{black}{
E. I. Vorobyov and A. M. Sobolev acknowledges support from the Russian Science Foundation grant 18-12-00193. 
}
%


\bibliographystyle{mn2e}

\footnotesize{
\bibliography{grid}

\begin{thebibliography}{}

\bibitem[\protect\citeauthoryear{{Abdo}, {Ackermann} \& {Ajello}}{{Abdo}
  et~al.}{2010}]{abdo_apj_722_2010}
{Abdo} A.~A.,  {Ackermann} M.,    {Ajello} M.,  2010, \apj, 722, 1303

\bibitem[\protect\citeauthoryear{{Arnaud}}{{Arnaud}}{1996}]{arnaud_aspc_101_19%
96}
{Arnaud} K.~A.,  1996, in {Jacoby} G.~H.,  {Barnes} J.,  eds, Astronomical Data
  Analysis Software and Systems V Vol.~101 of Astronomical Society of the
  Pacific Conference Series, {XSPEC: The First Ten Years}.
p.~17

\bibitem[\protect\citeauthoryear{{Aschenbach} \& {Leahy}}{{Aschenbach} \&
  {Leahy}}{1999}]{aschenbach_aa_341_1999}
{Aschenbach} B.,  {Leahy} D.~A.,  1999, \aap, 341, 602

\bibitem[\protect\citeauthoryear{{Asplund}, {Grevesse}, {Sauval} \&
  {Scott}}{{Asplund} et~al.}{2009}]{asplund_araa_47_2009}
{Asplund} M.,  {Grevesse} N.,  {Sauval} A.~J.,    {Scott} P.,  2009, \araa, 47,
  481

\bibitem[\protect\citeauthoryear{{Baade}}{{Baade}}{1938}]{baade_apj_88_1938}
{Baade} W.,  1938, \apj, 88, 285

\bibitem[\protect\citeauthoryear{{Badenes}, {Maoz} \& {Draine}}{{Badenes}
  et~al.}{2010}]{badenes_mnras_407_2010}
{Badenes} C.,  {Maoz} D.,    {Draine} B.~T.,  2010, \mnras, 407, 1301

\bibitem[\protect\citeauthoryear{{Baranov}, {Krasnobaev} \&
  {Kulikovskii}}{{Baranov} et~al.}{1971}]{baranov_sphd_15_1971}
{Baranov} V.~B.,  {Krasnobaev} K.~V.,    {Kulikovskii} A.~G.,  1971, Soviet
  Physics Doklady, 15, 791

\bibitem[\protect\citeauthoryear{{Bedogni} \& {D'Ercole}}{{Bedogni} \&
  {D'Ercole}}{1988}]{bedogni_190_aa_1988}
{Bedogni} R.,  {D'Ercole} A.,  1988, \aap, 190, 320

\bibitem[\protect\citeauthoryear{{Blaauw}}{{Blaauw}}{1993}]{blau1993ASPC...35.%
.207B}
{Blaauw} A.,  1993, in {Cassinelli} J.~P.,  {Churchwell} E.~B.,  eds, Massive
  Stars: Their Lives in the Interstellar Medium Vol.~35 of Astronomical Society
  of the Pacific Conference Series, {Massive Runaway Stars}.
p.~207

\bibitem[\protect\citeauthoryear{{Blandford}, {Kennel}, {McKee} \&
  {Ostriker}}{{Blandford} et~al.}{1983}]{blandford_301_natur_1983}
{Blandford} R.~D.,  {Kennel} C.~F.,  {McKee} C.~F.,    {Ostriker} J.~P.,  1983,
  \nat, 301, 586

\bibitem[\protect\citeauthoryear{{Blondin} \& {Koerwer}}{{Blondin} \&
  {Koerwer}}{1998}]{blondin_na_57_1998}
{Blondin} J.~M.,  {Koerwer} J.~F.,  1998, \na, 3, 571

\bibitem[\protect\citeauthoryear{{Blondin}, {Lundqvist} \&
  {Chevalier}}{{Blondin} et~al.}{1996}]{blondin_apj_472_1996}
{Blondin} J.~M.,  {Lundqvist} P.,    {Chevalier} R.~A.,  1996, \apj, 472, 257

\bibitem[\protect\citeauthoryear{{Borkowski}, {Blondin} \&
  {Sarazin}}{{Borkowski} et~al.}{1992}]{borkowski_apj_400_1992}
{Borkowski} K.~J.,  {Blondin} J.~M.,    {Sarazin} C.~L.,  1992, \apj, 400, 222

\bibitem[\protect\citeauthoryear{{Brighenti} \& {D'Ercole}}{{Brighenti} \&
  {D'Ercole}}{1994}]{brighenti_mnras_270_1994}
{Brighenti} F.,  {D'Ercole} A.,  1994, \mnras, 270, 65

\bibitem[\protect\citeauthoryear{{Brighenti} \& {D'Ercole}}{{Brighenti} \&
  {D'Ercole}}{1995a}]{brighenti_mnras_277_1995}
{Brighenti} F.,  {D'Ercole} A.,  1995a, \mnras, 277, 53

\bibitem[\protect\citeauthoryear{{Brighenti} \& {D'Ercole}}{{Brighenti} \&
  {D'Ercole}}{1995b}]{brighenti_mnras_273_1995}
{Brighenti} F.,  {D'Ercole} A.,  1995b, \mnras, 273, 443

\bibitem[\protect\citeauthoryear{{Brott}, {de Mink}, {Cantiello}, {Langer}, {de
  Koter}, {Evans}, {Hunter}, {Trundle} \& {Vink}}{{Brott}
  et~al.}{2011}]{brott_aa_530_2011a}
{Brott} I.,  {de Mink} S.~E.,  {Cantiello} M.,  {Langer} N.,  {de Koter} A.,
  {Evans} C.~J.,  {Hunter} I.,  {Trundle} C.,    {Vink} J.~S.,  2011, \aap,
  530, A115

\bibitem[\protect\citeauthoryear{{Bucciantini}, {Blondin}, {Del Zanna} \&
  {Amato}}{{Bucciantini} et~al.}{2003}]{bucciantini_aa_405_2003}
{Bucciantini} N.,  {Blondin} J.~M.,  {Del Zanna} L.,    {Amato} E.,  2003,
  \aap, 405, 617

\bibitem[\protect\citeauthoryear{{Chevalier}}{{Chevalier}}{1982}]{chevalier_ap%
j_258_1982}
{Chevalier} R.~A.,  1982, \apj, 258, 790

\bibitem[\protect\citeauthoryear{{Chevalier} \& {Liang}}{{Chevalier} \&
  {Liang}}{1989}]{chevalier_apj_344_1989}
{Chevalier} R.~A.,  {Liang} E.~P.,  1989, \apj, 344, 332

\bibitem[\protect\citeauthoryear{{Chiotellis}, {Schure} \& {Vink}}{{Chiotellis}
  et~al.}{2012}]{chiotellis_aa_537_2012}
{Chiotellis} A.,  {Schure} K.~M.,    {Vink} J.,  2012, \aap, 537, A139

\bibitem[\protect\citeauthoryear{{Chita}, {Langer}, {van Marle},
  {Garc{\'{\i}}a-Segura} \& {Heger}}{{Chita} et~al.}{2008}]{chita_aa_488_2008}
{Chita} S.~M.,  {Langer} N.,  {van Marle} A.~J.,  {Garc{\'{\i}}a-Segura} G.,
  {Heger} A.,  2008, \aap, 488, L37

\bibitem[\protect\citeauthoryear{{Ciotti} \& {D'Ercole}}{{Ciotti} \&
  {D'Ercole}}{1989}]{ciotti_aa_215_1989}
{Ciotti} L.,  {D'Ercole} A.,  1989, \aap, 215, 347

\bibitem[\protect\citeauthoryear{{Comer\'{o}n} \& {Kaper}}{{Comer\'{o}n} \&
  {Kaper}}{1998}]{comeron_aa_338_1998}
{Comer\'{o}n} F.,  {Kaper} L.,  1998, \aap, 338, 273

\bibitem[\protect\citeauthoryear{{Cowie} \& {McKee}}{{Cowie} \&
  {McKee}}{1977}]{cowie_apj_211_1977}
{Cowie} L.~L.,  {McKee} C.~F.,  1977, \apj, 211, 135

\bibitem[\protect\citeauthoryear{{Cox}, {Gull} \& {Green}}{{Cox}
  et~al.}{1991}]{cox_mnras_250_1991}
{Cox} C.~I.,  {Gull} S.~F.,    {Green} D.~A.,  1991, \mnras, 250, 750

\bibitem[\protect\citeauthoryear{{Cox}, {Kerschbaum}, {van Marle}, {Decin},
  {Ladjal} \& {Mayer}}{{Cox} et~al.}{2012}]{cox_aa_537_2012}
{Cox} N.~L.~J.,  {Kerschbaum} F.,  {van Marle} A.~J.,  {Decin} L.,  {Ladjal}
  D.,    {Mayer} A.,  2012, \aap, 543, C1

\bibitem[\protect\citeauthoryear{{Decin}, { }, {Royer}, {Van Marle},
  {Vandenbussche}, {Ladjal}, {Kerschbaum}, {Ottensamer}, {Barlow}, {Blommaert},
  {Gomez}, {Groenewegen}, {Lim}, {Swinyard}, {Waelkens} \& {Tielens}}{{Decin}
  et~al.}{2012}]{decin_aa_548_2012}
{Decin} L.,  { } N.~L.~J.,  {Royer} P.,  {Van Marle} A.~J.,  {Vandenbussche}
  B.,  {Ladjal} D.,  {Kerschbaum} F.,  {Ottensamer} R.,  {Barlow} M.~J.,
  {Blommaert} J.~A.~D.~L.,  {Gomez} H.~L.,  {Groenewegen} M.~A.~T.,  {Lim} T.,
  {Swinyard} B.~M.,  {Waelkens} C.,    {Tielens} A.~G.~G.~M.,  2012, \aap, 548,
  A113

\bibitem[\protect\citeauthoryear{{Dgani}, {van Buren} \&
  {Noriega-Crespo}}{{Dgani} et~al.}{1996}]{dgani_apj_461_1996}
{Dgani} R.,  {van Buren} D.,    {Noriega-Crespo} A.,  1996, \apj, 461, 927

\bibitem[\protect\citeauthoryear{{Dopita}}{{Dopita}}{1973}]{dopita_aa_29_1973}
{Dopita} M.~A.,  1973, \aap, 29, 387

\bibitem[\protect\citeauthoryear{{Dwarkadas}}{{Dwarkadas}}{2005}]{dwarkadas_ap%
j_630_2005}
{Dwarkadas} V.~V.,  2005, \apj, 630, 892

\bibitem[\protect\citeauthoryear{{Dwarkadas}}{{Dwarkadas}}{2007}]{dwarkadas_ap%
j_667_2007}
{Dwarkadas} V.~V.,  2007, \apj, 667, 226

\bibitem[\protect\citeauthoryear{{Eldridge}, {Langer} \& {Tout}}{{Eldridge}
  et~al.}{2011}]{eldridge_mnras_414_2011}
{Eldridge} J.~J.,  {Langer} N.,    {Tout} C.~A.,  2011, \mnras, 414, 3501

\bibitem[\protect\citeauthoryear{{Ferreira} \& {de Jager}}{{Ferreira} \& {de
  Jager}}{2008}]{ferreira_478_aa_2008}
{Ferreira} S.~E.~S.,  {de Jager} O.~C.,  2008, \aap, 478, 17

\bibitem[\protect\citeauthoryear{{Filippenko}}{{Filippenko}}{1997}]{filippenko%
_araa_35_1997}
{Filippenko} A.~V.,  1997, \araa, 35, 309

\bibitem[\protect\citeauthoryear{{Frail}, {Goss}, {Reynoso}, {Giacani}, {Green}
  \& {Otrupcek}}{{Frail} et~al.}{1996}]{frail_aj_111_1996}
{Frail} D.~A.,  {Goss} W.~M.,  {Reynoso} E.~M.,  {Giacani} E.~B.,  {Green}
  A.~J.,    {Otrupcek} R.,  1996, \aj, 111, 1651

\bibitem[\protect\citeauthoryear{{Gaensler}}{{Gaensler}}{1998}]{gaensler_apj_4%
93_1998}
{Gaensler} B.~M.,  1998, \apj, 493, 781

\bibitem[\protect\citeauthoryear{{Gaensler}}{{Gaensler}}{1999}]{gaensler_phd_1%
999}
{Gaensler} B.~M.,  1999, PhD thesis, University of Sydney

\bibitem[\protect\citeauthoryear{{Garcia-Segura}, {Langer} \& {Mac
  Low}}{{Garcia-Segura} et~al.}{1996}]{garciasegura_1996_aa_316}
{Garcia-Segura} G.,  {Langer} N.,    {Mac Low} M.-M.,  1996, \aap, 316, 133

\bibitem[\protect\citeauthoryear{{Gies}}{{Gies}}{1987}]{gies_apjs_64_1987}
{Gies} D.~R.,  1987, \apjs, 64, 545

\bibitem[\protect\citeauthoryear{{Gonz{\'a}lez-Casanova}, {De Colle},
  {Ramirez-Ruiz} \& {Lopez}}{{Gonz{\'a}lez-Casanova}
  et~al.}{2014}]{gonzalezcasanova_apj_781_2014}
{Gonz{\'a}lez-Casanova} D.~F.,  {De Colle} F.,  {Ramirez-Ruiz} E.,    {Lopez}
  L.~A.,  2014, \apjl, 781, L26

\bibitem[\protect\citeauthoryear{{Green}}{{Green}}{2009}]{green_cat_2009}
{Green} D.~A.,  2009, VizieR Online Data Catalog, 7253, 0

\bibitem[\protect\citeauthoryear{{Green} \& {Stephenson}}{{Green} \&
  {Stephenson}}{2003}]{green_lnp_598_2003}
{Green} D.~A.,  {Stephenson} F.~R.,  2003, in {Weiler} K.,  ed., Supernovae and
  Gamma-Ray Bursters Vol.~598 of Lecture Notes in Physics, Berlin Springer
  Verlag, {Historical Supernovae}.
pp 7--19

\bibitem[\protect\citeauthoryear{{Gvaramadze}, {Menten}, {Kniazev}, {Langer},
  {Mackey}, {Kraus}, {Meyer} \& {Kami{\'n}ski}}{{Gvaramadze}
  et~al.}{2014}]{Gvaramadze_2013}
{Gvaramadze} V.~V.,  {Menten} K.~M.,  {Kniazev} A.~Y.,  {Langer} N.,  {Mackey}
  J.,  {Kraus} A.,  {Meyer} D.~M.-A.,    {Kami{\'n}ski} T.,  2014, \mnras, 437,
  843

\bibitem[\protect\citeauthoryear{{Hobbs}, {Lorimer}, {Lyne} \&
  {Kramer}}{{Hobbs} et~al.}{2005}]{hibbs_360_mnras_2005}
{Hobbs} G.,  {Lorimer} D.~R.,  {Lyne} A.~G.,    {Kramer} M.,  2005, \mnras,
  360, 974

\bibitem[\protect\citeauthoryear{{Huthoff} \& {Kaper}}{{Huthoff} \&
  {Kaper}}{2002}]{huthoff_aa_383_2002}
{Huthoff} F.,  {Kaper} L.,  2002, \aap, 383, 999

\bibitem[\protect\citeauthoryear{{Kane}, {Drake} \& {Remington}}{{Kane}
  et~al.}{1999}]{kane_apj_511_1999}
{Kane} J.,  {Drake} R.~P.,    {Remington} B.~A.,  1999, \apj, 511, 335

\bibitem[\protect\citeauthoryear{{Katsuda}, {Tsunemi}, {Mori}, {Uchida},
  {Petre}, {Yamada} \& {Tamagawa}}{{Katsuda}
  et~al.}{2012}]{katsuda_apj_754_2012}
{Katsuda} S.,  {Tsunemi} H.,  {Mori} K.,  {Uchida} H.,  {Petre} R.,  {Yamada}
  S.,    {Tamagawa} T.,  2012, \apjl, 754, L7

\bibitem[\protect\citeauthoryear{Kaufman \& Kaufman}{Kaufman \&
  Kaufman}{2009}]{kaufman_EAS2009}
Kaufman M.,  Kaufman M.,  2009, EAS Publications Series, 34, 151

\bibitem[\protect\citeauthoryear{{Kothes}, {Fedotov}, {Foster} \&
  {Uyan{\i}ker}}{{Kothes} et~al.}{2006}]{kothes_aa_457_2006}
{Kothes} R.,  {Fedotov} K.,  {Foster} T.~J.,    {Uyan{\i}ker} B.,  2006, \aap,
  457, 1081

\bibitem[\protect\citeauthoryear{{Langer}}{{Langer}}{2012}]{langer_araa_50_201%
2}
{Langer} N.,  2012, \araa, 50, 107

\bibitem[\protect\citeauthoryear{{Langer}, {Garc{\'{\i}}a-Segura} \& {Mac
  Low}}{{Langer} et~al.}{1999}]{langer_ApJ_520_1999}
{Langer} N.,  {Garc{\'{\i}}a-Segura} G.,    {Mac Low} M.-M.,  1999, \apjl, 520,
  L49

\bibitem[\protect\citeauthoryear{{Lockett}, {Gauthier} \& {Elitzur}}{{Lockett}
  et~al.}{1999}]{lockett_apj_511_1999}
{Lockett} P.,  {Gauthier} E.,    {Elitzur} M.,  1999, \apj, 511, 235

\bibitem[\protect\citeauthoryear{{Lodders}}{{Lodders}}{2003}]{lodders_apj_591_%
2003}
{Lodders} K.,  2003, \apj, 591, 1220

\bibitem[\protect\citeauthoryear{{Lyne} \& {Lorimer}}{{Lyne} \&
  {Lorimer}}{1994}]{lyne_natur_369_1994}
{Lyne} A.~G.,  {Lorimer} D.~R.,  1994, \nat, 369, 127

\bibitem[\protect\citeauthoryear{{MacDonald} \& {Bailey}}{{MacDonald} \&
  {Bailey}}{1981}]{macdomald_mnras_197_1981}
{MacDonald} J.,  {Bailey} M.~E.,  1981, \mnras, 197, 995

\bibitem[\protect\citeauthoryear{{Mackey}, {Gvaramadze}, {Mohamed} \&
  {Langer}}{{Mackey} et~al.}{2014}]{mackey_sept_2014}
{Mackey} J.,  {Gvaramadze} V.~V.,  {Mohamed} S.,    {Langer} N.,  2014, ArXiv
  e-prints

\bibitem[\protect\citeauthoryear{{Mackey}, {Mohamed}, {Neilson}, {Langer} \&
  {Meyer}}{{Mackey} et~al.}{2012}]{mackey_apjlett_751_2012}
{Mackey} J.,  {Mohamed} S.,  {Neilson} H.~R.,  {Langer} N.,    {Meyer}
  D.~M.-A.,  2012, \apjl, 751, L10

\bibitem[\protect\citeauthoryear{{Manchester}}{{Manchester}}{1987}]{manchester%
_aa_171_1987}
{Manchester} R.~N.,  1987, \aap, 171, 205

\bibitem[\protect\citeauthoryear{{Medina}, {Raymond}, {Edgar}, {Caldwell},
  {Fesen} \& {Milisavljevic}}{{Medina} et~al.}{2014}]{medina_791_apj_2014}
{Medina} A.~A.,  {Raymond} J.~C.,  {Edgar} R.~J.,  {Caldwell} N.,  {Fesen}
  R.~A.,    {Milisavljevic} D.,  2014, \apj, 791, 30

\bibitem[\protect\citeauthoryear{{Meyer}, {Gvaramadze}, {Langer}, {Mackey},
  {Boumis} \& {Mohamed}}{{Meyer} et~al.}{2014}]{meyer_mnras_2013}
{Meyer} D.~M.-A.,  {Gvaramadze} V.~V.,  {Langer} N.,  {Mackey} J.,  {Boumis}
  P.,    {Mohamed} S.,  2014, \mnras, 439, L41

\bibitem[\protect\citeauthoryear{{Meyer}, {Mackey}, {Langer}, {Gvaramadze},
  {Mignone}, {Izzard} \& {Kaper}}{{Meyer} et~al.}{2014}]{meyer}
{Meyer} D.~M.-A.,  {Mackey} J.,  {Langer} N.,  {Gvaramadze} V.~V.,  {Mignone}
  A.,  {Izzard} R.~G.,    {Kaper} L.,  2014, \mnras, 444, 2754

\bibitem[\protect\citeauthoryear{{Mignone}, {Bodo}, {Massaglia}, {Matsakos},
  {Tesileanu}, {Zanni} \& {Ferrari}}{{Mignone}
  et~al.}{2007}]{mignone_apj_170_2007}
{Mignone} A.,  {Bodo} G.,  {Massaglia} S.,  {Matsakos} T.,  {Tesileanu} O.,
  {Zanni} C.,    {Ferrari} A.,  2007, \apjs, 170, 228

\bibitem[\protect\citeauthoryear{{Mignone}, {Zanni}, {Tzeferacos}, {van
  Straalen}, {Colella} \& {Bodo}}{{Mignone}
  et~al.}{2012}]{migmone_apjs_198_2012}
{Mignone} A.,  {Zanni} C.,  {Tzeferacos} P.,  {van Straalen} B.,  {Colella} P.,
     {Bodo} G.,  2012, \apjs, 198, 7

\bibitem[\protect\citeauthoryear{{Mohamed}, {Mackey} \& {Langer}}{{Mohamed}
  et~al.}{2012}]{mohamed_aa_541_2012}
{Mohamed} S.,  {Mackey} J.,    {Langer} N.,  2012, \aap, 541, A1

\bibitem[\protect\citeauthoryear{{Neufeld}, {Gusdorf}, {G{\"u}sten}, {Herczeg},
  {Kristensen}, {Melnick}, {Nisini}, {Ossenkopf}, {Tafalla} \& {van
  Dishoeck}}{{Neufeld} et~al.}{2014}]{neufeld_apj_781_2014}
{Neufeld} D.~A.,  {Gusdorf} A.,  {G{\"u}sten} R.,  {Herczeg} G.~J.,
  {Kristensen} L.,  {Melnick} G.~J.,  {Nisini} B.,  {Ossenkopf} V.,  {Tafalla}
  M.,    {van Dishoeck} E.~F.,  2014, \apj, 781, 102

\bibitem[\protect\citeauthoryear{{Noriega-Crespo}, {van Buren}, {Cao} \&
  {Dgani}}{{Noriega-Crespo} et~al.}{1997}]{noriegacrespo_aj_114_1997}
{Noriega-Crespo} A.,  {van Buren} D.,  {Cao} Y.,    {Dgani} R.,  1997, \aj,
  114, 837

\bibitem[\protect\citeauthoryear{{Orlando}, {Bocchino}, {Miceli}, {Petruk} \&
  {Pumo}}{{Orlando} et~al.}{2012}]{orlando_apj_749_2012}
{Orlando} S.,  {Bocchino} F.,  {Miceli} M.,  {Petruk} O.,    {Pumo} M.~L.,
  2012, \apj, 749, 156

\bibitem[\protect\citeauthoryear{{Orlando}, {Bocchino}, {Reale}, {Peres} \&
  {Pagano}}{{Orlando} et~al.}{2008}]{orlando_apj_678_2008}
{Orlando} S.,  {Bocchino} F.,  {Reale} F.,  {Peres} G.,    {Pagano} P.,  2008,
  \apj, 678, 274

\bibitem[\protect\citeauthoryear{{Orlando}, {Bocchino}, {Reale}, {Peres} \&
  {Petruk}}{{Orlando} et~al.}{2007}]{orlando_aa_470_2007}
{Orlando} S.,  {Bocchino} F.,  {Reale} F.,  {Peres} G.,    {Petruk} O.,  2007,
  \aap, 470, 927

\bibitem[\protect\citeauthoryear{{Osterbrock} \& {Bochkarev}}{{Osterbrock} \&
  {Bochkarev}}{1989}]{osterbrock_1989}
{Osterbrock} D.~E.,  {Bochkarev} N.~G.,  1989, \sovast, 33, 694

\bibitem[\protect\citeauthoryear{{Pannuti}, {Rho}, {Heinke} \&
  {Moffitt}}{{Pannuti} et~al.}{2014}]{pannuti_147_aj_2014}
{Pannuti} T.~G.,  {Rho} J.,  {Heinke} C.~O.,    {Moffitt} W.~P.,  2014, \aj,
  147, 55

\bibitem[\protect\citeauthoryear{{Parker}, {Phillipps}, {Pierce}, {Hartley},
  {Hambly}, {Read} \& {MacGillivray}}{{Parker}
  et~al.}{2005}]{parker_mnras_362_2005}
{Parker} Q.~A.,  {Phillipps} S.,  {Pierce} M.~J.,  {Hartley} M.,  {Hambly}
  N.~C.,  {Read} M.~A.,    {MacGillivray} 2005, \mnras, 362, 689

\bibitem[\protect\citeauthoryear{{P{\'e}rez-Rend{\'o}n}, {Garc{\'{\i}}a-Segura}
  \& {Langer}}{{P{\'e}rez-Rend{\'o}n} et~al.}{2009}]{peresrendon_aa_506_2009}
{P{\'e}rez-Rend{\'o}n} B.,  {Garc{\'{\i}}a-Segura} G.,    {Langer} N.,  2009,
  \aap, 506, 1249

\bibitem[\protect\citeauthoryear{{Petruk}, {Dubner}, {Castelletti}, {Bocchino},
  {Iakubovskyi}, {Kirsch}, {Miceli}, {Orlando} \& {Telezhinsky}}{{Petruk}
  et~al.}{2009}]{petruk_393_mnras_2009}
{Petruk} O.,  {Dubner} G.,  {Castelletti} G.,  {Bocchino} F.,  {Iakubovskyi}
  D.,  {Kirsch} M.~G.~F.,  {Miceli} M.,  {Orlando} S.,    {Telezhinsky} I.,
  2009, \mnras, 393, 1034

\bibitem[\protect\citeauthoryear{{Reach} \& {Rho}}{{Reach} \&
  {Rho}}{1999}]{reach_apj_511_1999}
{Reach} W.~T.,  {Rho} J.,  1999, \apj, 511, 836

\bibitem[\protect\citeauthoryear{{Reach}, {Rho}, {Tappe}, {Pannuti}, {Brogan},
  {Churchwell}, {Meade}, {Babler}, {Indebetouw} \& {Whitney}}{{Reach}
  et~al.}{2006}]{reach_aj_131_2006}
{Reach} W.~T.,  {Rho} J.,  {Tappe} A.,  {Pannuti} T.~G.,  {Brogan} C.~L.,
  {Churchwell} E.~B.,  {Meade} M.~R.,  {Babler} B.,  {Indebetouw} R.,
  {Whitney} B.~A.,  2006, \aj, 131, 1479

\bibitem[\protect\citeauthoryear{{Rozyczka} \& {Tenorio-Tagle}}{{Rozyczka} \&
  {Tenorio-Tagle}}{1995}]{rozyczka_274_MNRAS_1995}
{Rozyczka} M.,  {Tenorio-Tagle} G.,  1995, \mnras, 274, 1157

\bibitem[\protect\citeauthoryear{{Rozyczka}, {Tenorio-Tagle}, {Franco} \&
  {Bodenheimer}}{{Rozyczka} et~al.}{1993}]{rozyczka_mnras_261_1993}
{Rozyczka} M.,  {Tenorio-Tagle} G.,  {Franco} J.,    {Bodenheimer} P.,  1993,
  \mnras, 261, 674

\bibitem[\protect\citeauthoryear{{Schlegel}}{{Schlegel}}{1990}]{schlegel_mnras%
_244_1990}
{Schlegel} E.~M.,  1990, \mnras, 244, 269

\bibitem[\protect\citeauthoryear{{Schneiter}, {Vel{\'a}zquez}, {Reynoso} \& {de
  Colle}}{{Schneiter} et~al.}{2010}]{schneiter_mnras_408_2010}
{Schneiter} E.~M.,  {Vel{\'a}zquez} P.~F.,  {Reynoso} E.~M.,    {de Colle} F.,
  2010, \mnras, 408, 430

\bibitem[\protect\citeauthoryear{{Schure} \& {Bell}}{{Schure} \&
  {Bell}}{2013}]{schure_mnras_435_2013}
{Schure} K.~M.,  {Bell} A.~R.,  2013, \mnras, 435, 1174

\bibitem[\protect\citeauthoryear{{Spitzer}}{{Spitzer}}{1962}]{spitzer_1962}
{Spitzer} L.,  1962, {Physics of Fully Ionized Gases}

\bibitem[\protect\citeauthoryear{{Stevens}, {Blondin} \& {Pollock}}{{Stevens}
  et~al.}{1992}]{stevens_apj_386_1992}
{Stevens} I.~R.,  {Blondin} J.~M.,    {Pollock} A.~M.~T.,  1992, \apj, 386, 265

\bibitem[\protect\citeauthoryear{{Stone} \& {Norman}}{{Stone} \&
  {Norman}}{1992}]{stone_apjs_80_1992}
{Stone} J.~M.,  {Norman} M.~L.,  1992, \apjs, 80, 753

\bibitem[\protect\citeauthoryear{{Sutherland} \& {Dopita}}{{Sutherland} \&
  {Dopita}}{1993}]{sutherland_apjs_88_1993}
{Sutherland} R.~S.,  {Dopita} M.~A.,  1993, \apjs, 88, 253

\bibitem[\protect\citeauthoryear{{Tenorio-Tagle}, {Bodenheimer}, {Franco} \&
  {Rozyczka}}{{Tenorio-Tagle} et~al.}{1990}]{tenoriotagle_mnras_244_1990}
{Tenorio-Tagle} G.,  {Bodenheimer} P.,  {Franco} J.,    {Rozyczka} M.,  1990,
  \mnras, 244, 563

\bibitem[\protect\citeauthoryear{{Tenorio-Tagle}, {Rozyczka}, {Franco} \&
  {Bodenheimer}}{{Tenorio-Tagle} et~al.}{1991}]{tenoriotagle_mnras_251_1991}
{Tenorio-Tagle} G.,  {Rozyczka} M.,  {Franco} J.,    {Bodenheimer} P.,  1991,
  \mnras, 251, 318

\bibitem[\protect\citeauthoryear{{Tenorio-Tagle}, {Rozyczka} \&
  {Yorke}}{{Tenorio-Tagle} et~al.}{1985}]{tenoriotagle_aa_148_1985}
{Tenorio-Tagle} G.,  {Rozyczka} M.,    {Yorke} H.~W.,  1985, \aap, 148, 52

\bibitem[\protect\citeauthoryear{{Toledo-Roy}, {Esquivel}, {Vel{\'a}zquez} \&
  {Reynoso}}{{Toledo-Roy} et~al.}{2014}]{toledo_mnras_442_2014}
{Toledo-Roy} J.~C.,  {Esquivel} A.,  {Vel{\'a}zquez} P.~F.,    {Reynoso} E.~M.,
   2014, \mnras, 442, 229

\bibitem[\protect\citeauthoryear{{Truelove} \& {McKee}}{{Truelove} \&
  {McKee}}{1999}]{truelove_apjs_120_1999}
{Truelove} J.~K.,  {McKee} C.~F.,  1999, \apjs, 120, 299

\bibitem[\protect\citeauthoryear{{Uchida}, {Tsunemi}, {Katsuda}, {Kimura} \&
  {Kosugi}}{{Uchida} et~al.}{2009}]{uchida_pasj_61_2009}
{Uchida} H.,  {Tsunemi} H.,  {Katsuda} S.,  {Kimura} M.,    {Kosugi} H.,  2009,
  \pasj, 61, 301

\bibitem[\protect\citeauthoryear{{van Dishoeck}, {Jansen} \& {Phillips}}{{van
  Dishoeck} et~al.}{1993}]{vandishoeck_aa_279_1993}
{van Dishoeck} E.~F.,  {Jansen} D.~J.,    {Phillips} T.~G.,  1993, \aap, 279,
  541

\bibitem[\protect\citeauthoryear{{van Marle}, {Decin} \& {Meliani}}{{van Marle}
  et~al.}{2014}]{vanmarle_aa_561_2014}
{van Marle} A.~J.,  {Decin} L.,    {Meliani} Z.,  2014, \aap, 561, A152

\bibitem[\protect\citeauthoryear{{van Marle}, {Langer}, {Yoon} \&
  {Garc{\'{\i}}a-Segura}}{{van Marle} et~al.}{2008}]{vanmarle_aa_478_2008}
{van Marle} A.~J.,  {Langer} N.,  {Yoon} S.-C.,    {Garc{\'{\i}}a-Segura} G.,
  2008, \aap, 478, 769

\bibitem[\protect\citeauthoryear{{van Marle}, {Meliani}, {Keppens} \&
  {Decin}}{{van Marle} et~al.}{2011}]{vanmarle_apj_734_2011}
{van Marle} A.~J.,  {Meliani} Z.,  {Keppens} R.,    {Decin} L.,  2011, \apjl,
  734, L26

\bibitem[\protect\citeauthoryear{{van Marle}, {Smith}, {Owocki} \& {van
  Veelen}}{{van Marle} et~al.}{2010}]{vanmarle_mnras_407_2010}
{van Marle} A.~J.,  {Smith} N.,  {Owocki} S.~P.,    {van Veelen} B.,  2010,
  \mnras, 407, 2305

\bibitem[\protect\citeauthoryear{{van Veelen}, {Langer}, {Vink},
  {Garc{\'{\i}}a-Segura} \& {van Marle}}{{van Veelen}
  et~al.}{2009}]{vanveelen_aa_50._2009}
{van Veelen} B.,  {Langer} N.,  {Vink} J.,  {Garc{\'{\i}}a-Segura} G.,    {van
  Marle} A.~J.,  2009, \aap, 503, 495

\bibitem[\protect\citeauthoryear{{Vel{\'a}zquez}, {Martinell}, {Raga} \&
  {Giacani}}{{Vel{\'a}zquez} et~al.}{2004}]{velazquez_apj_601_2004}
{Vel{\'a}zquez} P.~F.,  {Martinell} J.~J.,  {Raga} A.~C.,    {Giacani} E.~B.,
  2004, \apj, 601, 885

\bibitem[\protect\citeauthoryear{{Vel{\'a}zquez}, {Vigh}, {Reynoso},
  {G{\'o}mez} \& {Schneiter}}{{Vel{\'a}zquez}
  et~al.}{2006}]{velazquez_apj_649_2006}
{Vel{\'a}zquez} P.~F.,  {Vigh} C.~D.,  {Reynoso} E.~M.,  {G{\'o}mez} D.~O.,
  {Schneiter} E.~M.,  2006, \apj, 649, 779

\bibitem[\protect\citeauthoryear{{Vigh}, {Vel{\'a}zquez}, {G{\'o}mez},
  {Reynoso}, {Esquivel} \& {Matias Schneiter}}{{Vigh}
  et~al.}{2011}]{vigh_apj_727_2011}
{Vigh} C.~D.,  {Vel{\'a}zquez} P.~F.,  {G{\'o}mez} D.~O.,  {Reynoso} E.~M.,
  {Esquivel} A.,    {Matias Schneiter} E.,  2011, \apj, 727, 32

\bibitem[\protect\citeauthoryear{{Vink}}{{Vink}}{2012}]{vink_aarv_20_2012}
{Vink} J.,  2012, \aapr, 20, 49

\bibitem[\protect\citeauthoryear{{Vink}, {Kaastra} \& {Bleeker}}{{Vink}
  et~al.}{1996}]{vink_aa_307_1996}
{Vink} J.,  {Kaastra} J.~S.,    {Bleeker} J.~A.~M.,  1996, \aap, 307, L41

\bibitem[\protect\citeauthoryear{{Vink}, {Kaastra} \& {Bleeker}}{{Vink}
  et~al.}{1997}]{vink_aa_328_1997}
{Vink} J.,  {Kaastra} J.~S.,    {Bleeker} J.~A.~M.,  1997, \aap, 328, 628

\bibitem[\protect\citeauthoryear{{Vink}}{{Vink}}{2006}]{vink_asp_353_2006}
{Vink} J.~S.,  2006, in {Lamers} H.~J.~G.~L.~M.,  {Langer} N.,  {Nugis} T.,
  {Annuk} K.,  eds, Stellar Evolution at Low Metallicity: Mass Loss,
  Explosions, Cosmology Vol.~353 of Astronomical Society of the Pacific
  Conference Series, {Massive star feedback -- from the first stars to the
  present}.
p.~113

\bibitem[\protect\citeauthoryear{{Vishniac}}{{Vishniac}}{1994}]{vishniac_apj_4%
28_1994}
{Vishniac} E.~T.,  1994, \apj, 428, 186

\bibitem[\protect\citeauthoryear{{Wang}, {Dyson} \& {Kahn}}{{Wang}
  et~al.}{1993}]{wang_MNRAS_261_1993}
{Wang} L.,  {Dyson} J.~E.,    {Kahn} F.~D.,  1993, \mnras, 261, 391

\bibitem[\protect\citeauthoryear{{Weaver}, {McCray}, {Castor}, {Shapiro} \&
  {Moore}}{{Weaver} et~al.}{1977}]{weaver_apj_218_1977}
{Weaver} R.,  {McCray} R.,  {Castor} J.,  {Shapiro} P.,    {Moore} R.,  1977,
  \apj, 218, 377

\bibitem[\protect\citeauthoryear{{Whalen}, {van Veelen}, {O'Shea} \&
  {Norman}}{{Whalen} et~al.}{2008}]{whalen_apj_682_2008}
{Whalen} D.,  {van Veelen} B.,  {O'Shea} B.~W.,    {Norman} M.~L.,  2008, \apj,
  682, 49

\bibitem[\protect\citeauthoryear{{Whiteoak} \& {Green}}{{Whiteoak} \&
  {Green}}{1996}]{whiteoak_aas_118_1996}
{Whiteoak} J.~B.~Z.,  {Green} A.~J.,  1996, \aaps, 118, 329

\bibitem[\protect\citeauthoryear{{Wolfire}, {McKee}, {Hollenbach} \&
  {Tielens}}{{Wolfire} et~al.}{2003}]{wolfire_apj_587_2003}
{Wolfire} M.~G.,  {McKee} C.~F.,  {Hollenbach} D.,    {Tielens} A.~G.~G.~M.,
  2003, \apj, 587, 278

\bibitem[\protect\citeauthoryear{{Woosley}, {Heger} \& {Weaver}}{{Woosley}
  et~al.}{2002}]{woosley_rvmp_74_2002}
{Woosley} S.~E.,  {Heger} A.,    {Weaver} T.~A.,  2002, Reviews of Modern
  Physics, 74, 1015

\bibitem[\protect\citeauthoryear{{Yusef-Zadeh}, {Wardle}, {Rho} \&
  {Sakano}}{{Yusef-Zadeh} et~al.}{2003}]{yusefzadeh_apj_585_2003}
{Yusef-Zadeh} F.,  {Wardle} M.,  {Rho} J.,    {Sakano} M.,  2003, \apj, 585,
  319

\bibitem[\protect\citeauthoryear{{Zavlin}, {Pavlov} \& {Trumper}}{{Zavlin}
  et~al.}{1998}]{zavlin_aa_331_1998}
{Zavlin} V.~E.,  {Pavlov} G.~G.,    {Trumper} J.,  1998, \aap, 331, 821

\end{thebibliography}


\begin{thebibliography}{}

\bibitem[\protect\citeauthoryear{{Adams}, {Ruden} \& {Shu}}{{Adams}
  et~al.}{1989}]{adam_apj_347_1989}
{Adams} F.~C.,  {Ruden} S.~P.,    {Shu} F.~H.,  1989, \apj, 347, 959

\bibitem[\protect\citeauthoryear{{Banerjee}, {Pudritz} \&
  {Anderson}}{{Banerjee} et~al.}{2006}]{banerjee_mnras_373_2006}
{Banerjee} R.,  {Pudritz} R.~E.,    {Anderson} D.~W.,  2006, \mnras, 373, 1091

\bibitem[\protect\citeauthoryear{{Beuther}, {Walsh}, {Johnston}, {Henning},
  {Kuiper}, {Longmore} \& {Walmsley}}{{Beuther}
  et~al.}{2017}]{beuther_aa_603_2017}
{Beuther} H.,  {Walsh} A.~J.,  {Johnston} K.~G.,  {Henning} T.,  {Kuiper} R.,
  {Longmore} S.~N.,    {Walmsley} C.~M.,  2017, \aap, 603, A10

\bibitem[\protect\citeauthoryear{{Bitsch}, {Morbidelli}, {Lega} \&
  {Crida}}{{Bitsch} et~al.}{2014}]{bitsch_aa_564_2014}
{Bitsch} B.,  {Morbidelli} A.,  {Lega} E.,    {Crida} A.,  2014, \aap, 564,
  A135

\bibitem[\protect\citeauthoryear{{Black} \& {Bodenheimer}}{{Black} \&
  {Bodenheimer}}{1975}]{black_apj_199_1975}
{Black} D.~C.,  {Bodenheimer} P.,  1975, \apj, 199, 619

\bibitem[\protect\citeauthoryear{{Bonnell} \& {Bate}}{{Bonnell} \&
  {Bate}}{1994}]{bonnell_mnras_271_1994}
{Bonnell} I.~A.,  {Bate} M.~R.,  1994, \mnras, 271

\bibitem[\protect\citeauthoryear{{Bonnell}, {Bate} \& {Zinnecker}}{{Bonnell}
  et~al.}{1998}]{1998MNRAS.298...93B}
{Bonnell} I.~A.,  {Bate} M.~R.,    {Zinnecker} H.,  1998, \mnras, 298, 93

\bibitem[\protect\citeauthoryear{{Breen}, {Ellingsen}, {Contreras}, {Green},
  {Caswell}, {Stevens}, {Dawson} \& {Voronkov}}{{Breen}
  et~al.}{2013}]{breen_mnras_435_2013}
{Breen} S.~L.,  {Ellingsen} S.~P.,  {Contreras} Y.,  {Green} J.~A.,  {Caswell}
  J.~L.,  {Stevens} J.~B.,  {Dawson} J.~R.,    {Voronkov} M.~A.,  2013, \mnras,
  435, 524

\bibitem[\protect\citeauthoryear{{Brogan}, {Hunter}, {Cyganowski}, {Chibueze},
  {Friesen}, {Hirota}, {MacLeod}, {McGuire} \& {Sobolev}}{{Brogan}
  et~al.}{2018}]{2018arXiv180904178B}
{Brogan} C.~L.,  {Hunter} T.~R.,  {Cyganowski} C.~J.,  {Chibueze} J.~O.,
  {Friesen} R.~K.,  {Hirota} T.,  {MacLeod} G.~C.,  {McGuire} B.~A.,
  {Sobolev} A.~M.,  2018, \apj, 866, 87

\bibitem[\protect\citeauthoryear{{Burns}}{{Burns}}{2018}]{arXiv180102211B}
{Burns} R.~A.,  2018, in {Tarchi} A.,  {Reid} M.~J.,   {Castangia} P.,  eds,
  Astrophysical Masers: Unlocking the Mysteries of the Universe Vol.~336 of IAU
  Symposium, {Water masers in bowshocks: Addressing the radiation pressure
  problem of massive star formation}.
pp 263--266

\bibitem[\protect\citeauthoryear{{Burns}, {Handa}, {Imai}, {Nagayama},
  {Omodaka}, {Hirota}, {Motogi}, {van Langevelde} \& {Baan}}{{Burns}
  et~al.}{2017}]{burns_mnras_467_2017}
{Burns} R.~A.,  {Handa} T.,  {Imai} H.,  {Nagayama} T.,  {Omodaka} T.,
  {Hirota} T.,  {Motogi} K.,  {van Langevelde} H.~J.,    {Baan} W.~A.,  2017,
  \mnras, 467, 2367

\bibitem[\protect\citeauthoryear{{Burns}, {Handa}, {Nagayama}, {Sunada} \&
  {Omodaka}}{{Burns} et~al.}{2016}]{burns_mnras_460_2016}
{Burns} R.~A.,  {Handa} T.,  {Nagayama} T.,  {Sunada} K.,    {Omodaka} T.,
  2016, \mnras, 460, 283

\bibitem[\protect\citeauthoryear{{Caratti o Garatti}, {Stecklum}, {Garcia
  Lopez}, {Eisloffel}, {Ray}, {Sanna}, {Cesaroni}, {Walmsley}, {Oudmaijer}, {de
  Wit}, {Moscadelli}, {Greiner}, {Krabbe}, {Fischer}, {Klein} \&
  {Ibanez}}{{Caratti o Garatti} et~al.}{2017}]{caratti_nature_2016}
{Caratti o Garatti} A.,  {Stecklum} B.,  {Garcia Lopez} R.,  {Eisloffel} J.,
  {Ray} T.~P.,  {Sanna} A.,  {Cesaroni} R.,  {Walmsley} C.~M.,  {Oudmaijer}
  R.~D.,  {de Wit} W.~J.,  {Moscadelli} L.,  {Greiner} J.,  {Krabbe} A.,
  {Fischer} C.,  {Klein} R.,    {Ibanez} J.~M.,  2017, Nature Physics, 13, 276

\bibitem[\protect\citeauthoryear{{Caratti o Garatti}, {Stecklum}, {Linz},
  {Garcia Lopez} \& {Sanna}}{{Caratti o Garatti}
  et~al.}{2015}]{caratti_aa_573_2015}
{Caratti o Garatti} A.,  {Stecklum} B.,  {Linz} H.,  {Garcia Lopez} R.,
  {Sanna} A.,  2015, \aap, 573, A82

\bibitem[\protect\citeauthoryear{{Cesaroni}, {Galli}, {Lodato}, {Walmsley} \&
  {Zhang}}{{Cesaroni} et~al.}{2006}]{cesaroni_natur_444_2006}
{Cesaroni} R.,  {Galli} D.,  {Lodato} G.,  {Walmsley} M.,    {Zhang} Q.,  2006,
  \nat, 444, 703

\bibitem[\protect\citeauthoryear{{Cesaroni}, {Hofner}, {Araya} \&
  {Kurtz}}{{Cesaroni} et~al.}{2010}]{cesaroni_aa_509_2010}
{Cesaroni} R.,  {Hofner} P.,  {Araya} E.,    {Kurtz} S.,  2010, \aap, 509, A50

\bibitem[\protect\citeauthoryear{{Cesaroni}, {Moscadelli}, {Neri}, {Sanna},
  {Caratti o Garatti}, {Eisloffel}, {Stecklum}, {Ray} \& {Walmsley}}{{Cesaroni}
  et~al.}{2018}]{cesaroni_aa_612_2018}
{Cesaroni} R.,  {Moscadelli} L.,  {Neri} R.,  {Sanna} A.,  {Caratti o Garatti}
  A.,  {Eisloffel} J.,  {Stecklum} B.,  {Ray} T.,    {Walmsley} C.~M.,  2018,
  \aap, 612, A103

\bibitem[\protect\citeauthoryear{{Cesaroni}, {S{\'a}nchez-Monge},
  {Beltr{\'a}n}, {Johnston}, {Maud}, {Moscadelli} \& {Mottram}}{{Cesaroni}
  et~al.}{2017}]{cesaroni_aa_602_2017}
{Cesaroni} R.,  {S{\'a}nchez-Monge} {\'A}.,  {Beltr{\'a}n} M.~T.,  {Johnston}
  K.~G.,  {Maud} L.~T.,  {Moscadelli} L.,    {Mottram} J.~C.,  2017, \aap, 602,
  A59

\bibitem[\protect\citeauthoryear{{Chen}, {Ren}, {Zhang}, {Shen} \&
  {Qiu}}{{Chen} et~al.}{2017}]{chen_apj_835_2017}
{Chen} X.,  {Ren} Z.,  {Zhang} Q.,  {Shen} Z.,    {Qiu} K.,  2017, \apj, 835,
  227

\bibitem[\protect\citeauthoryear{{Chini}, {Hoffmeister}, {Nasseri}, {Stahl} \&
  {Zinnecker}}{{Chini} et~al.}{2012}]{chini_424_mnras_2012}
{Chini} R.,  {Hoffmeister} V.~H.,  {Nasseri} A.,  {Stahl} O.,    {Zinnecker}
  H.,  2012, \mnras, 424, 1925

\bibitem[\protect\citeauthoryear{{Commer{\c c}on}, {Hennebelle} \&
  {Henning}}{{Commer{\c c}on} et~al.}{2011}]{commercon_apj_742_2011}
{Commer{\c c}on} B.,  {Hennebelle} P.,    {Henning} T.,  2011, \apjl, 742, L9

\bibitem[\protect\citeauthoryear{{Commer{\c c}on}, {Teyssier}, {Audit},
  {Hennebelle} \& {Chabrier}}{{Commer{\c c}on}
  et~al.}{2011}]{commercon_aa_529_2011}
{Commer{\c c}on} B.,  {Teyssier} R.,  {Audit} E.,  {Hennebelle} P.,
  {Chabrier} G.,  2011, \aap, 529, A35

\bibitem[\protect\citeauthoryear{{Cunningham}, {Moeckel} \&
  {Bally}}{{Cunningham} et~al.}{2009}]{Cunningham_apj_692_2009}
{Cunningham} N.~J.,  {Moeckel} N.,    {Bally} J.,  2009, \apj, 692, 943

\bibitem[\protect\citeauthoryear{{Flock}, {Fromang}, {Gonz{\'a}lez} \&
  {Commer{\c c}on}}{{Flock} et~al.}{2013}]{flock_aa_560_2013}
{Flock} M.,  {Fromang} S.,  {Gonz{\'a}lez} M.,    {Commer{\c c}on} B.,  2013,
  \aap, 560, A43

\bibitem[\protect\citeauthoryear{{Forgan}, {Ilee}, {Cyganowski}, {Brogan} \&
  {Hunter}}{{Forgan} et~al.}{2016}]{forgan_mnras_463_2016}
{Forgan} D.~H.,  {Ilee} J.~D.,  {Cyganowski} C.~J.,  {Brogan} C.~L.,
  {Hunter} T.~R.,  2016, \mnras, 463, 957

\bibitem[\protect\citeauthoryear{{Fuente}, {Neri}, {Mart{\'{\i}}n-Pintado},
  {Bachiller}, {Rodr{\'{\i}}guez-Franco} \& {Palla}}{{Fuente}
  et~al.}{2001}]{fuente_aa_366_2001}
{Fuente} A.,  {Neri} R.,  {Mart{\'{\i}}n-Pintado} J.,  {Bachiller} R.,
  {Rodr{\'{\i}}guez-Franco} A.,    {Palla} F.,  2001, \aap, 366, 873

\bibitem[\protect\citeauthoryear{{Fujisawa}, {Yonekura}, {Sugiyama},
  {Horiuchi}, {Hayashi}, {Hachisuka}, {Matsumoto} \& {Niinuma}}{{Fujisawa}
  et~al.}{2015}]{fujisawa_atel_2015}
{Fujisawa} K.,  {Yonekura} Y.,  {Sugiyama} K.,  {Horiuchi} H.,  {Hayashi} T.,
  {Hachisuka} K.,  {Matsumoto} N.,    {Niinuma} K.,  2015, The Astronomer's
  Telegram, 8286

\bibitem[\protect\citeauthoryear{{Gammie}}{{Gammie}}{1996}]{gammie_apj_462_1996}
{Gammie} C.~F.,  1996, \apj, 462, 725

\bibitem[\protect\citeauthoryear{{Ginsburg}, {Bally}, {Goddi}, {Plambeck} \&
  {Wright}}{{Ginsburg} et~al.}{2018}]{2018arXiv180410622G}
{Ginsburg} A.,  {Bally} J.,  {Goddi} C.,  {Plambeck} R.,    {Wright} M.,  2018,
  \apj, 860, 119

\bibitem[\protect\citeauthoryear{{Goedhart}, {Gaylard} \& {van der
  Walt}}{{Goedhart} et~al.}{2003}]{goedhart_mnras_339_2003}
{Goedhart} S.,  {Gaylard} M.~J.,    {van der Walt} D.~J.,  2003, \mnras, 339,
  L33

\bibitem[\protect\citeauthoryear{{Haemmerl{\'e}}, {Eggenberger}, {Meynet},
  {Maeder} \& {Charbonnel}}{{Haemmerl{\'e}}
  et~al.}{2016}]{haemmerle_585_aa_2016}
{Haemmerl{\'e}} L.,  {Eggenberger} P.,  {Meynet} G.,  {Maeder} A.,
  {Charbonnel} C.,  2016, \aap, 585, A65

\bibitem[\protect\citeauthoryear{{Haemmerl{\'e}}, {Eggenberger}, {Meynet},
  {Maeder}, {Charbonnel} \& {Klessen}}{{Haemmerl{\'e}}
  et~al.}{2017}]{haemmerle_602_aap_2017}
{Haemmerl{\'e}} L.,  {Eggenberger} P.,  {Meynet} G.,  {Maeder} A.,
  {Charbonnel} C.,    {Klessen} R.~S.,  2017, \aap, 602, A17

\bibitem[\protect\citeauthoryear{{Haemmerl{\'e}} \& {Peters}}{{Haemmerl{\'e}}
  \& {Peters}}{2016}]{haemmerle_458_mnras_2016}
{Haemmerl{\'e}} L.,  {Peters} T.,  2016, \mnras, 458, 3299

\bibitem[\protect\citeauthoryear{{Harries}}{{Harries}}{2015}]{harries_mnras_448_2015}
{Harries} T.~J.,  2015, \mnras, 448, 3156

\bibitem[\protect\citeauthoryear{{Harries}, {Douglas} \& {Ali}}{{Harries}
  et~al.}{2017}]{harries_2017}
{Harries} T.~J.,  {Douglas} T.~A.,    {Ali} A.,  2017, \mnras, 471, 4111

\bibitem[\protect\citeauthoryear{{Hirano}, {Hosokawa}, {Yoshida} \&
  {Kuiper}}{{Hirano} et~al.}{2017}]{Hirano_2017Science}
{Hirano} S.,  {Hosokawa} T.,  {Yoshida} N.,    {Kuiper} R.,  2017, Science,
  357, 1375

\bibitem[\protect\citeauthoryear{{Hosokawa}, {Hirano}, {Kuiper}, {Yorke},
  {Omukai} \& {Yoshida}}{{Hosokawa} et~al.}{2016}]{hosokawa_2015}
{Hosokawa} T.,  {Hirano} S.,  {Kuiper} R.,  {Yorke} H.~W.,  {Omukai} K.,
  {Yoshida} N.,  2016, \apj, 824, 119

\bibitem[\protect\citeauthoryear{{Hosokawa} \& {Omukai}}{{Hosokawa} \&
  {Omukai}}{2009}]{hosokawa_apj_691_2009}
{Hosokawa} T.,  {Omukai} K.,  2009, \apj, 691, 823

\bibitem[\protect\citeauthoryear{{Hunter}, {Brogan}, {MacLeod}, {Cyganowski},
  {Chandler}, {Chibueze}, {Friesen}, {Indebetouw}, {Thesner} \&
  {Young}}{{Hunter} et~al.}{2017}]{2017arXiv170108637H}
{Hunter} T.~R.,  {Brogan} C.~L.,  {MacLeod} G.,  {Cyganowski} C.~J.,
  {Chandler} C.~J.,  {Chibueze} J.~O.,  {Friesen} R.,  {Indebetouw} R.,
  {Thesner} C.,    {Young} K.~H.,  2017, \apjl, 837, L29

\bibitem[\protect\citeauthoryear{{Hunter}, {Brogan}, {MacLeod}, {Cyganowski},
  {Chibueze}, {Friesen}, {Hirota}, {Smits}, {Chandler} \&
  {Indebetouw}}{{Hunter} et~al.}{2018}]{2018arXiv180102141H}
{Hunter} T.~R.,  {Brogan} C.~L.,  {MacLeod} G.~C.,  {Cyganowski} C.~J.,
  {Chibueze} J.~O.,  {Friesen} R.,  {Hirota} T.,  {Smits} D.~P.,  {Chandler}
  C.~J.,    {Indebetouw} R.,  2018, \apj, 854, 170

\bibitem[\protect\citeauthoryear{{Ilee}, {Cyganowski}, {Nazari}, {Hunter},
  {Brogan}, {Forgan} \& {Zhang}}{{Ilee} et~al.}{2016}]{ilee_mnras_462_2016}
{Ilee} J.~D.,  {Cyganowski} C.~J.,  {Nazari} P.,  {Hunter} T.~R.,  {Brogan}
  C.~L.,  {Forgan} D.~H.,    {Zhang} Q.,  2016, \mnras, 462, 4386

\bibitem[\protect\citeauthoryear{{Johnston}, {Robitaille}, {Beuther}, {Linz},
  {Boley}, {Kuiper}, {Keto}, {Hoare} \& {van Boekel}}{{Johnston}
  et~al.}{2015}]{johnston_apj_813_2015}
{Johnston} K.~G.,  {Robitaille} T.~P.,  {Beuther} H.,  {Linz} H.,  {Boley} P.,
  {Kuiper} R.,  {Keto} E.,  {Hoare} M.~G.,    {van Boekel} R.,  2015, \apjl,
  813, L19

\bibitem[\protect\citeauthoryear{{Keto} \& {Wood}}{{Keto} \&
  {Wood}}{2006}]{keto_apj_637_2006}
{Keto} E.,  {Wood} K.,  2006, \apj, 637, 850

\bibitem[\protect\citeauthoryear{{Klassen}, {Pudritz}, {Kuiper}, {Peters} \&
  {Banerjee}}{{Klassen} et~al.}{2016}]{klassen_apj_823_2016}
{Klassen} M.,  {Pudritz} R.~E.,  {Kuiper} R.,  {Peters} T.,    {Banerjee} R.,
  2016, \apj, 823, 28

\bibitem[\protect\citeauthoryear{{Kobulnicky}, {Kiminki}, {Lundquist}, {Burke},
  {Chapman}, {Keller}, {Lester}, {Rolen}, {Topel}, {Bhattacharjee}, {Smullen},
  {Vargas {\'A}lvarez}, {Runnoe}, {Dale} \& {Brotherton}}{{Kobulnicky}
  et~al.}{2014}]{2014ApJS..213...34K}
{Kobulnicky} H.~A.,  {Kiminki} D.~C.,  {Lundquist} M.~J.,  {Burke} J.,
  {Chapman} J.,  {Keller} E.,  {Lester} K.,  {Rolen} E.~K.,  {Topel} E.,
  {Bhattacharjee} A.,  {Smullen} R.~A.,  {Vargas {\'A}lvarez} C.~A.,  {Runnoe}
  J.~C.,  {Dale} D.~A.,    {Brotherton} M.~M.,  2014, \apjs, 213, 34

\bibitem[\protect\citeauthoryear{{Kolb}, {Stute}, {Kley} \& {Mignone}}{{Kolb}
  et~al.}{2013}]{kolb_aa_559_2013}
{Kolb} S.~M.,  {Stute} M.,  {Kley} W.,    {Mignone} A.,  2013, \aap, 559, A80

\bibitem[\protect\citeauthoryear{{Kratter}, {Matzner}, {Krumholz} \&
  {Klein}}{{Kratter} et~al.}{2010}]{kratter_apj_708_2010}
{Kratter} K.~M.,  {Matzner} C.~D.,  {Krumholz} M.~R.,    {Klein} R.~I.,  2010,
  \apj, 708, 1585

\bibitem[\protect\citeauthoryear{{Kraus}, {Kluska}, {Kreplin}, {Bate},
  {Harries}, {Hofmann}, {Hone}, {Monnier}, {Weigelt}, {Anugu}, {de Wit} \&
  {Wittkowski}}{{Kraus} et~al.}{2017}]{kraus_apj_835_2017}
{Kraus} S.,  {Kluska} J.,  {Kreplin} A.,  {Bate} M.,  {Harries} T.~J.,
  {Hofmann} K.-H.,  {Hone} E.,  {Monnier} J.~D.,  {Weigelt} G.,  {Anugu} A.,
  {de Wit} W.~J.,    {Wittkowski} M.,  2017, \apjl, 835, L5

\bibitem[\protect\citeauthoryear{{Krumholz}, {Klein} \& {McKee}}{{Krumholz}
  et~al.}{2007}]{krumholz_apj_656_2007}
{Krumholz} M.~R.,  {Klein} R.~I.,    {McKee} C.~F.,  2007, \apj, 656, 959

\bibitem[\protect\citeauthoryear{{Langer}}{{Langer}}{2012}]{langer_araa_50_2012}
{Langer} N.,  2012, \araa, 50, 107

\bibitem[\protect\citeauthoryear{{Lodato} \& {Clarke}}{{Lodato} \&
  {Clarke}}{2011}]{lodato_mnras_413_2011}
{Lodato} G.,  {Clarke} C.~J.,  2011, \mnras, 413, 2735

\bibitem[\protect\citeauthoryear{{Machida} \& {Matsumoto}}{{Machida} \&
  {Matsumoto}}{2011}]{machida_mnras_413_2011}
{Machida} M.~N.,  {Matsumoto} T.,  2011, \mnras, 413, 2767

\bibitem[\protect\citeauthoryear{{MacLeod}, {Smits}, {Goedhart}, {Hunter},
  {Brogan}, {Chibueze}, {van den Heever}, {Thesner}, {Banda} \&
  {Paulsen}}{{MacLeod} et~al.}{2018}]{macleoad_MNRAS_478_2018}
{MacLeod} G.~C.,  {Smits} D.~P.,  {Goedhart} S.,  {Hunter} T.~R.,  {Brogan}
  C.~L.,  {Chibueze} J.~O.,  {van den Heever} S.~P.,  {Thesner} C.~J.,  {Banda}
  P.~J.,    {Paulsen} J.~D.,  2018, \mnras, 478, 1077

\bibitem[\protect\citeauthoryear{{Mahy}, {Rauw}, {De Becker}, {Eenens} \&
  {Flores}}{{Mahy} et~al.}{2013}]{2013A&A...550A..27M}
{Mahy} L.,  {Rauw} G.,  {De Becker} M.,  {Eenens} P.,    {Flores} C.~A.,  2013,
  \aap, 550, A27

\bibitem[\protect\citeauthoryear{{Matsumoto}, {Machida} \&
  {Inutsuka}}{{Matsumoto} et~al.}{2017}]{matsumoto_apj_839_2017}
{Matsumoto} T.,  {Machida} M.~N.,    {Inutsuka} S.-i.,  2017, \apj, 839, 69

\bibitem[\protect\citeauthoryear{{Maud}, {Hoare}, {Galv{\'a}n-Madrid}, {Zhang},
  {de Wit}, {Keto}, {Johnston} \& {Pineda}}{{Maud}
  et~al.}{2017}]{maud_467_mnras_2017}
{Maud} L.~T.,  {Hoare} M.~G.,  {Galv{\'a}n-Madrid} R.,  {Zhang} Q.,  {de Wit}
  W.~J.,  {Keto} E.,  {Johnston} K.~G.,    {Pineda} J.~E.,  2017, \mnras, 467,
  L120

\bibitem[\protect\citeauthoryear{{Meru} \& {Bate}}{{Meru} \&
  {Bate}}{2010}]{meru_mnras_406_2010}
{Meru} F.,  {Bate} M.~R.,  2010, \mnras, 406, 2279

\bibitem[\protect\citeauthoryear{{Meyer}, {Kuiper}, {Kley}, {Johnston} \&
  {Vorobyov}}{{Meyer} et~al.}{2018}]{meyer_mnras_473_2018}
{Meyer} D.~M.-A.,  {Kuiper} R.,  {Kley} W.,  {Johnston} K.~G.,    {Vorobyov}
  E.,  2018, \mnras, 473, 3615

\bibitem[\protect\citeauthoryear{{Meyer}, {Langer}, {Mackey}, {Vel{\'a}zquez}
  \& {Gusdorf}}{{Meyer} et~al.}{2015}]{meyer_mnras_450_2015}
{Meyer} D.~M.-A.,  {Langer} N.,  {Mackey} J.,  {Vel{\'a}zquez} P.~F.,
  {Gusdorf} A.,  2015, \mnras, 450, 3080

\bibitem[\protect\citeauthoryear{{Meyer}, {Mackey}, {Langer}, {Gvaramadze},
  {Mignone}, {Izzard} \& {Kaper}}{{Meyer} et~al.}{2014}]{meyer}
{Meyer} D.~M.-A.,  {Mackey} J.,  {Langer} N.,  {Gvaramadze} V.~V.,  {Mignone}
  A.,  {Izzard} R.~G.,    {Kaper} L.,  2014, \mnras, 444, 2754

\bibitem[\protect\citeauthoryear{{Meyer}, {Vorobyov}, {Kuiper} \&
  {Kley}}{{Meyer} et~al.}{2017}]{meyer_mnras_464_2017}
{Meyer} D.~M.-A.,  {Vorobyov} E.~I.,  {Kuiper} R.,    {Kley} W.,  2017, \mnras,
  464, L90

\bibitem[\protect\citeauthoryear{{Mignone}, {Bodo}, {Massaglia}, {Matsakos},
  {Tesileanu}, {Zanni} \& {Ferrari}}{{Mignone}
  et~al.}{2007}]{mignone_apj_170_2007}
{Mignone} A.,  {Bodo} G.,  {Massaglia} S.,  {Matsakos} T.,  {Tesileanu} O.,
  {Zanni} C.,    {Ferrari} A.,  2007, \apjs, 170, 228

\bibitem[\protect\citeauthoryear{{Mignone}, {Zanni}, {Tzeferacos}, {van
  Straalen}, {Colella} \& {Bodo}}{{Mignone}
  et~al.}{2012}]{migmone_apjs_198_2012}
{Mignone} A.,  {Zanni} C.,  {Tzeferacos} P.,  {van Straalen} B.,  {Colella} P.,
     {Bodo} G.,  2012, \apjs, 198, 7

\bibitem[\protect\citeauthoryear{{Moscadelli}, {Sanna}, {Goddi}, {Walmsley},
  {Cesaroni}, {Caratti o Garatti}, {Stecklum}, {Menten} \&
  {Kraus}}{{Moscadelli} et~al.}{2017}]{moscadelli_aa_600_2017}
{Moscadelli} L.,  {Sanna} A.,  {Goddi} C.,  {Walmsley} M.~C.,  {Cesaroni} R.,
  {Caratti o Garatti} A.,  {Stecklum} B.,  {Menten} K.~M.,    {Kraus} A.,
  2017, \aap, 600, L8

\bibitem[\protect\citeauthoryear{{Nayakshin} \& {Lodato}}{{Nayakshin} \&
  {Lodato}}{2012}]{nayakshin_mnras_426_2012}
{Nayakshin} S.,  {Lodato} G.,  2012, \mnras, 426, 70

\bibitem[\protect\citeauthoryear{{Peters}, {Banerjee}, {Klessen}, {Mac Low},
  {Galv{\'a}n-Madrid} \& {Keto}}{{Peters} et~al.}{2010}]{peters_apj_711_2010}
{Peters} T.,  {Banerjee} R.,  {Klessen} R.~S.,  {Mac Low} M.-M.,
  {Galv{\'a}n-Madrid} R.,    {Keto} E.~R.,  2010, \apj, 711, 1017

\bibitem[\protect\citeauthoryear{{Purser}, {Lumsden}, {Hoare} \&
  {Cunningham}}{{Purser} et~al.}{2018}]{purser_mnras_475_2018}
{Purser} S.~J.~D.,  {Lumsden} S.~L.,  {Hoare} M.~G.,    {Cunningham} N.,  2018,
  \mnras, 475, 2

\bibitem[\protect\citeauthoryear{{Purser}, {Lumsden}, {Hoare}, {Urquhart},
  {Cunningham}, {Purcell}, {Brooks}, {Garay}, {G{\'u}zman} \&
  {Voronkov}}{{Purser} et~al.}{2016}]{purser_mnras_460_2016}
{Purser} S.~J.~D.,  {Lumsden} S.~L.,  {Hoare} M.~G.,  {Urquhart} J.~S.,
  {Cunningham} N.,  {Purcell} C.~R.,  {Brooks} K.~J.,  {Garay} G.,
  {G{\'u}zman} A.~E.,    {Voronkov} M.~A.,  2016, \mnras, 460, 1039

\bibitem[\protect\citeauthoryear{{Reg{\'a}ly} \& {Vorobyov}}{{Reg{\'a}ly} \&
  {Vorobyov}}{2017}]{regaly_aa_601_2017}
{Reg{\'a}ly} Z.,  {Vorobyov} E.,  2017, \aap, 601, A24

\bibitem[\protect\citeauthoryear{{Reiter}, {Kiminki}, {Smith} \&
  {Bally}}{{Reiter} et~al.}{2017}]{reiter_mnras_470_2017}
{Reiter} M.,  {Kiminki} M.~M.,  {Smith} N.,    {Bally} J.,  2017, \mnras, 470,
  4671

\bibitem[\protect\citeauthoryear{{Rice}, {Lodato} \& {Armitage}}{{Rice}
  et~al.}{2005}]{rice_mnras_364_2005}
{Rice} W.~K.~M.,  {Lodato} G.,    {Armitage} P.~J.,  2005, \mnras, 364, L56

\bibitem[\protect\citeauthoryear{{Richling} \& {Yorke}}{{Richling} \&
  {Yorke}}{1997}]{richling_aa_327_1997}
{Richling} S.,  {Yorke} H.~W.,  1997, \aap, 327, 317

\bibitem[\protect\citeauthoryear{{Rogers} \& {Wadsley}}{{Rogers} \&
  {Wadsley}}{2012}]{roger_mnras_423_2012}
{Rogers} P.~D.,  {Wadsley} J.,  2012, \mnras, 423, 1896

\bibitem[\protect\citeauthoryear{{Rosen}, {Krumholz}, {McKee} \&
  {Klein}}{{Rosen} et~al.}{2016}]{rosen_mnras_463_2016}
{Rosen} A.~L.,  {Krumholz} M.~R.,  {McKee} C.~F.,    {Klein} R.~I.,  2016,
  \mnras, 463, 2553

\bibitem[\protect\citeauthoryear{{Rosen}, {Krumholz}, {Oishi}, {Lee} \&
  {Klein}}{{Rosen} et~al.}{2017}]{rosen_jcp_330_2017}
{Rosen} A.~L.,  {Krumholz} M.~R.,  {Oishi} J.~S.,  {Lee} A.~T.,    {Klein}
  R.~I.,  2017, Journal of Computational Physics, 330, 924

\bibitem[\protect\citeauthoryear{{Samal}, {Chen}, {Takami}, {Jose} \&
  {Froebrich}}{{Samal} et~al.}{2018}]{2018arXiv180311413S}
{Samal} M.~R.,  {Chen} W.~P.,  {Takami} M.,  {Jose} J.,    {Froebrich} D.,
  2018, \mnras, 477, 4577

\bibitem[\protect\citeauthoryear{{Sana}, {de Mink}, {de Koter}, {Langer},
  {Evans}, {Gieles}, {Gosset}, {Izzard}, {Le Bouquin} \& {Schneider}}{{Sana}
  et~al.}{2012}]{sana_sci_337_2012}
{Sana} H.,  {de Mink} S.~E.,  {de Koter} A.,  {Langer} N.,  {Evans} C.~J.,
  {Gieles} M.,  {Gosset} E.,  {Izzard} R.~G.,  {Le Bouquin} J.-B.,
  {Schneider} F.~R.~N.,  2012, Science, 337, 444

\bibitem[\protect\citeauthoryear{{Schneider}, {Izzard}, {Langer} \& {de
  Mink}}{{Schneider} et~al.}{2015}]{schneider_apj_805_2015}
{Schneider} F.~R.~N.,  {Izzard} R.~G.,  {Langer} N.,    {de Mink} S.~E.,  2015,
  \apj, 805, 20

\bibitem[\protect\citeauthoryear{{Scholz}, {Froebrich} \& {Wood}}{{Scholz}
  et~al.}{2013}]{2013MNRAS.430.2910S}
{Scholz} A.,  {Froebrich} D.,    {Wood} K.,  2013, \mnras, 430, 2910

\bibitem[\protect\citeauthoryear{{Seifried}, {Banerjee}, {Klessen}, {Duffin} \&
  {Pudritz}}{{Seifried} et~al.}{2011}]{seifried_mnras_417_2011}
{Seifried} D.,  {Banerjee} R.,  {Klessen} R.~S.,  {Duffin} D.,    {Pudritz}
  R.~E.,  2011, \mnras, 417, 1054

\bibitem[\protect\citeauthoryear{{Sobolev}, {Cragg} \& {Godfrey}}{{Sobolev}
  et~al.}{1997}]{sobolev_aa_324_1997}
{Sobolev} A.~M.,  {Cragg} D.~M.,    {Godfrey} P.~D.,  1997, \aap, 324, 211

\bibitem[\protect\citeauthoryear{{Stecklum}, {Caratti o Garatti}, {Cardenas},
  {Greiner}, {Kruehler}, {Klose} \& {Eisloeffel}}{{Stecklum}
  et~al.}{2016}]{stecklum_ATel_2016}
{Stecklum} B.,  {Caratti o Garatti} A.,  {Cardenas} M.~C.,  {Greiner} J.,
  {Kruehler} T.,  {Klose} S.,    {Eisloeffel} J.,  2016, The Astronomer's
  Telegram, 8732

\bibitem[\protect\citeauthoryear{{Stecklum}, {Caratti o Garatti}, {Hodapp},
  {Linz}, {Moscadelli} \& {Sanna}}{{Stecklum} et~al.}{2018}]{stecklum_2017b}
{Stecklum} B.,  {Caratti o Garatti} A.,  {Hodapp} K.,  {Linz} H.,  {Moscadelli}
  L.,    {Sanna} A.,  2018, in {Tarchi} A.,  {Reid} M.~J.,   {Castangia} P.,
  eds, Astrophysical Masers: Unlocking the Mysteries of the Universe Vol.~336
  of IAU Symposium, {Infrared variability, maser activity, and accretion of
  massive young stellar objects}.
pp 37--40

\bibitem[\protect\citeauthoryear{{Stecklum}, {Heese}, {Wolf}, {Garatti},
  {Ibanez} \& {Linz}}{{Stecklum} et~al.}{2017}]{stecklum_2017a}
{Stecklum} B.,  {Heese} S.,  {Wolf} S.,  {Garatti} A.~C.~o.,  {Ibanez} J.~M.,
   {Linz} H.,  2017, ArXiv e-prints

\bibitem[\protect\citeauthoryear{{Szymczak}, {Olech}, {Sarniak}, {Wolak} \&
  {Bartkiewicz}}{{Szymczak} et~al.}{2018}]{szymczak_mnras_474_2018}
{Szymczak} M.,  {Olech} M.,  {Sarniak} R.,  {Wolak} P.,    {Bartkiewicz} A.,
  2018, \mnras, 474, 219

\bibitem[\protect\citeauthoryear{{Szymczak}, {Olech}, {Wolak}, {G{\'e}rard} \&
  {Bartkiewicz}}{{Szymczak} et~al.}{2018}]{2018arXiv180707334S}
{Szymczak} M.,  {Olech} M.,  {Wolak} P.,  {G{\'e}rard} E.,    {Bartkiewicz} A.,
   2018, \aap, 617, A80

\bibitem[\protect\citeauthoryear{{Tanaka}, {Tan} \& {Zhang}}{{Tanaka}
  et~al.}{2017}]{tanaka_apj_835_2017}
{Tanaka} K.~E.~I.,  {Tan} J.~C.,    {Zhang} Y.,  2017, \apj, 835, 32

\bibitem[\protect\citeauthoryear{{Tapia}, {Persi}, {Bohigas}, {Roth} \&
  {G{\'o}mez}}{{Tapia} et~al.}{2006}]{tapia_mnras_367_2006}
{Tapia} M.,  {Persi} P.,  {Bohigas} J.,  {Roth} M.,    {G{\'o}mez} M.,  2006,
  \mnras, 367, 513

\bibitem[\protect\citeauthoryear{{Tapia}, {Roth} \& {Persi}}{{Tapia}
  et~al.}{2015a}]{tapia_mnras_448_2015}
{Tapia} M.,  {Roth} M.,    {Persi} P.,  2015a, \mnras, 448, 1402

\bibitem[\protect\citeauthoryear{{Tapia}, {Roth} \& {Persi}}{{Tapia}
  et~al.}{2015b}]{tapia_mnras_446_2015}
{Tapia} M.,  {Roth} M.,    {Persi} P.,  2015b, \mnras, 446, 4088

\bibitem[\protect\citeauthoryear{{Testi}}{{Testi}}{2003}]{testi_2003}
{Testi} L.,  2003, in {De Buizer} J.~M.,  {van der Bliek} N.~S.,  eds, Galactic
  Star Formation Across the Stellar Mass Spectrum Vol.~287 of Astronomical
  Society of the Pacific Conference Series, {Intermediate Mass Stars (Invited
  Review)}.
pp 163--173

\bibitem[\protect\citeauthoryear{{Truelove}, {Klein}, {McKee}, {Holliman} II,
  {Howell}, {Greenough} \& {Woods}}{{Truelove}
  et~al.}{1998}]{truelove_apj_495_1998}
{Truelove} J.~K.,  {Klein} R.~I.,  {McKee} C.~F.,  {Holliman} II J.~H.,
  {Howell} L.~H.,  {Greenough} J.~A.,    {Woods} D.~T.,  1998, \apj, 495, 821

\bibitem[\protect\citeauthoryear{{Vaidya}, {Fendt}, {Beuther} \&
  {Porth}}{{Vaidya} et~al.}{2011}]{vaidya_apj_742_2011}
{Vaidya} B.,  {Fendt} C.,  {Beuther} H.,    {Porth} O.,  2011, \apj, 742, 56

\bibitem[\protect\citeauthoryear{{Vorobyov}}{{Vorobyov}}{2010}]{vorobyov_apj_713_2010}
{Vorobyov} E.~I.,  2010, \apj, 713, 1059

\bibitem[\protect\citeauthoryear{{Vorobyov} \& {Basu}}{{Vorobyov} \&
  {Basu}}{2005}]{vorobyov_apj_633_2005}
{Vorobyov} E.~I.,  {Basu} S.,  2005, \apjl, 633, L137

\bibitem[\protect\citeauthoryear{{Vorobyov} \& {Basu}}{{Vorobyov} \&
  {Basu}}{2015}]{vorobyov_apj_805_2015}
{Vorobyov} E.~I.,  {Basu} S.,  2015, \apj, 805, 115

\bibitem[\protect\citeauthoryear{{Vorobyov}, {Elbakyan}, {Hosokawa}, {Sakurai},
  {Guedel} \& {Yorke}}{{Vorobyov} et~al.}{2017}]{vorobyov_aa_605_2017}
{Vorobyov} E.~I.,  {Elbakyan} V.,  {Hosokawa} T.,  {Sakurai} Y.,  {Guedel} M.,
    {Yorke} H.,  2017, \aap, 605, A77

\bibitem[\protect\citeauthoryear{{Vorobyov} \& {Elbakyan}}{{Vorobyov} \&
  {Elbakyan}}{2018}]{2018arXiv180607675V}
{Vorobyov} E.~I.,  {Elbakyan} V.~G.,  2018, \aap, 618, A7

\bibitem[\protect\citeauthoryear{{Vorobyov}, {Elbakyan}, {Plunkett}, {Dunham},
  {Audard}, {Guedel} \& {Dionatos}}{{Vorobyov}
  et~al.}{2018}]{2018arXiv180106707V}
{Vorobyov} E.~I.,  {Elbakyan} V.~G.,  {Plunkett} A.~L.,  {Dunham} M.~M.,
  {Audard} M.,  {Guedel} M.,    {Dionatos} O.,  2018, \aap, 613, A18

\bibitem[\protect\citeauthoryear{{Wang}, {Beuther}, {Bik}, {Vasyunina},
  {Jiang}, {Puga}, {Linz}, {Rod{\'o}n}, {Henning} \& {Tamura}}{{Wang}
  et~al.}{2011}]{wang_aa_527_2011}
{Wang} Y.,  {Beuther} H.,  {Bik} A.,  {Vasyunina} T.,  {Jiang} Z.,  {Puga} E.,
  {Linz} H.,  {Rod{\'o}n} J.~A.,  {Henning} T.,    {Tamura} M.,  2011, \aap,
  527, A32

\bibitem[\protect\citeauthoryear{{Yorke} \& {Sonnhalter}}{{Yorke} \&
  {Sonnhalter}}{2002}]{2002ApJ...569..846Y}
{Yorke} H.~W.,  {Sonnhalter} C.,  2002, \apj, 569, 846

\bibitem[\protect\citeauthoryear{{Zhao}, {Caselli}, {Li} \&
  {Krasnopolsky}}{{Zhao} et~al.}{2018}]{zhao_mnras_473_2018}
{Zhao} B.,  {Caselli} P.,  {Li} Z.-Y.,    {Krasnopolsky} R.,  2018, \mnras,
  473, 4868

\bibitem[\protect\citeauthoryear{{Zinchenko}, {Liu}, {Su}, {Salii}, {Sobolev},
  {Zemlyanukha}, {Beuther}, {Ojha}, {Samal} \& {Wang}}{{Zinchenko}
  et~al.}{2015}]{zinchenko_apj_810_2015}
{Zinchenko} I.,  {Liu} S.-Y.,  {Su} Y.-N.,  {Salii} S.~V.,  {Sobolev} A.~M.,
  {Zemlyanukha} P.,  {Beuther} H.,  {Ojha} D.~K.,  {Samal} M.~R.,    {Wang} Y.,
   2015, \apj, 810, 10

\bibitem[\protect\citeauthoryear{{Zinchenko}, {Liu}, {Su} \&
  {Sobolev}}{{Zinchenko} et~al.}{2017}]{zinchenko_aa_606_2017}
{Zinchenko} I.,  {Liu} S.-Y.,  {Su} Y.-N.,    {Sobolev} A.~M.,  2017, \aap,
  606, L6

\bibitem[\protect\citeauthoryear{{Zinchenko}, {Liu}, {Su} \&
  {Zemlyanukha}}{{Zinchenko} et~al.}{2018}]{2018IAUS..332..270Z}
{Zinchenko} I.,  {Liu} S.-Y.,  {Su} Y.-N.,    {Zemlyanukha} P.,  2018, in
  {Cunningham} M.,  {Millar} T.,   {Aikawa} Y.,  eds, IAU Symposium Vol.~332 of
  IAU Symposium, {Multiline observations of S255IR with ALMA}.
pp 270--273

\end{thebibliography}
}


\end{document}